\documentclass[onecolumn,
nofootinbib,
 amsmath,amssymb,
 aps,
notitlepage,prd
]{revtex4-2}

\usepackage[margin=1.5in]{geometry}
\usepackage{graphicx}
\usepackage{dcolumn}
\usepackage{bm}

\usepackage{amsmath}
\usepackage{amsthm}
\usepackage{tensor}
\usepackage{hyperref}
\usepackage{color}
\usepackage{soul}
\setstcolor{red}
\usepackage[normalem]{ulem}
\usepackage{float}

\newtheorem{theorem}{Theorem}

\begin{document}

\preprint{}

\title{On turbulence for spacetimes with stable trapping}

\author{Gabriele Benomio}
\email{gabriele.benomio@gssi.it}
\affiliation{Gran Sasso Science Institute, Viale Francesco Crispi 7, L'Aquila (AQ), 67100, Italy}

\author{Alejandro C\'ardenas-Avenda\~no}
\email{cardenas-avendano@lanl.gov}
\affiliation{Computational Physics and Methods (CCS-2) \& Center for Nonlinear Studies (CNLS), Los Alamos National Laboratory, Los Alamos NM 87545, U.S.A.}
\affiliation{Department of Physics \& Princeton Gravity Initiative, Princeton University, Washington Road, Princeton NJ 08544, U.S.A.}

\author{Frans Pretorius}
\email{fpretori@princeton.edu}
\affiliation{Department of Physics \& Princeton Gravity Initiative, Princeton University, Washington Road, Princeton NJ 08544, U.S.A.}

\author{Andrew Sullivan}
\affiliation{Department of Physics \& Princeton Gravity Initiative, Princeton University, Washington Road, Princeton NJ 08544, U.S.A.}


\begin{abstract}
\vspace{0.5cm}
Motivated by understanding the nonlinear gravitational dynamics of spacetimes admitting \emph{stably} trapped null geodesics, such as ultracompact objects and black string solutions to general relativity, we explore the dynamics of nonlinear scalar waves on a simple (fixed) model geometry with stable trapping.~More specifically, we consider the time evolution of solutions to the cubic (defocusing) wave equation on a four-dimensional static, spherically symmetric, and asymptotically flat (horizonless) spacetime admitting a stable photon sphere.

Unlike the known results for scalar waves on spacetimes with \emph{un}stable trapping, our study shows fundamental differences between \emph{linear} and \emph{nonlinear} scalar dynamics.~The local energy, as well as all local higher-order energies, of solutions to the \emph{linear} wave equation on our model spacetime can be rigorously proven to remain uniformly bounded and to decay uniformly in time.~However, due to the presence of stable trapping, the uniform decay rate is \emph{slow}.~To help elucidate how the slow linear decay affects solutions to the \emph{nonlinear} wave equation considered, we examine numerical solutions of the latter, restricting to axisymmetric initial data in this work.~In contrast to the linear dynamics, we exhibit a family of nonlinear solutions with {\em turbulent behaviour}.~Within the region of stable trapping, the slow linear decay allows local \emph{higher-order} energies of the nonlinear solution to {\em grow} over the time interval that we numerically evolve.~The growth is induced by a \emph{direct energy cascade}:~Beginning with initial data containing a small number of low-order multipole modes, a spectrum of high-order multipole modes are populated in time by the nonlinear interactions and eventually dominate over the low-modes in the evolution.

That the system exhibits a direct cascade limits the time over which our numerical scheme can provide convergent solutions for the short-wavelength structure that develops, and hence we can only speculate what this intermediate-time dynamics implies for the nonlinear stability of the motivating spacetimes in general relativity.~Nevertheless, we provide a heuristic argument suggesting that, if a similar behaviour occurs for gravitational wave perturbations of these spacetimes, it would likely {\em not} generically lead to black hole or singularity formation.
\end{abstract}

\maketitle

\newpage

\tableofcontents

\section{Introduction} \label{sec_intro}

The dynamics of nonlinear waves is a classical theme in the mathematical study of partial differential equations. From the physical perspective, nonlinear wave equations describe a vast and diverse range of phenomena \cite{Courant_Hilbert_pde_book}, whose domains include fluid dynamics, continuum mechanics, elasticity theory, nonlinear optics, and plasma physics, and often serve as models for more complicated physical systems. In the present paper, we consider nonlinear waves on a fixed, curved spacetime geometry as a model for the gravitational wave dynamics of certain classes of solutions to the Einstein equations. The latter are notoriously challenging to analyse, both rigorously and numerically, and there is a long history of attempting to gain insights into their putative dynamics using simpler nonlinear wave models.

\medskip

The motivating spacetimes for this work are \emph{ultracompact objects} in four-dimensional, asymptotically flat spacetimes and higher-dimensional black hole spacetimes with a compact extra-dimension known as \emph{black strings}. By an ultracompact object, we refer to any linearly stable, horizonless soliton or star-like solution to the relevant non-vacuum Einstein equations that is sufficiently compact to possess a region of \emph{stably} trapped \emph{null} geodesics. In spherical symmetry, this requires the surface of the star (or the majority of its energy, if it does not have a well-defined surface) to be within an areal radius of $\sim3M$, with $M$ being the ADM mass of the spacetime. Such compact objects need to be composed of ``exotic'' matter (ordinary matter at nuclear densities, as occurs in neutron stars, does not suffice), hypothetical examples of which include certain kinds of boson stars~\cite{Ruffini:1969qy,Colpi:1986ye}, black shells~\cite{Danielsson:2017riq}, fuzzballs~\cite{Mathur:2005zp}, or gravastars~\cite{Mazur:2004fk} (see~\cite{Cardoso:2019rvt} for a review with a comprehensive list of candidates, and current observational constraints on their existence). Black strings and their geometry will be described in Section \ref{sec_intro_motivation_strings}, where their relation to the present work is also discussed. As with ultracompact objects, black strings possess a region of \emph{stably} trapped \emph{null} geodesics~\cite{Benomio_thesis_doi_2, Benomio_Rings}, including those black strings with sufficiently small compact extra-dimension so as to be free of the long-wavelength Gregory--Laflamme instability~\cite{GLinstability}.

\medskip

Stable null geodesic trapping fundamentally alters the dispersive properties of spacetime. In regions where such trapping occurs, the underlying geometry inhibits the standard energy decay, allowing nonlinear effects to persist and potentially trigger instabilities and richer phenomena. The nonlinear scalar waves studied in this paper are designed to replicate these key features, thereby providing an effective model for the dynamics of gravitational wave perturbations in spacetimes exhibiting stable trapping.

\medskip

In the following Sections \ref{sec_intro_linear_waves} and \ref{sec_intro_nonlinear_waves}, we describe some aspects of the dynamics of both linear and nonlinear waves in the presence of null geodesic trapping. These sections provide the relevant context for the numerical results presented in this paper, which are outlined in Section \ref{sec_intro_this_work}. In Section \ref{sec_intro_motivation_strings}, we elaborate on the motivation for the present work in connection to the dynamics of black strings. Section \ref{sec_intro_outline} concludes the introduction and provides an outline of the bulk of the paper.

\subsection{Linear waves in the presence of trapping} \label{sec_intro_linear_waves}

The dynamics of scalar waves is very informative, already at the linear level, of the interaction between the geometry of a spacetime and the dispersion of its perturbations. In particular, geometric properties of null geodesics of the spacetime, such as the possible presence of \emph{trapped} null geodesics, play a key role.

\medskip

Provided that solutions to the linear wave equation\footnote{For the present introduction, the reader can think of solutions to the wave equation arising from smooth, compactly supported initial data. We also remark that we always consider the \emph{massless} wave equation.} 
\begin{equation} \label{eqn_intro_LW}
\Box_g\phi=0
\end{equation}
can be shown to decay uniformly in time on a given (fixed) stationary spacetime, if trapped null geodesics are present, then the uniform decay \emph{rate} may crucially depend on the nature of trapping.~\emph{Unstable} trapping is more favorable for dispersion, and has been shown, in various cases, to allow for \emph{fast} (e.g., polynomial or exponential) decay.~Instances include linear waves on Schwarzschild(-de Sitter) and sub-extremal Kerr(-de Sitter, slowly rotating) black hole exteriors \cite{FullKerr, Dyatlov_decay_linear_waves_kerr_deSitter}, which uniformly decay with polynomial (exponential) rates.~On the other hand, one expects \emph{slow} decay in the presence of \emph{stable} trapping.~Indeed, high-frequency linear waves localised in the proximity of a stably trapped null geodesic at initial time remain confined by an effective potential well for long timescales, eventually (only slowly) dispersing through the potential barrier (the property allowing for tunneling in quantum mechanics).~As a result, the evolution of such localised waves on these timescales is well approximated by time-periodic functions (i.e.,~\emph{quasimodes}, localised around a stably trapped null geodesic).~For several cases of stationary spacetimes possessing stable trapping, quasimode constructions have been employed to prove that linear waves cannot uniformly decay faster than logarithmically in time \cite{SharpLogHolz, KeirLogLog, Benomio_Rings, Kunduri_slow_decay_solitons} (see also \cite{MicroKeir, Idelon-Riton_Lower_Bound_Dirac} for related results).

\subsection{Nonlinear waves in the presence of trapping} \label{sec_intro_nonlinear_waves}

Typically, fast enough decay of linear waves is necessary to prove global existence and uniform decay of \emph{small-data} solutions to nonlinear wave equations with \emph{derivative} nonlinearities, i.e.,
\begin{equation} \label{eqn_intro_NLW_deriv_nonlinearity}
    \Box_g\phi=\mathcal{N}(\phi, \partial\phi) \, .
\end{equation}
For example, on Schwarzschild(-de Sitter) and Kerr(-de Sitter, slowly rotating) black hole exteriors, works \cite{Luk_Nonlinear_Wave_Kerr, DHRT_quasilinear_waves_full_kerr, Hintz_Vasy_quasilinear_waves_Kerr_deSitter} exploit the fast linear decay mentioned in Section \ref{sec_intro_linear_waves} to prove uniform decay for small-data solutions to nonlinear wave equations of the form \eqref{eqn_intro_NLW_deriv_nonlinearity}, as well as for all their (higher-order) derivatives and local (higher-order) energies. In these works, the derivative nonlinearities satisfy a version of the so-called \emph{null condition},\footnote{The null condition was originally introduced as a sufficient condition to prove small-data global existence for nonlinear waves on Minkowski space \cite{Klainerman_null_cond, Christ_null_cond}. Nonlinear wave equations with derivative nonlinearities which do \emph{not} satisfy the null condition may admit, already on Minkowski space, solutions arising from arbitrarily small (smooth, compactly supported) initial data which blow-up in finite time \cite{John_blow_up_quasilinear_wave_eqns}.} which, in the case of \emph{quadratic} derivative nonlinearities of the form
\begin{equation} \label{eqn_intro_NLW_null_condition}
\Box_g\phi= \mathcal{Q}(\partial\phi) \, ,
\end{equation}
may be viewed as modelling the nonlinear structure of the Einstein equations.\footnote{The Einstein equations possess quadratic derivative nonlinearities in the metric $g$, corresponding to terms of the form $g\cdot(\partial g)^2$. In appropriate gauges, such nonlinearities satisfy a variant of the null condition.} More recent works \cite{DHRT, Klainerman_Szeftel_Kerr_small_a_1, StabKerrdS} demonstrate that the nonlinear scalar dynamics of the model equation \eqref{eqn_intro_NLW_null_condition} is indeed inherited by small gravitational wave perturbations of these black hole spacetimes as solutions to the full Einstein equations.

\medskip

When the linear decay is slow, proving small-data global existence of nonlinear solutions may become problematic. In particular, linear logarithmic decay does \emph{not} suffice for the known nonlinear arguments to yield small-data global existence of solutions to nonlinear wave equations of the form \eqref{eqn_intro_NLW_null_condition}. As a result, the rigorous global analysis of equations of the form \eqref{eqn_intro_NLW_null_condition} on spacetimes admitting stable trapping (as, for instance, on the spacetimes considered in \cite{SharpLogHolz, KeirLogLog, Benomio_Rings, Kunduri_slow_decay_solitons}) remains, to large extent, \emph{terra incognita}. The study of the dynamics of spacetimes with stable trapping as solutions to the Einstein equations is even more challenging and its possible outcomes even more unclear.

\medskip

Of particular relevance for the present work will be another class of nonlinear wave equations, namely those with power nonlinearities
\begin{equation} \label{eqn_intro_NLW_general_p}
\Box_g\phi=\pm \phi|\phi|^{p-1} 
\end{equation}
with $p>1$, which, in the context of general relativity, arise as the equations of motion for the Einstein--nonlinear scalar field system
\begin{gather}
    R_{\mu\nu}-\frac{1}{2}Rg_{\mu\nu}=2T_{\mu\nu} \, , \label{intro_Einstein_scalar_field}\\
    T_{\mu\nu}=\partial_{\mu}\phi\partial_{\nu}\phi-\frac{1}{2}g_{\mu\nu}\left(g^{\alpha\beta}\partial_{\alpha}\phi\partial_{\beta}\phi\right) \mp g_{\mu\nu}\frac{|\phi|^{p+1}}{p+1} \, . \nonumber
\end{gather}
When considered on a (fixed) stationary spacetime, a fundamental property of equation \eqref{eqn_intro_NLW_general_p} is that the energy of its solutions is \emph{conserved} over time. The fact that this property plays an important role in the phenomenology of the equation is already apparent in the special case of the cubic (defocusing) wave equation\footnote{The term \emph{defocusing} refers to the choice of sign of the cubic nonlinearity. We note that, in our convention, the signature of the metric $g$ is $(-,+,\dots,+)$.} ($p=3$)
\begin{equation} \label{eqn_intro_NLW}
\Box_g\phi=\phi|\phi|^2 
\end{equation}
on a static spacetime, as for instance on the Schwarzschild black hole exterior. In this setting, the energy is moreover a coercive quantity,\footnote{Meaning that control over the energy yields control over the individual terms of the form $(\partial\phi)^2$ which appear in the energy. This property relies on the defocusing sign of the nonlinearity, which forces the energy to be a sum of positive terms.} and its conservation turns out to immediately imply global existence of \emph{large-data} solutions.\footnote{In addition to the conservation and coercivity of the energy, the specific power of the nonlinearity in \eqref{eqn_intro_NLW} (i.e., $p=3$) and a standard Sobolev embedding are also exploited.} In fact, the global existence of solutions does \emph{not} rely on linear decay, and is indeed already established in the early works \cite{Bachelot_Nicolas_nonlinear_KG_schwarzschild, Nicolas_nonlinear_KG_schwarzschild}. In particular, the unstable nature of trapping on the Schwarzschild exterior plays no role. Fast linear decay, on the other hand, has to be exploited to prove uniform decay of (large-data) solutions, as well as for all their associated higher-order quantities (see later works \cite{Blue_Soffer_semilinear_waves_Schwarzschild_large_data, Blue_Soffer_semilinear_waves_Schwarzschild}).

\medskip

In addition to its connection with the equation of motion of the system \eqref{intro_Einstein_scalar_field} (for $p=3$), one can view the nonlinear wave equation \eqref{eqn_intro_NLW} as an even simpler model for the nonlinear terms $g\cdot(\partial g)^2$ of the Einstein equations, in which the quadratic derivative nonlinearity is replaced by a quadratic nonlinearity in the solution.

\subsection{Stable trapping as a source of turbulence} \label{sec_intro_this_work}

The aim of this paper is to shed light on the dynamics of nonlinear waves in the presence of stable trapping. To this end, we consider a model static, spherically symmetric and asymptotically flat spacetime $(\mathcal{M},g)$ in $3+1$-dimensions, with
\begin{align*}
\mathcal{M} & \cong \mathbb{R}^{3+1} \, , & g&=-f(r)\,dt^2+f^{-1}(r)\, dr^2+r^2\,(d\vartheta^2+\sin^2\vartheta\,d\varphi^2)
\end{align*}
and smooth function $f(r)>0$ such that the spacetime admits a \emph{stable} photon sphere (see Section \ref{sec_model_spacetime}, and note that no black hole or ergoregion are present). The spirit of our choice of model spacetime is that of considering the simplest possible geometry which isolates the stable trapping feature and moreover allows for rigorous results on the uniform dynamics of \emph{linear} waves. Indeed, as an application of the general result \cite{KeirLogLog}, uniform decay of smooth solutions to the linear wave equation \eqref{eqn_intro_LW} on our spacetime can be proven, with sharp-logarithmic uniform decay rate (see Section \ref{sec_lin_we}). The slow decay already affects \emph{axisymmetric} solutions, whereas spherically symmetric solutions are not effectively confined and decay faster (see the related remarks of Section \ref{sec_lin_solution_proto_data}).

\medskip

As a nonlinear wave model, we consider the cubic wave equation \eqref{eqn_intro_NLW} on $(\mathcal{M},g)$. Exploiting the fact that the energy of solutions
\begin{align*}
\mathbb{E}_{\text{nl}}[\phi](\tau)=\int_{t=\tau} &dr \, d\vartheta \, d\varphi \, \frac{r^2\sin\vartheta}{2 \, f(r)} \\
&\times \left[  (\partial_t\phi)^2+f^2(r)\, (\partial_r\phi)^2 +\frac{f(r)}{r^2}\,(\partial_{\vartheta}\phi)^2+\frac{f(r)}{r^2\sin^2\vartheta}\,(\partial_{\varphi}\phi)^2+f(r)\,\frac{|\phi|^4}{2}\right]  
\end{align*}
is a coercive quantity and is conserved over time, one can establish global existence of \emph{large-data} smooth solutions. The conservation and coercivity of the energy also imply that first derivatives of solutions remain uniformly bounded in time, whereas they are compatible with \emph{higher-order} derivatives (and \emph{higher-order} energies) of solutions growing in time (see Section \ref{sec_nonlin_we} for further details).

\medskip

In this work, we present numerical solutions to \eqref{eqn_intro_NLW} arising from a family of compactly supported initial data which are (see Section \ref{sec_proto_data} for further details)
\begin{itemize}
\item \emph{axisymmetric}, with the symmetry being then propagated in the evolution by the equation \ref{eqn_intro_NLW},
\item \emph{radially localised} around a stably trapped null geodesic,
\item supported on a \emph{small number of low-order multipole modes}.
\end{itemize}
For such initial data, the numerical solutions develop the following \textbf{turbulent dynamics}: Over the time interval that we numerically evolve, we observe the \emph{growth of higher-order derivatives} (and \emph{energies}) of solutions within the region of stable trapping (Section \ref{sec_growth_derivatives}). The growth is driven by a \emph{direct angular-mode} (and \emph{energy}) \emph{cascade}, where higher-order multiple modes are populated in time by the nonlinear interactions and eventually dominate the low-modes in the evolution (Section \ref{sec_mode_cascade}). Our numerical results moreover suggest the persistence of the turbulent behaviour for our class of initial data when one also assumes that the dynamical quantities that are numerically evolved are initially \emph{small} (Section \ref{sec_small_data}).

\medskip

One can expect the above kind of turbulent dynamics to occur in a non-linear wave equation that is confined, such as on a compact, finite geometry. The novelty of our setup is we can interpret the turbulent dynamics as being induced by an {\em effective} confinement of solutions within a small, localized region of an infinite domain due to the presence of stable trapping there. The slow dispersion through the potential barrier favours nonlinear angular-mode interactions, which result in the excitement of higher-order (and more effectively trapped) angular modes. The energy of the solution, initially supported on the low-order angular modes contained in the initial data, flows to the higher modes, which eventually dominate in the evolution as the low modes more quickly disperse. Since the effective confinement only occurs if the solutions are supported on (possibly axisymmetric) angular modes, capturing it necessarily requires a (at least) $2+1$-dimensional numerical scheme. For all spherically symmetric solutions to the equation, the growth of higher-order derivatives (and energies) is absent.

\medskip

The turbulent dynamics that we observe numerically is in stark contrast with the linear dynamics on our model spacetime, and with the discussed decay results for the same nonlinear wave equation \eqref{eqn_intro_NLW} in the presence of \emph{un}stable trapping. On the other hand, it relates to a number of both rigorous and numerical results for nonlinear wave (and dispersive) equations in the presence of (perfect or effective) confinement (see Section \ref{sec_discussion} for more details).

\medskip

One may expect that the turbulent dynamics observed in this work is a phenomenon affecting general spacetimes with stable trapping, and which would persist for \emph{gravitational} perturbations of our motivating spacetimes (i.e., ultracompact objects and black strings). The genericity and possible endstates of the turbulent gravitational dynamics would remain open questions. Nonetheless, for small gravitational perturbations of our motivating spacetimes, we already speculate that the turbulent cascade, although possibly generic, would likely {\em not} generically lead to the formation of naked singularities or (new) black holes. See Section \ref{sec_discussion} for a discussion.

\subsection{Toward turbulent black strings} \label{sec_intro_motivation_strings}

Our motivation for this work comes primarily from the study of the dynamics of ultracompact objects and black strings. For what concerns the latter spacetimes, one may view the numerical results presented in this work as pointing toward an intriguing, yet to be explored, role of turbulence in higher dimensional general relativity, as we shall describe in this section.

\medskip

\emph{Black strings} are families of higher dimensional black holes \cite{Horowitz_Strominger_strings_branes}. Static black string exterior spacetimes\footnote{We restrict here to the simplest example of such spacetimes. Similar considerations apply to static \emph{black branes}.} $$(\mathcal{M},g_{r_+,L})$$ are obtained by taking the product of Schwarzschild exterior spacetimes and a compact extra-dimension, i.e.,
\begin{gather*}
\mathcal{M} \cong (-\infty,\infty)_t \times [r_+,\infty)_r \times \mathbb{S}^2_{(\vartheta,\varphi)} \times \mathbb{S}^1_{\psi} \, , \\
g_{r_+,L} =-f(r)\,dt^2+f^{-1}(r) \, dr^2 +r^2(d\vartheta^2+\sin^2\vartheta\, d\varphi^2)+L^2 d\psi^2
\end{gather*}
with $f(r)=1-r_+/r$, and form a two-parameter family of solutions to the five-dimensional vacuum Einstein equations (with zero cosmological constant). The spacetime parameters $r_+$ and $L$ are positive and correspond to the event horizon radius and the size of the extra-dimension respectively. Black strings are asymptotically Kaluza--Klein spacetimes, meaning that their asymptotic structure is that of the product of Minkowski space and a compact extra-dimension.

\medskip

It is well-known that the linear gravitational dynamics of black strings is strongly affected by the choice of spacetime parameters. For parameters $L/r_+\geq c_{\star}$ with $c_{\star}$ some critical constant (i.e., size of the extra-dimension sufficiently \emph{large} as compared to the horizon radius), black strings suffer from the celebrated Gregory--Laflamme instability \cite{GLinstability}, which arises as the existence of exponentially growing (fixed-frequency) mode-solutions
\begin{align}
h_{\mu\nu}(t,r,\psi)&=e^{\omega t} H_{\mu\nu}(r,\psi) \, ,  & \omega &\in\mathbb{R}_+
\end{align}
to the linearised vacuum Einstein equations (see the recent work \cite{Collingbourne_GL} for a rigorous proof). The instability and its (both linear and nonlinear) dynamics have been the subject of a vast literature \cite{Gubser_mitra_black_string_1, Gubser_mitra_black_string_2, Hovd_Myers_String_Ring_GL, Reall_stability_black_branes, Figueras_murata_reall_penrose_inequalities, Hollands_wald_black_branes}, with some numerical works showing that its nonlinear dynamics may generically lead to a violation of the weak cosmic censorship conjecture \cite{Choptuik_et_al_black_string, Garfinkle_et_al_black_string, Lehner_Pret_Instab_String}. On the other hand, the linear gravitational dynamics of black strings with parameters $L/r_+ < c_{\star}$ (i.e., size of the extra-dimension sufficiently \emph{small} as compared to the horizon radius) does \emph{not} appear to suffer from the Gregory--Laflamme instability, and is therefore widely expected to be \emph{stable}.\footnote{No mathematically rigorous result is however known concerning the linear stability (already at the level of \emph{mode} stability) of black strings within the parameter range $L/r_+ < c_{\star}$ as solutions to the linearised vacuum Einstein equations.} Such an expectation is often extended to the nonlinear gravitational dynamics of these spacetimes, which would effectively resemble the four-dimensional dynamics of Schwarzschild black holes. We argue that the present work questions the validity of this latter expectation, already for nonlinear \emph{scalar} perturbations. To explain why, we shall first describe some aspects of the \emph{scalar} dynamics of black strings.

\medskip

By the standard Fourier decomposition
\begin{equation*}
    \phi(t,r,\vartheta,\varphi,\psi)= \frac{1}{\sqrt{2\pi}}\int_{-\infty}^{\infty}\sum_{\ell\geq 0, |m|\leq \ell} \, \sum_{|k|\geq 0}  e^{-i\omega t} \, \mathfrak{u}(r) \, Y^{\ell}_{m}(\vartheta,\varphi) \, e^{ik\psi} d\omega \, ,
\end{equation*}
of solutions to the linear wave equation on black strings, with $\omega\in\mathbb{R}$, $\ell\in\mathbb{N}_0$ and $m,k\in\mathbb{Z}_0$, the wave equation can be fully separated and reduced to the radial o.d.e.
\begin{equation*}
    \mathfrak{u}^{\prime\prime}(r)+\left(\omega^2-V_{\text{string}}(r) \right) \mathfrak{u}(r)=0
\end{equation*}
with non-negative radial potential
\begin{equation} \label{radial_potential_string}
    V_{\text{string}}(r)=\left( 1-\frac{r_+}{r}\right)\left( \frac{\ell(\ell+1)}{r^2}+\frac{k^2}{L^2} +\frac{r_+}{r^3}\right) \, .
\end{equation}
For \emph{any} choice of spacetime parameters $r_+,L$, there exist frequencies
\begin{equation*}
1\ll k^2 < \left(\frac{L^2}{3 \, r_+^2}\right)\ell(\ell+1) 
\end{equation*}
such that the potential \eqref{radial_potential_string} admits a local \emph{minimum} (together with a local maximum, located at smaller radius). In the high-frequency limit
\begin{equation*}
    1\ll \omega^2 \sim \ell(\ell+1) \sim k^2 \, ,
\end{equation*}
this property of the potential \eqref{radial_potential_string} encodes the existence of \emph{stably} trapped null geodesics on black string spacetimes, marking a fundamental difference between black string and Schwarzschild geometries. We emphasise that, to experience the effective confinement due to stable trapping, linear waves necessarily need non-trivial dependence on the angular coordinate associated to the extra-dimension \emph{and} (at least) one of the angular coordinates on $\mathbb{S}^2$,\footnote{In other words, spherically symmetric and $\mathbb{S}^1_{\psi}$-symmetric linear waves do \emph{not} experience stable trapping.} e.g., $\phi=\phi(t,r,\vartheta,\psi)$.

\medskip

As discussed, the linear dynamics of scalar waves is sensitive to the presence of stable trapping. Upcoming work by the first author proves that, for any choice of spacetime parameters, uniform logarithmic decay holds for scalar linear waves on black strings. In view of \cite{Benomio_Rings}, the logarithmic uniform decay rate has to be sharp, thus implying that, in analogy with the present work, uniform decay of linear waves is \emph{slow}.\footnote{Slow uniform decay is expected to also characterize the linear \emph{gravitational} dynamics of the (putatively) linearly stable black strings within the parameter range $L/r_+ < c_{\star}$.}

\medskip

Future numerical work by the authors will investigate the \emph{nonlinear} dynamics of black strings in a novel direction:
\begin{itemize}
\item As a first step, we will study numerical solutions to nonlinear wave equations on fixed black string spacetimes. We will employ an analogous model wave equation and numerical scheme as those introduced in this work. As an effect of the slow linear decay due to stable trapping, we expect that the turbulent dynamics described in this work will also occur for all black strings, \emph{regardless of the choice of spacetime parameters}. Some important differences and additional technical difficulties will however arise. First, the background spacetime will possess an event horizon and a different asymptotic structure, which are to be dealt with in the numerical implementation. Moreover, since a certain non-trivial dependence on the angular coordinates is necessary for scalar waves to be affected by the presence of stable trapping, the numerical scheme will have to be augmented with (at least) one more spatial dimension, and thus be implemented in (at least) $3+1$-dimensions.\footnote{As in the present work, turbulence would already arise for a family of $\mathbb{S}^1_{\varphi}$-symmetric initial data. Spherically symmetric and $\mathbb{S}^1_{\psi}$-symmetric solutions would \emph{not} suffer from the turbulent behaviour. Similarly, spherically symmetric and $\mathbb{S}^1_{\psi}$-symmetric gravitational perturbations are \emph{not} expected to exhibit a turbulent dynamics.}
\item As a second step, we will examine the possible turbulent dynamics of black strings as solutions to the five-dimensional vacuum Einstein equations. To this end, we will consider black strings which are \emph{not} affected by the Gregory--Laflamme instability (i.e., with \emph{small} parameter ratio $L/r_+$), but whose gravitational perturbations may nonetheless suffer from the turbulent dynamics.\footnote{We point out that the turbulent dynamics would be expected to occur for \emph{all} black strings. For black strings also affected by the Gregory--Laflamme instability, the Gregory--Laflamme and turbulent dynamics would coexist, making the features of the latter more difficult to isolate.} As for scalar perturbations, the gravitational problem will require a $3+1$-dimensional numerical scheme, now for the full vacuum Einstein equations. The numerical implementation will constitute one of the main challenges of the problem, and will require to go beyond the previously developed numerical schemes to evolve the Gregory--Laflamme dynamics of black strings (for which $2+1$ dimensions are sufficient). The global gravitational dynamics of turbulent black strings would also be an interesting problem to consider in future work. See also the related discussion in Section \ref{sec_discussion_GW}.
\end{itemize}

The turbulent behaviour of nonlinear gravitational perturbations of black strings would constitute a dynamical signature of the presence of a (arbitrarily) small extra-dimension. A similar scenario as the one described for black strings may arise for more general Kaluza--Klein black holes. In this sense, \textbf{the turbulence of black strings would question the more general idea that black holes with small extra-dimensions exhibit an effectively four-dimensional gravitational dynamics}.

\subsection{Outline of the paper} \label{sec_intro_outline}

The remainder of the paper is organized as follows. In Section \ref{sec_model_spacetime}, we introduce our model spacetime and its properties. In Section \ref{sec_lin_we}, we recall the relevant mathematical results for linear waves on our model spacetime. In Section \ref{sec_nonlin_we}, we introduce the model nonlinear wave equation. Section \ref{sec_numerical_implementation} describes the numerical implementation of the problem and the scheme adopted to evolve numerical solutions to the nonlinear wave equation. Section \ref{sec_numerical_results} is the main section of the paper and presents our numerical results. In Section \ref{sec_discussion}, we further discuss our results and their relation with the existing literature. The Appendices \ref{app:ICs} and \ref{app:convergence} contain supplementary material which complements the bulk of the paper.

\section{The model spacetime} \label{sec_model_spacetime}

\begin{figure}
    \includegraphics[scale=.6]{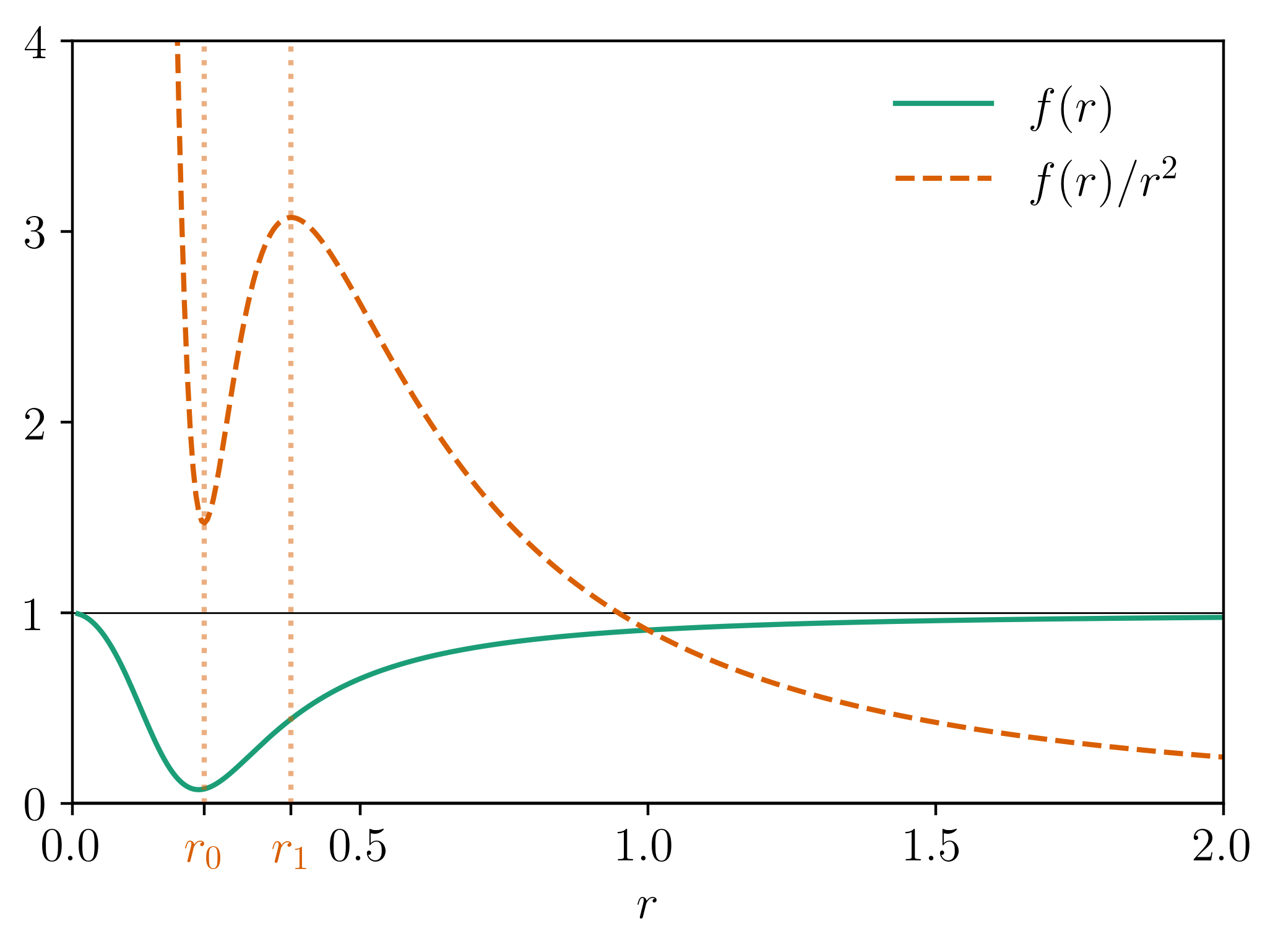}
    \caption{We choose the scalar function $f(r)= 1-r^2(0.026+11.21\, r^4)^{-1}$ (solid green line). The choice of $f(r)$ is such that the local minimum of the geodesic potential $V(r)=f(r)r^{-2}$ (dashed orange line) is located sufficiently close to the origin (i.e., $r_0\sim 0.22$) to guarantee the necessary numerical resolution within the stable trapping region (see Section \ref{sec_numerical_implementation}).}
    \label{fig:finfo}
\end{figure}

We consider the ambient manifold $$\mathcal{M} \cong \mathbb{R}^{3+1}$$ equipped with standard time and spherical coordinates
\begin{align*}
t&\in (-\infty,\infty) \, , &    r&\in (0,\infty) \, , & (\vartheta,\varphi)&\in \mathbb{S}^2 \, . 
\end{align*}
We introduce the Lorentzian metric
\begin{equation} \label{metric_study}
 g=-f(r)\,dt^2+f^{-1}(r)\, dr^2+r^2\,(d\vartheta^2+\sin^2\vartheta\,d\varphi^2) 
\end{equation}
on $\mathcal{M}$, with smooth scalar function $f(r)$ such that
\begin{itemize}
    \item $f(r)$ extends to an everywhere smooth (including at the coordinate origin) spacetime function, with $f(0)=1$.
    \item $0<f(r)\leq 1$ for any $r> 0$. Moreover, $f(r)$ satisfies the limit $f(r)\rightarrow 1$ as $r\rightarrow \infty$.
    \item $f(r)r^{-2}$ possesses a local \emph{minimum} at $r=r_0$ for some $r_0>0$.
\end{itemize}
The explicit choice of $f(r)$ employed in this work is stated and depicted in Figure \ref{fig:finfo}. The spacetime $(\mathcal{M},g)$ is smooth, static, spherically symmetric and asymptotically flat. Moreover, no black hole (horizon or) region and no ergoregion (or ergosurface) are present.

\medskip

For any null geodesic
\begin{align*} 
    \gamma:[0,\infty) & \rightarrow \mathcal{M} \, ,\\
    \lambda &\mapsto \gamma(\lambda)=\left(t(\lambda),r(\lambda),\vartheta(\lambda),\varphi(\lambda)\right)
\end{align*}
of the spacetime, one can choose angular coordinates on the foliation spheres such that the geodesic is initially tangent to the equatorial plane,\footnote{Without loss of generality, we assume that such angular coordinates coincide with the angular coordinates $(\vartheta,\varphi)$ appearing in the metric \eqref{metric_study}.} i.e., 
\begin{align*}
    \vartheta(0)&=\frac{\pi}{2} \, , & \frac{d\vartheta}{d\lambda} (0)&=0 \, .
\end{align*}
Relative to such coordinates, the geodesic equation implies that $\gamma$ remains confined to the equatorial plane (i.e., $\vartheta(\lambda)=\pi/2$ for all $\lambda$) and satisfies the radial o.d.e. 
\begin{equation*}
\left(\frac{dr}{d\lambda}\right)^2+m^2V(r) =E^2 \, ,  
\end{equation*}
with associated conserved (for all $\lambda$) quantities 
\begin{align*}
  E&=f(r)\frac{dt}{d\lambda} \, , &  m&=r^2\frac{d\varphi}{d\lambda} 
\end{align*}
and radial potential $V(r)=f(r)r^{-2}$ (see Figure \ref{fig:finfo}). By virtue of our prescription of $f(r)$, the geodesic potential $V(r)$ possesses \emph{two} extremal points, namely a local \emph{minimum} (at $r=r_0$) and a local maximum (at $r=r_{1}$, with $r_1>r_0$). The local minimum (resp., maximum) corresponds to the existence of a null geodesic which is initially tangent to the timelike hypersurface
\begin{align*}
    \mathfrak{P}_{r_0}&=\mathbb{R}_t\times \mathbb{S}^2_{t,r_0} \, ,  & \text{(resp., $\mathfrak{P}_{r_1}$}&\text{$=\mathbb{R}_t\times \mathbb{S}^2_{t,r_1}$)}
\end{align*}
and remains tangent to $\mathfrak{P}_{r_0}$ (resp., $\mathfrak{P}_{r_1}$), i.e., with a circular orbit $r(\lambda)=r_0$ (resp., $r(\lambda)=r_1$) for all $\lambda$. Recalling that the original choice of angular coordinates was arbitrary, one may view $\mathfrak{P}_{r_0}$ and $\mathfrak{P}_{r_1}$ as \emph{photon spheres} of the spacetime, meaning that \emph{all} null geodesics which are initially tangent to $\mathfrak{P}_{r_0}$ (resp., $\mathfrak{P}_{r_1}$) remain tangent to $\mathfrak{P}_{r_0}$ (resp., $\mathfrak{P}_{r_1}$). The photon sphere $\mathfrak{P}_{r_0}$ is \emph{stable}, whereas the photon sphere $\mathfrak{P}_{r_1}$ is unstable.

\medskip

The spacetime $(\mathcal{M},g)$ is the model spacetime considered for the remainder of this work.\footnote{For the sake of our later arguments, one could instead think of considering a smooth, static, spherically symmetric spacetime which admits stably trapped null geodesics and exactly coincides with Minkowski space outside of a spherical shell. However, our choice of model spacetime is more convenient for the numerical implementation.}

\section{Linear waves on the model spacetime} \label{sec_lin_we}

We consider the scalar linear wave equation
\begin{equation} \label{WE_th}
    \Box_g\phi=0
\end{equation}
for scalar functions $\phi:\mathcal{M}\rightarrow \mathbb{R}$ on the model spacetime $(\mathcal{M},g)$ of Section \ref{sec_model_spacetime}. 

\medskip

Let $k\in\mathbb{N}$, with $k\geq 1$. We introduce the \emph{$k$-th order energies}
\begin{align}
\mathbb{E}^k[\phi](\tau)=& \, \int_{t=\tau} dr \, d\vartheta \, d\varphi \, \frac{r^2\sin\vartheta}{2 \, f(r)} \label{energy_k_lin_waves}\\
&\times \sum_{0\leq|\alpha|\leq k-1}\left[  (\partial_t\partial^{\alpha}\phi)^2+f^2(r)\, (\partial_r\partial^{\alpha}\phi)^2 +\frac{f(r)}{r^2}\,(\partial_{\vartheta}\partial^{\alpha}\phi)^2+\frac{f(r)}{r^2\sin^2\vartheta}\,(\partial_{\varphi}\partial^{\alpha}\phi)^2\right]  \, , \nonumber
\end{align}
where we use the standard multi-index notation for $\alpha$. The order $k$ of the energy coincides with the order of the top-order derivatives of the solution appearing in the energy. We will denote $\mathbb{E}^1[\phi](t)=\mathbb{E}[\phi](t)$, with
\begin{align}
\mathbb{E}[\phi](\tau)=\int_{t=\tau} &dr \, d\vartheta \, d\varphi \, \frac{r^2\sin\vartheta}{2 \, f(r)} \label{energy_lin_waves}\\
&\times \left[  (\partial_t\phi)^2+f^2(r)\, (\partial_r\phi)^2 +\frac{f(r)}{r^2}\,(\partial_{\vartheta}\phi)^2+\frac{f(r)}{r^2\sin^2\vartheta}\,(\partial_{\varphi}\phi)^2\right]  \, . \nonumber
\end{align}
For any $t$-invariant set $\Omega\subset \mathcal{M}$ such that $\Omega\cap \left\lbrace t=\tau \right\rbrace$ is non-empty and bounded (for all $\tau\in\mathbb{R}$), we introduce the $k$-th order \emph{local}  energies $\mathbb{E}^k_{\Omega}[\phi](t)$, for which the integral \eqref{energy_k_lin_waves} is now taken over the bounded region $\Omega\cap \left\lbrace t=\tau \right\rbrace$. We will denote $\mathbb{E}^1_{\Omega}[\phi](t)=\mathbb{E}_{\Omega}[\phi](t)$.

\medskip

The following theorem summarises the uniform dynamics of linear waves on $(\mathcal{M},g)$. The set $\Omega$ invoked in the theorem is assumed to satisfy the properties described above.

\medskip

\begin{theorem} \label{th_linear_waves}
For any $k\in\mathbb{N}$, with $k\geq 1$, and any set $\Omega$, there exist real constants $B_k,C_{\Omega , k}>0$ such that, for any smooth,\footnote{The statement continues to hold for initial data of finite regularity. In particular, the theorem holds for the (finite-regularity) initial data considered in Section \ref{sec_proto_data}.} compactly supported initial data $(\phi,\partial_t\phi)|_{t=0}$ prescribed at $t=0$, the smooth solution $\phi$ to \eqref{WE_th} satisfies the following properties:
\begin{itemize}
    \item (Global existence) The solution $\phi$ exists and is unique for all $t\geq 0$.
    \item (Conservation of energy) The equality $$\mathbb{E}[\phi](t)=\mathbb{E}[\phi](0)$$ holds for all $t\geq 0$.
    \item (Uniform energy boundedness) The inequality
    \begin{equation} \label{unif_bddness_linear_waves}
         \mathbb{E}^k[\phi](t)\leq B_{k} \, \mathbb{E}^k[\phi](0)
    \end{equation}
    holds for all $t\geq 0$.
    \item (Uniform local energy decay, \cite{KeirLogLog}) The inequality
    \begin{equation} \label{log_decay}
        \mathbb{E}^k_{\Omega}[\phi](t)\leq \frac{C_{\Omega , k}}{[\log(2+t)]^{2}}  \, \mathbb{E}^{k+1}[\phi](0)
    \end{equation}
    holds for all $t\geq 0$. Moreover, the logarithmic uniform energy decay rate is sharp.\footnote{The word \emph{sharp} means that no uniform local energy decay of the form \eqref{log_decay} with decay rate faster than logarithmic can possibly hold. We also note that the fact that the energy on the right hand side of \eqref{log_decay} is of higher order than the one on the left hand side is a standard feature of decay inequalities of this form when the background spacetime possesses (unstably or stably) trapped null geodesics.}
\end{itemize}
Furthermore, uniform boundedness and uniform (sharp-)logarithmic decay for the solution and (higher-order) derivatives of the solution hold pointwise.
\end{theorem}

\medskip
 
The first three bullet points in Theorem \ref{th_linear_waves} are rather straightforward to prove. We note, in particular, that the uniform energy boundedness claimed in the third bullet point is, for the case $k=1$,\footnote{We remark that the third bullet point states uniform boundedness for energies \emph{of all orders}.} an immediate consequence of the conservation of energy from the second bullet point. The decay statement of the last bullet point can be recovered as a special case of the more general result \cite{KeirLogLog}, which establishes an analogous energy decay statement for a class of static, spherically symmetric and asymptotically flat spacetimes (see Theorems 4.7 and 5.3 of \cite{KeirLogLog}, and Corollary 5.4 therein for a pointwise decay statement). The sharpness of the logarithmic uniform energy decay rate is to be directly ascribed to the presence of stably trapped null geodesics on $(\mathcal{M},g)$ and can be proven by applying the \emph{quasimode} construction of \cite{KeirLogLog}. Finally, we note that, for $k=1$, the \emph{local} energy decay is compatible with the conservation of energy stated in the second bullet point, and in fact it encodes the \emph{dispersion} of solutions.

\medskip

While the local energy of the generic (possibly axisymmetric) solution decays only logarithmically in time, the local energy decay rate of solutions supported on a finite number of angular modes (including, in particular, spherically symmetric solutions) is faster. Indeed, by decomposing solutions into spherical harmonics $Y^{\ell}_m(\vartheta,\varphi)$, i.e.,
\begin{equation*}
    \phi(t,r,\vartheta,\varphi) = \frac{1}{\sqrt{2\pi}}\sum_{\ell\geq 0,|m|\leq \ell}\phi_{\ell m}(t,r,\vartheta,\varphi)  
\end{equation*}
with $\phi_{\ell m}(t,r,\vartheta,\varphi)=b_{\ell m}(t,r)Y^{\ell}_m(\vartheta,\varphi)$, one can prove that the local energy of each angular mode $\phi_{\ell m}$ decays with an \emph{integrable} rate, i.e., the integrated local (for any $\Omega$) energy decay estimate 
\begin{equation} \label{lin_exp_decay}
   \int_0^\infty  \mathbb{E}_{\Omega}[\phi_{\ell m}](t) \, dt \leq C_1 e^{C_2\sqrt{\ell(\ell+1)}} \, \mathbb{E}[\phi_{\ell m}](0)
\end{equation}
holds, with positive constants $C_1$ and $C_2$ independent of $\ell$ (cf. Theorem 5.1 in \cite{KeirLogLog}). Similar considerations hold for local higher-order energies. Crucially, the inequality \eqref{lin_exp_decay} degenerates as $\ell\rightarrow\infty$, encoding the fact that (i) the decay of the local energy is slower for high-frequency angular modes and (ii) the inequality degenerates for the generic solution (i.e., for infinite sums of angular modes). Indeed, the logarithmic decay rate of the generic solution is non-integrable.

\section{The model nonlinear wave equation} \label{sec_nonlin_we}

We consider the cubic (defocusing) wave equation 
\begin{equation} \label{WE_nonlin_th}
\Box_{g}\phi=\phi|\phi|^2  
\end{equation}
for scalar functions $\phi:\mathcal{M}\rightarrow\mathbb{R}$ on the model spacetime $(\mathcal{M},g)$ of Section \ref{sec_model_spacetime}.

\medskip

We introduce the (first-order) energy
\begin{align}
\mathbb{E}_{\text{nl}}[\phi](\tau)=\int_{t=\tau} &dr \, d\vartheta \, d\varphi \, \frac{r^2\sin\vartheta}{2 \, f(r)} \label{energy_nonlin_waves}\\
&\times \left[  (\partial_t\phi)^2+f^2(r)\, (\partial_r\phi)^2 +\frac{f(r)}{r^2}\,(\partial_{\vartheta}\phi)^2+\frac{f(r)}{r^2\sin^2\vartheta}\,(\partial_{\varphi}\phi)^2+f(r)\,\frac{|\phi|^4}{2}\right]  \, . \nonumber
\end{align}
We state a large-data global existence theorem for solutions to the equation \eqref{WE_nonlin_th}. As in the linear case (cf. Theorem \ref{th_linear_waves}), the equation \eqref{WE_nonlin_th} admits global-in-time smooth solutions and a conserved energy.

\medskip

\begin{theorem} \label{th_nonlinear_waves}
For any smooth,\footnote{The statement continues to hold for initial data of finite regularity. In particular, the theorem holds for the (finite-regularity) initial data considered in the later Section \ref{sec_proto_data}.} compactly supported initial data $(\phi,\partial_t\phi)|_{t=0}$ prescribed at $t=0$, the smooth solution $\phi$ to \eqref{WE_nonlin_th} satisfies the following properties:
\begin{itemize}
    \item (Global existence) The solution $\phi$ exists and is unique for all $t\geq 0$.
    \item (Conservation of energy) The equality $$\mathbb{E}_{\textup{nl}}[\phi](t)=\mathbb{E}_{\textup{nl}}[\phi](0)$$ holds for all $t\geq 0$.
\end{itemize}
\end{theorem}

\medskip

The second bullet point in Theorem \ref{th_nonlinear_waves} is an easy check, whereas the first bullet point can be, for instance, recovered as a special case of the more general result \cite{Cagnac_choquet_bruhat_global_wp_nlw}.\footnote{The work \cite{Cagnac_choquet_bruhat_global_wp_nlw} establishes global existence of smooth solutions to the equation \eqref{WE_nonlin_th} (with possibly an additional linear Klein-Gordon term) on four-dimensional, globally hyperbolic spacetimes admitting uniformly timelike time-coordinate curves.} We emphasise that Theorem \ref{th_nonlinear_waves} does \emph{not} require any smallness assumption on the size of the initial data. For equation \eqref{WE_nonlin_th}, global existence is directly implied by the conservation (and coercivity) of the energy (combined with a standard Sobolev embedding). In particular, linear decay does \emph{not} play any role. We also remark that the conservation of the energy \eqref{energy_nonlin_waves} implies a uniform bound on \emph{first} derivatives of solutions, but remains compatible with (possibly unbounded) growth of high-order derivatives.

\medskip

For future convenience, we define the Fourier coefficient
\begin{equation}
\label{Eq:SD}
c_{\ell}(t,r) = 2\pi r^2 \int_{0}^{\pi}\phi\left(t,r,\vartheta\right)\cdot Y_{0}^{\ell}\left(\vartheta \right)\sin\vartheta \, d\vartheta  
\end{equation}
for any given axisymmetric solution to equation \eqref{WE_nonlin_th}. We also define the Fourier coefficient
\begin{equation}
\label{Eq:SD_3}
\tilde{c}_{\ell}(t,r) = 2\pi r^2 \int_{0}^{\pi}\phi^3\left(t,r,\vartheta\right)\cdot Y_{0}^{\ell}\left(\vartheta \right)\sin\vartheta \, d\vartheta  
\end{equation}
for the cubic power of the solution. By Fourier-decomposing the solution, one can write the energy \eqref{energy_nonlin_waves} as
\begin{align}
\mathbb{E}_{\text{nl}}[\phi](t)=& \,\frac{1}{2\pi}\sum_{\ell}\int_{0}^{\infty} dr  \, \frac{r^2}{2 \, f(r)}  \left[  (\partial_tc_{\ell})^2+f^2(r)\, (\partial_rc_{\ell})^2 +b\,\frac{f(r)}{r^2} \, \ell(\ell+1) c^2_{\ell} \right] \label{nonlinear_energy_l_mode}\\
&+\frac{1}{8\pi}\sum_{\ell}\int_{0}^{\infty} dr \,  r^2 \,\tilde{c}_{\ell}\, c_{\ell}  \nonumber
\end{align}
for some constant $b>0$ independent of $\ell$, where the term in the second line of \eqref{nonlinear_energy_l_mode} is positive (as it arises from the positive last term in \eqref{energy_nonlin_waves}). By defining the energy quantity
\begin{align}
    \mathbb{E}_{\ell}[\phi](t)=& \, \frac{1}{2\pi}\int_{0}^{\infty} dr  \, \frac{r^2}{2 \, f(r)}  \left[  (\partial_tc_{\ell})^2+f^2(r)\, (\partial_rc_{\ell})^2 +b\,\frac{f(r)}{r^2} \, \ell(\ell+1) c^2_{\ell} \right] \label{nonlinear_energy_l_mode_b}\\
    &+\frac{1}{8\pi}\int_{0}^{\infty} dr \,  r^2 \,\tilde{c}_{\ell}\, c_{\ell}  \, , \nonumber
\end{align}
which may be interpreted as the ``energy of the $\ell$-angular mode,'' the conservation of the energy \eqref{energy_nonlin_waves} yields
\begin{equation} \label{conservation_fourier_energies}
    \sum_{\ell}\mathbb{E}_{\ell}[\phi](t)= \mathbb{E}_{\text{nl}}[\phi](0)
\end{equation}
for all $t\geq 0$.

\section{Numerical implementation} \label{sec_numerical_implementation}

To numerically solve equation \eqref{WE_nonlin_th}, we start by re-writing the equation in the form\footnote{Greek indices run over all the coordinates, Latin indices are restricted to spatial coordinates. For brevity, $g$ will denote the determinant of the metric, and the commas partial derivatives.}
\begin{equation*}
    \frac{1}{\sqrt{-g}}\left(\sqrt{-g}g^{\mu\nu}\phi_{,\mu}\right)_{,\nu}= \phi^3 \, .
\end{equation*}
We consider the auxiliary scalar quantity 
\begin{equation*} 
    \Pi=-g^{t\mu}\phi_{,\mu} \, .
\end{equation*}
Equation \eqref{WE_nonlin_th} can then be recast as a system of two evolution equations which are of first order in the time variable, i.e.,
\begin{align}
  \phi_{,t} &=-\frac{\Pi+g^{ti}\phi_{,i}}{g^{tt}} \, , \label{eq:NumericalDecompositionWE_1}\\
  \Pi_{,t}  &= \frac{1}{\sqrt{-g}}\left[-\left(\sqrt{-g}\right)_{,t}\Pi+\left(\sqrt{-g}g^{ij}\phi_{,i}\right)_{,j} \right. \nonumber \\
            & \phantom{=} \left. -\left(\sqrt{-g}g^{it}\left(\frac{\Pi+g^{tk}\phi_{,k}}{g^{tt}}\right)\right)_{,i}-\sqrt{-g} \phi^3 \right]. \label{eq:NumericalDecompositionWE_2}
\end{align}
The quantity $\Pi_{,t}$ will be sometimes denoted by $\dot{\Pi}$.

\medskip

As a preliminary step, we transform our spatial coordinates from spherical ($r,\vartheta,\varphi$) to standard Cartesian coordinates $(x,y,z)$. In particular, $\varphi$ is treated as the azimuthal angle relative to the $z$-axis and the origins of the two coordinate systems coincide.
We discretise the spatial derivatives using standard fourth-order finite differences, evolve in time with a standard fourth-order Runge--Kutta method, and, for stability purposes, we add numerical dissipation through standard Kreiss--Oliger dissipation operators~\cite{kreiss1973methods} to the right hand sides of the equations \eqref{eq:NumericalDecompositionWE_1}-\eqref{eq:NumericalDecompositionWE_2}. 

\medskip

To achieve sufficiently high resolution for the numerical evolution with limited computational resources, we assume that the solution is \emph{axisymmetric}. Together with the axisymmetry of the background spacetime, the symmetry class of the solutions allows one to apply the so-called \emph{modified Cartoon method} \cite{Pretorius:2004jg,Alcubierre:1999ab} and evolve the system of equations \eqref{eq:NumericalDecompositionWE_1}-\eqref{eq:NumericalDecompositionWE_2} on a $2+1$ dimensional computational domain (the spacetime is still $3+1$ dimensional). The dimensional reduction is implemented as follows: 
\begin{itemize}
    \item We write the azimuthal Killing vector field $\xi$ of the background spacetime as
\begin{equation*}
    \xi= -y\,\frac{\partial}{\partial x} +x\,\frac{\partial}{\partial y} \, ,
\end{equation*}
meaning that we choose the axis of symmetry to coincide with the $z$-axis.
    \item Let
    \begin{equation*}
        \mathcal{P}\cong \left\lbrace 0 \right\rbrace_x  \times [0,\infty)_y \times (-\infty,\infty)_z \, .
    \end{equation*}
    Once initial data are prescribed at $t=0$, we numerically solve the equations \eqref{eq:NumericalDecompositionWE_1}-\eqref{eq:NumericalDecompositionWE_2} in the spacetime region $\mathcal{C}\subset \mathcal{M}$ such that 
    \begin{equation*}
       \mathcal{C} \cong [0,\infty)_t \times \mathcal{P} \, .
    \end{equation*}
    \item We reconstruct the derivatives with respect to $x$ on $\mathcal{C}$, which can, in fact, be written in terms of the derivatives with respect to $y$ and $z$ by exploiting the identities for Lie derivatives
    \begin{align*}
        \mathcal{L}_{\xi}g&=0 \, , & \mathcal{L}_{\xi}\phi &= 0 \, .
    \end{align*}
    The reconstruction procedure uses, in particular, the property that $\xi$ is nowhere tangent to $\mathcal{P}$ (see \cite{Pretorius:2004jg} for the relevance of this fact).
    \item Using the axisymmetry of the solution, the solution can then be extended from $\mathcal{C}$ to the whole spacetime $\mathcal{M}$.    
\end{itemize}

\medskip

As an auxiliary step, we also compactify the spatial coordinates while going through the dimensional-reduction procedure. Indeed, the equations \eqref{eq:NumericalDecompositionWE_1}-\eqref{eq:NumericalDecompositionWE_2} are evolved in compactified Cartesian coordinates $\left(t,X,Y,Z\right)$, with
\begin{align*}
    X &\in [-1,1] \, , & Y &\in [-1,1]  \, ,  & Z &\in [-1,1] \, ,
\end{align*}
obtained via the coordinate transformation
\begin{align}
\label{Eq:Compactification}
X =& \frac{2}{\pi}\arctan x \, , & Y =& \frac{2}{\pi}\arctan y \, , & Z =& \frac{2}{\pi}\arctan z \, . 
\end{align}
In the compactified coordinates, we have
\begin{equation} \label{compactified_spatial_domain}
    \mathcal{P}\cong \left\lbrace 0 \right\rbrace_X \times [0,1]_Y \times [-1,1]_Z \, .
\end{equation}
Since the compactification includes spacelike infinity, the asymptotically flat boundary condition is applied by enforcing the Minkowski space values for the metric, and that the field goes to zero. On the other hand, since we are working on the $X=0$ slice, on-axis regularity conditions are necessary and prescribed following \cite{Pretorius:2004jg}.

\section{Numerical results} \label{sec_numerical_results}

In this section, we present the results of our numerical simulations.\footnote{The code used for the simulations is publicly available at~\url{https://github.com/alejandroc137/ScalarWaveEvolution}.} We focus on three aspects of numerical solutions to equation \eqref{WE_nonlin_th}, each discussed in a distinct section: The growth of high-order derivatives (Section \ref{sec_growth_derivatives}), the direct angular-mode cascade (Section \ref{sec_mode_cascade}), and the evolution of small initial data (Section \ref{sec_small_data}).

\subsection{Initial data} \label{sec_proto_data}

The solutions considered arise as the future evolution of time-symmetric initial data of the form
\begin{align}
\phi|_{t=0}&=\epsilon \, u(r)\cdot \sum_{\left[\ell\right]}Y_{0}^{\ell}(\vartheta) \, ,  & \partial_{t}\phi|_{t=0}&=0 \, ,  \label{proto_data}
\end{align}
where $\epsilon$ is a positive real constant. The scalar function $u(r)$ is compactly supported on a finite interval $[r_{0,1},r_{0,2}]$ around $r=r_0$ (cf. Figure \ref{fig:finfo}) and will be chosen to be smooth on its support (but only finitely differentiable on $r\in (0,\infty)$). The scalar functions $Y_{0}^{\ell}(\vartheta)$ denote the standard spherical harmonics $Y_m^{\ell}(\vartheta,\varphi)$ of the unit $2$-sphere\footnote{The spherical harmonics are normalised so as to have unit $L^2(\mathbb{S}^2)$-norm.} with $m=0$, and $\left[  \ell \right]$ symbolizes the set of spherical harmonics considered (e.g., $\left[\ell_1,\ell_2 \right]$ in \eqref{proto_data} would denote the sum $Y_{0}^{\ell_1}(\vartheta)+Y_{0}^{\ell_2}(\vartheta)$). We will only consider initial data containing \emph{finite} sums of spherical harmonics. 

\medskip

The initial data \eqref{proto_data} are \emph{axisymmetric}. The axisymmetry of the initial data is preserved by equation \eqref{WE_nonlin_th}, and thus the solution remains axisymmetric for all future times.

\subsection{Intermezzo: Linear decay} \label{sec_lin_solution_proto_data}

The \emph{linear} solutions arising from initial data of the form prescribed in Section \ref{sec_proto_data} can be written as \emph{finite} sums of (axisymmetric) angular modes. As discussed in Section \ref{sec_lin_we}, the uniform decay rate of the local energy of these solutions, as well as of their local higher-order energies, is faster than for the generic solution. In this respect, since the nonlinear turbulent dynamics that we shall describe is favoured by slow linear decay, the turbulent behaviour of the initial data prescribed in Section \ref{sec_proto_data} may be expected to persist for more general, perhaps \emph{generic}, initial data. See the related discussion in Section \ref{sec_discussion_beyond_scalar_model}.

\subsection{Growth of high-order derivatives}  \label{sec_growth_derivatives}

\begin{figure}
    \includegraphics[scale=.5]{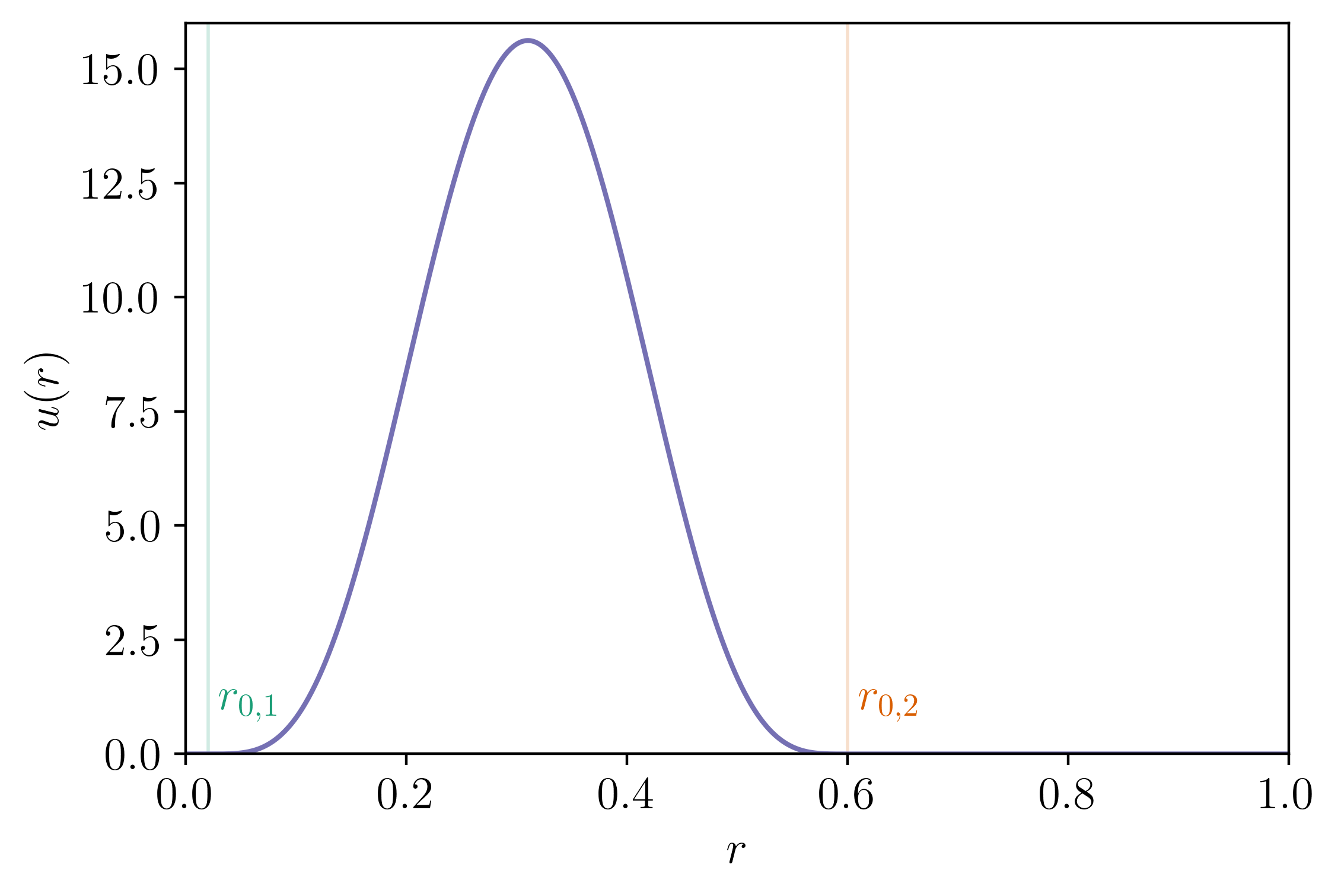}
    \caption{The function $u(r)$ as defined for the initial data in \eqref{Eq:piecewise_s}. We recall that the minimum of the radial geodesic potential $V(r)$ is located at $r_0\sim 0.22$ (cf. Figure \ref{fig:finfo}).}
    \label{fig:ur}
\end{figure}

We consider numerical solutions to equation \eqref{WE_nonlin_th}. For the numerical simulations of the present section, we choose our ``canonical'' initial data  \eqref{proto_data} to be 
\begin{equation}
    u(r)=4000 \cdot \frac{\left(r-r_{0,1}\right)^{4}\left(r-r_{0,2}\right)^{4}}{\left(r_{0,1}-r_{0,2}\right)^{8}}   \label{Eq:piecewise_s}
\end{equation}
for $r\in [r_{0,1}, r_{0,2}]$ (see Figure \ref{fig:ur}), and
with parameters
\begin{align} 
\epsilon &= 1 \, , &  r_{0,1}&=0.02 \, ,  &  r_{0,2}&=0.6 \, , & [\ell]&=[1,2]. \label{choices_ID}
\end{align}
The results shown in this (and the following) section(s) are run on a grid of resolution $(\Delta_Y,\Delta_Z)=0.003125$, with a Courant factor $C=0.5$ (i.e., $\Delta_t=0.5 \, \Delta_Y$). We have applied a sixth-order Kreiss--Oliger dissipation, with $\epsilon_{\rm KO}=0.3$ for numerical stability. The convergence tests of all the simulations presented are collected in Appendix \ref{app:convergence}.

\medskip

The selection of $\epsilon=1$ in \eqref{choices_ID} ensures that the nonlinearity of equation \eqref{WE_nonlin_th} significantly influences the dynamics of the solution during the initial stages, reducing the need for very long evolutions. This behavior is illustrated in Figure \ref{fig:Amplitudes} for one quantity of interest (\eqref{sup_2nd_deriv_theta} below), where we display the results from the evolution of $\epsilon=1$-initial data compared to the evolution of several different initial amplitudes around this value. A solution dominated by the linear dynamics should exhibit similar behavior for different initial amplitudes (modulo an overall scale proportional to $\epsilon$). This is evident for the $\epsilon=0.125$ vs. $\epsilon=0.25$ cases plotted in the figure, but notably begins to fail increasing $\epsilon$ to values close to $1$. The remainder of this section focuses on amplitudes around $\epsilon=1$; for a discussion of the evolution of smaller amplitude initial data, see Section \ref{sec_small_data}.

\medskip

\begin{figure}
\centering
    \includegraphics[scale=.5]{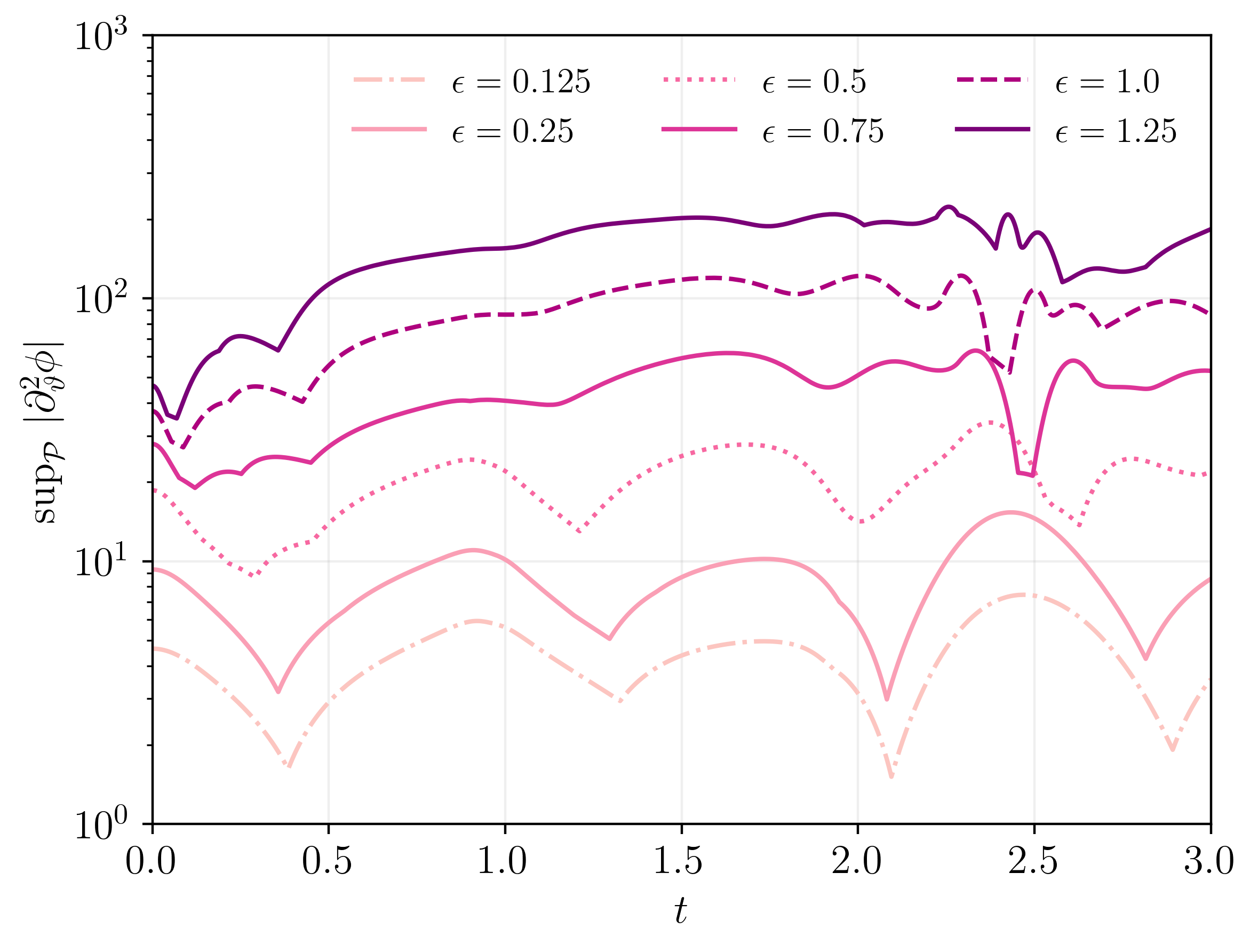}
    \caption{The evolution of the quantity \eqref{sup_2nd_deriv_theta} over the time interval $t\in [0,3]$, with a logarithmic scale on the vertical axis. Six different values of $\epsilon$ are examined, each corresponding to a different color in the plot. For $\epsilon=0.75$ or larger, the solution already exhibits a markedly nonlinear behaviour within the time interval considered. The convergence of these simulations is shown in Figure \ref{fig:ConvergenceLinToNonLin}.}
    \label{fig:Amplitudes}
\end{figure}

\begin{figure}
    \includegraphics[width=\textwidth]{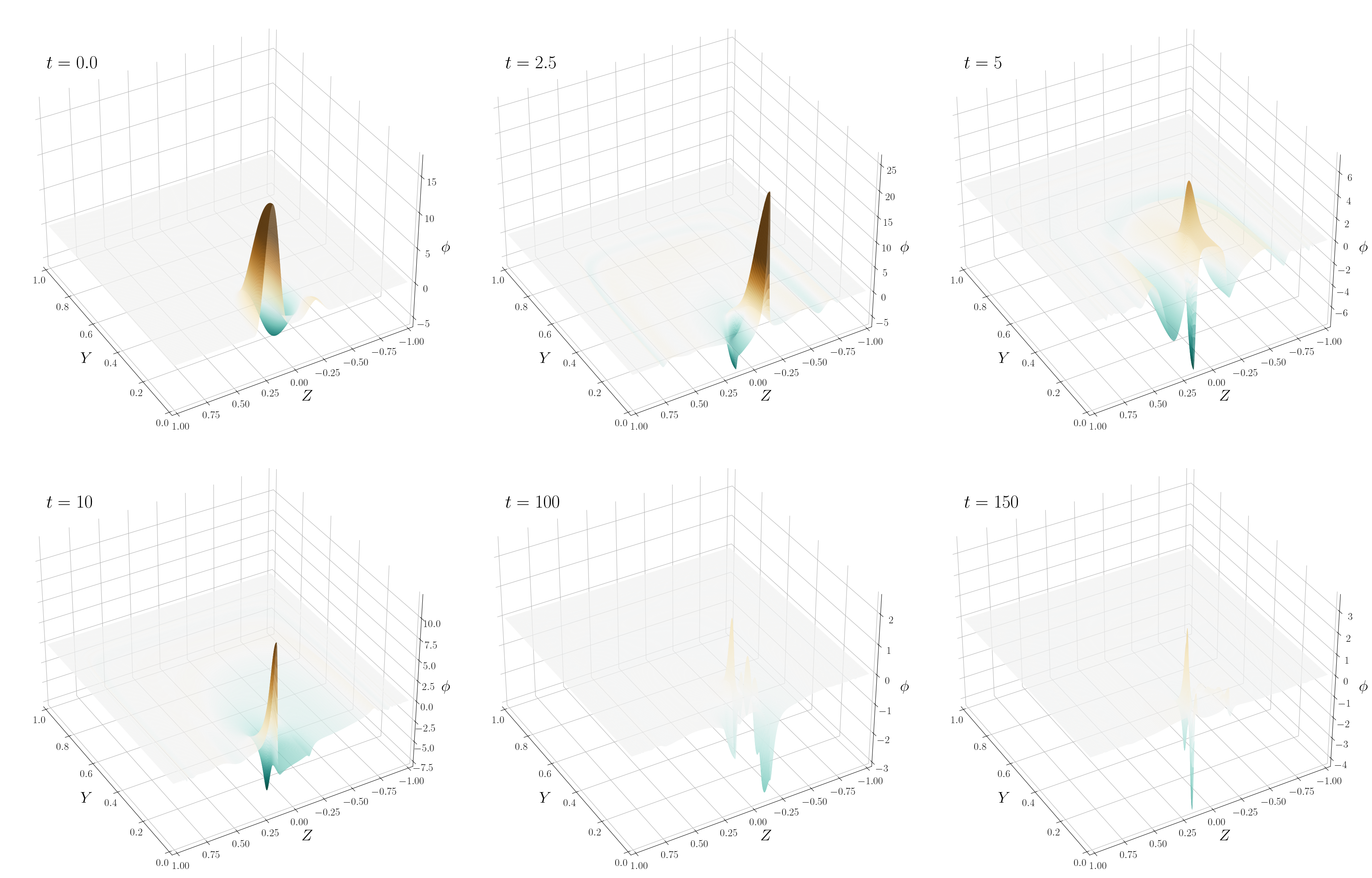}
    \caption{Time snapshots of the evolution of the solution $\phi$ on the spatial domain $\mathcal{P}$ (recall \eqref{compactified_spatial_domain}) over the time interval $t\in[0,150]$. For different snapshots, time increases from the left to the right (top to bottom). The vertical axis and color correspond to the amplitude of the solution, with orange depicting positive values of the solution and green negative values. The range of the vertical axis varies from one panel to another.}
    \label{fig:Snapshots}
\end{figure}

\begin{figure}
\centering
    \includegraphics[scale=.5]{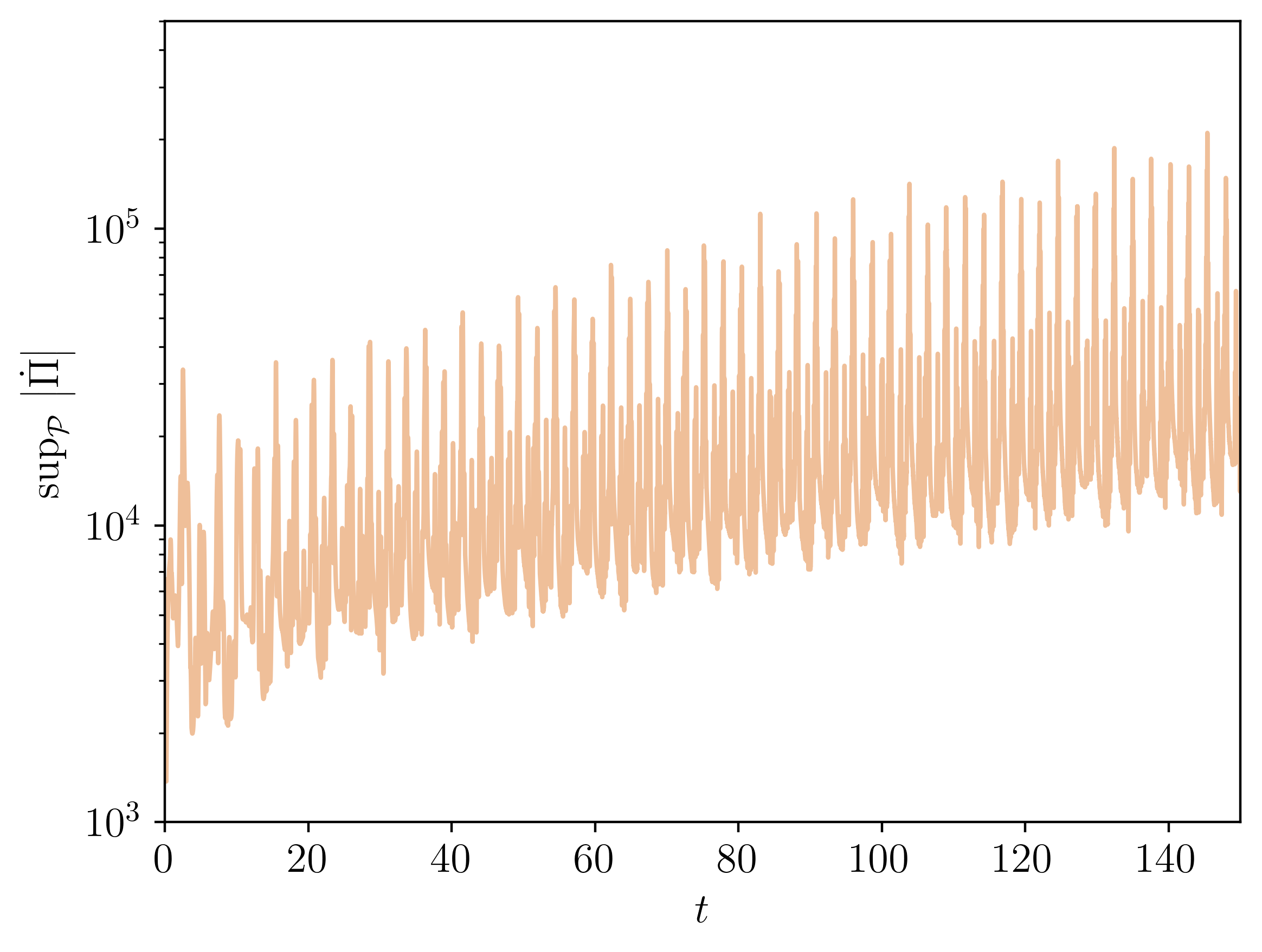}
        \caption{The evolution of the quantity \eqref{sup_pi_dot} over the time interval $t\in [0,150]$, with a logarithmic scale on the vertical axis. The point-wise relative numerical error in this quantity is generally sub-percent level (see Appendix~\ref{app:convergence} for more details).}
    \label{Fig:GrowthOfPidot_Single}
\end{figure}

\begin{figure}
\centering
    \includegraphics[scale=.5]{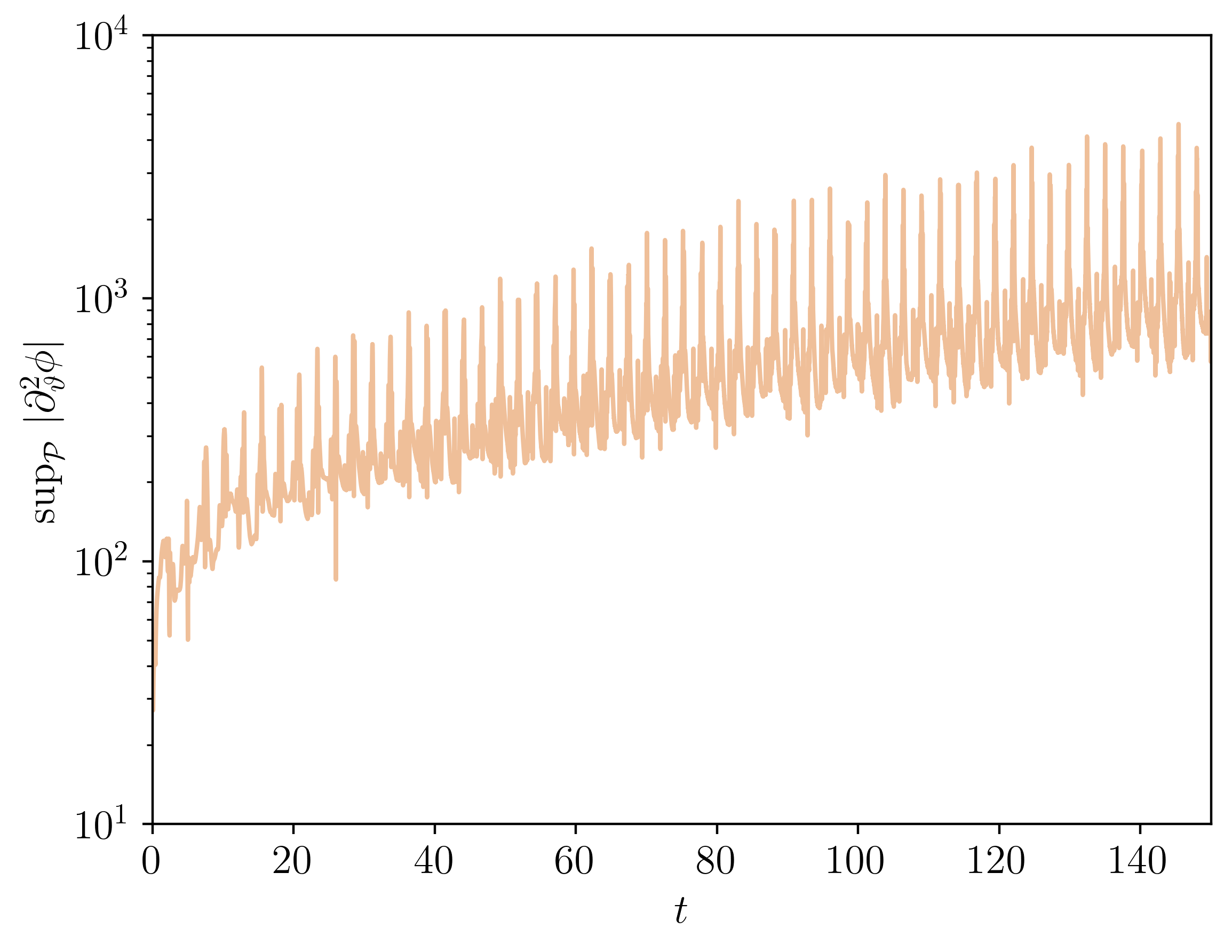}
    \caption{The evolution of the quantity \eqref{sup_2nd_deriv_theta} over the time interval $t\in [0,150]$, with a logarithmic scale on the vertical axis (note the different range of the axis from Figure \ref{Fig:GrowthOfPidot_Single}). The point-wise relative numerical error in this quantity is generally sub-percent level (see Appendix~\ref{app:convergence} for more details).}
    \label{Fig:GrowthOfTheta2_Single}
\end{figure}

\begin{figure}
\centering
    \includegraphics[scale=.5]{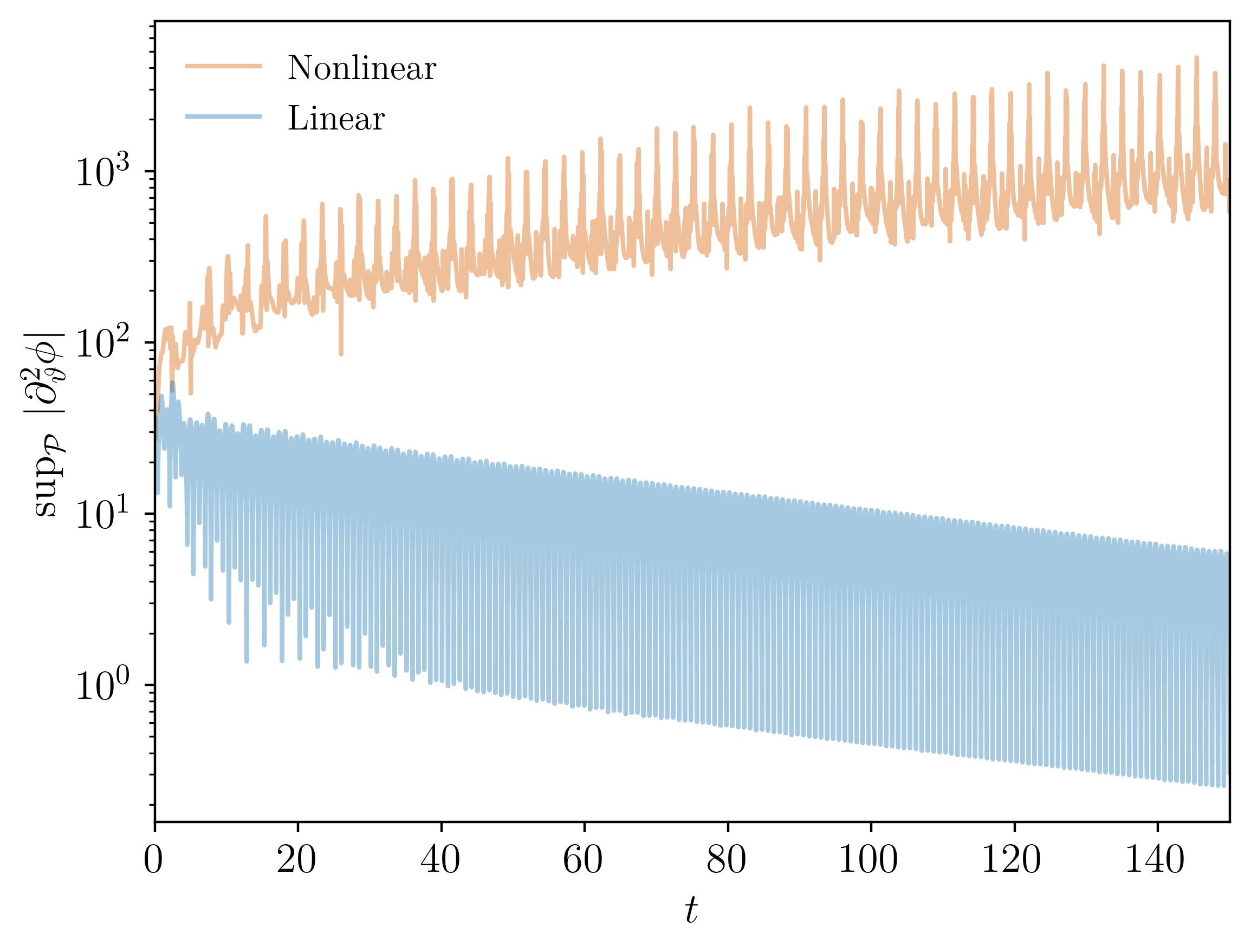}
    \caption{The evolution of the quantity \eqref{sup_2nd_deriv_theta} over the time interval $t\in [0,150]$, with a logarithmic scale on the vertical axis. We consider the solution to the linear wave equation \eqref{WE_th} (blue line) and the solution to the nonlinear wave equation \eqref{WE_nonlin_th} (orange line), both arising from the initial data considered in Section \ref{sec_growth_derivatives}. The orange line coincides with the one plotted in Figure \ref{Fig:GrowthOfTheta2_Single}.}
    \label{fig:NlinVsLin}
\end{figure}

\begin{figure}
\centering
    \includegraphics[scale=.5]{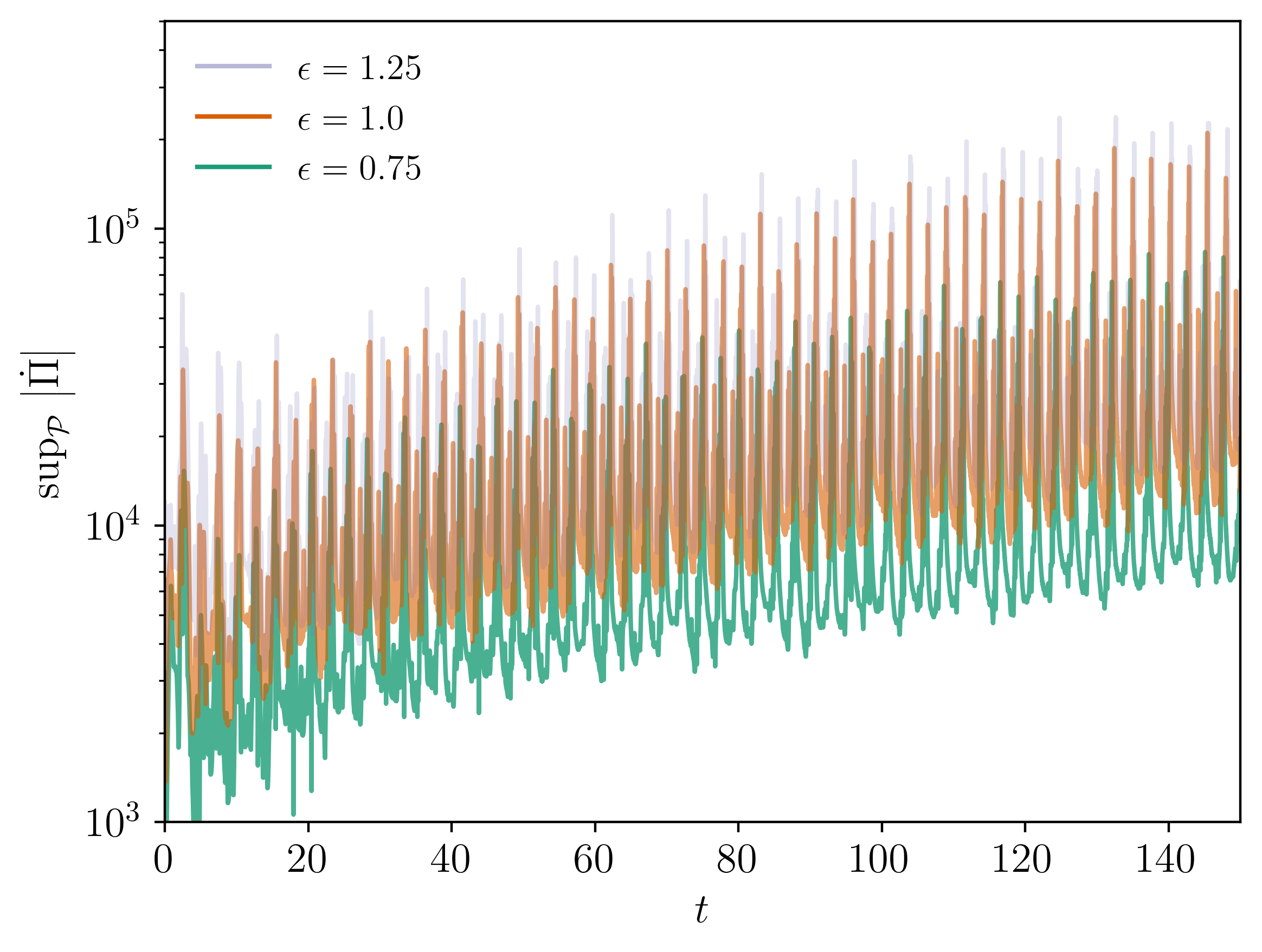}
    \caption{The evolution of the quantity \eqref{sup_pi_dot} over the time interval $t\in [0,150]$, with a logarithmic scale on the vertical axis. The three values $\epsilon=\lbrace 0.75, 1, 1.25 \rbrace$ are examined, each corresponding to a different color in the plot (green, orange and purple, respectively). The convergence of these simulations is shown in Figure \ref{fig:Convergence}.}
    \label{Fig:GrowthOfPidot}
\end{figure}

\begin{figure}
\centering
    \includegraphics[scale=.5]{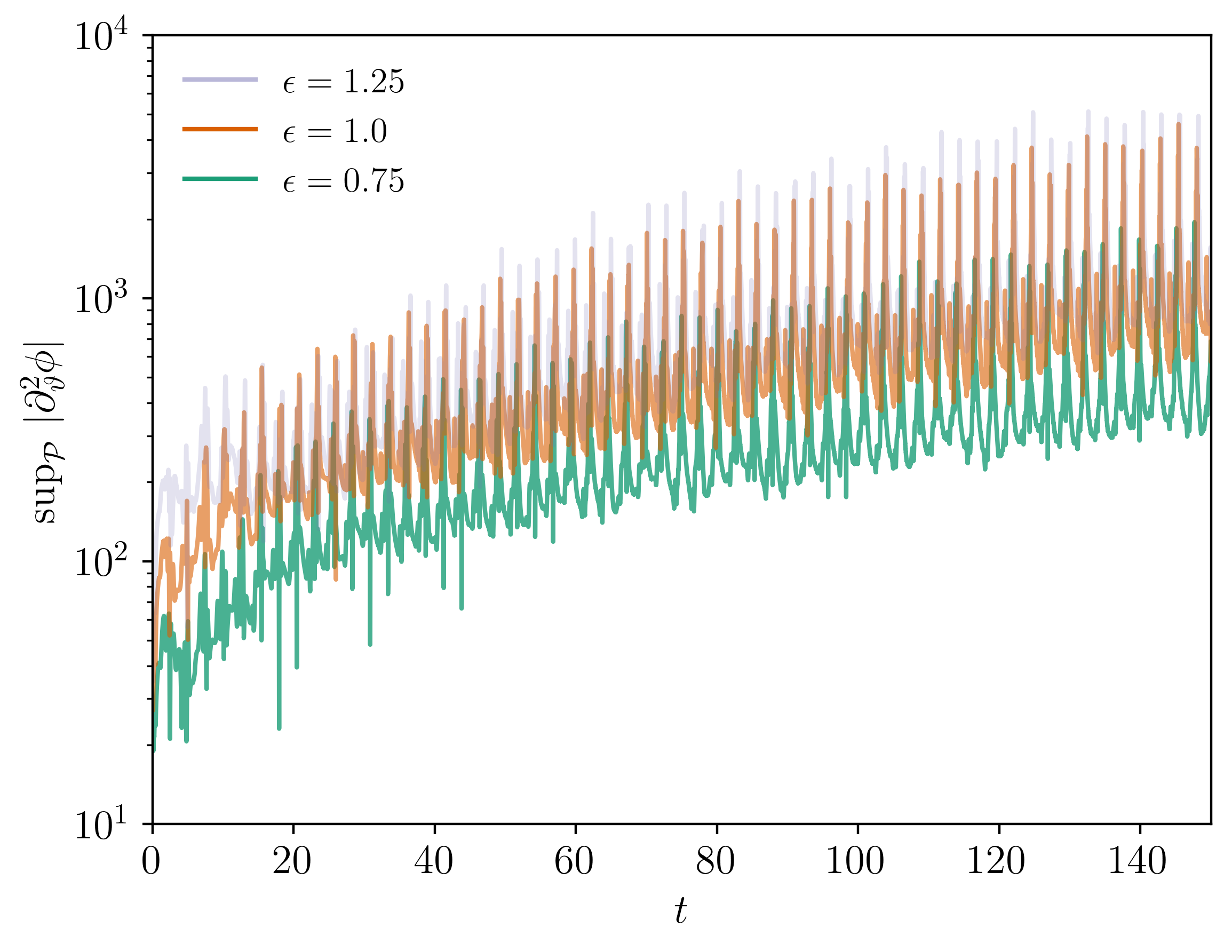}
    \caption{The evolution of the quantity \eqref{sup_2nd_deriv_theta} over the time interval $t\in [0,150]$, with a logarithmic scale on the vertical axis. The three values $\epsilon=\lbrace 0.75, 1, 1.25 \rbrace$ are examined, each corresponding to a different color in the plot (green, orange and purple, respectively).}
    \label{Fig:GrowthOfTheta2}
\end{figure}

Figure \ref{fig:Snapshots} shows the numerical evolution of the solution $\phi$. While the solution disperses (with its amplitude decaying over time), part of the solution remains confined within the stable trapping region over the entire interval of time that we evolve. Within the stable trapping region, the solution appears to develop high-frequency structure, already hinting at a possible growth of its high-order derivatives.

\medskip

Figure \ref{Fig:GrowthOfPidot_Single} tracks the numerical evolution of the quantity\footnote{The quantity $\dot{\Pi}$, defined in Section \ref{sec_numerical_implementation}, differs from $\partial^2_t\phi$ by an overall (time independent) positive factor.}
\begin{equation} \label{sup_pi_dot}
    \sup_{\mathcal{P}}|\dot{\Pi}|(t) \, ,
\end{equation}
where we recall the definition of $\mathcal{P}$ from \eqref{compactified_spatial_domain}.\footnote{Taking the supremum over $\mathcal{P}$ should be thought as being equivalent to taking the supremum over the entire spatial domain of our model spacetime.} The quantity \eqref{sup_pi_dot} is seen to \emph{grow} over the entire time interval considered. Second spatial derivatives exhibit an analogous behaviour as the one showed in Figure \ref{Fig:GrowthOfPidot_Single}. For instance, Figure \ref{Fig:GrowthOfTheta2_Single} examines the numerical evolution of the quantity
\begin{equation} \label{sup_2nd_deriv_theta}
    \sup_{\mathcal{P}}|\partial_\vartheta^2\phi|(t)  \, ,
\end{equation}
which also \emph{grows} over the entire time interval considered. We note that the growth of the quantities \eqref{sup_pi_dot} and \eqref{sup_2nd_deriv_theta} is intended upon averaging over an appropriate timescale, as it is apparent from the figures. From the dynamics of the solution observed in Figure \ref{fig:Snapshots} and the fact that an analogous behaviour for the quantities \eqref{sup_pi_dot} and \eqref{sup_2nd_deriv_theta} can be seen to persist when the supremum is only taken over the stable trapping region, we conclude that the growth of the second derivatives captured by Figures \ref{Fig:GrowthOfPidot_Single} and \ref{Fig:GrowthOfTheta2_Single} occurs within (and is tied to the presence of) the stable trapping region. Moreover, we remark that the growth of second-order derivatives is a genuinely \emph{nonlinear} phenomenon. The difference between linear and nonlinear evolution of second-order derivatives is illustrated in Figure \ref{fig:NlinVsLin} for the quantity \eqref{sup_2nd_deriv_theta}. For the same initial data prescribed for Figure \ref{Fig:GrowthOfTheta2_Single}, the quantity \eqref{sup_2nd_deriv_theta} associated to the linear solution can be seen to decay over the time interval considered (as predicted by Theorem \ref{th_linear_waves} and discussed in Section \ref{sec_lin_solution_proto_data}).

\medskip

For illustration, in Figures \ref{Fig:GrowthOfPidot} and \ref{Fig:GrowthOfTheta2} we show the quantities \eqref{sup_pi_dot} and \eqref{sup_2nd_deriv_theta} respectively from numerical evolution of the same form of initial data just discussed, but now including two additional choices of initial amplitude $\epsilon=0.75$ and $\epsilon=1.25$. Both plotted quantities grow throughout the time interval considered, with a similar growth rate for all three values of $\epsilon$. To go beyond a crude comparison of growth trends and investigate the possible presence of a scaling relation for the growing quantities \eqref{sup_pi_dot} and \eqref{sup_2nd_deriv_theta} within our class of initial data, one would need to carry out additional numerical simulations over a range of values of $\epsilon$ covering several orders of magnitude, and possibly over much longer timescales than presented here (see also Section \ref{sec_small_data} and the related comments therein).

\medskip

Although not displayed here, we have also examined some additional higher (than second) order derivatives of the solution (in the $\epsilon=1$-case) and observed a similar behaviour to the one described for second derivatives.

\subsection{Direct angular-mode cascade}  \label{sec_mode_cascade}

In this section, we show that the growth of high-order derivatives described in Section \ref{sec_growth_derivatives} is the result of a direct angular-mode cascade within our numerical solutions. To capture the mode cascade, we shall consider the Fourier coefficients $c_{\ell}(t,r_0)$ and $\tilde{c}_{\ell}(t,r_0)$, where we recall the definitions \eqref{Eq:SD}-\eqref{Eq:SD_3} and that $r=r_0$ corresponds to the location of the local minimum of the radial geodesic potential (see Figure \ref{fig:finfo}).

\medskip

\begin{figure}
\includegraphics[scale=.5]{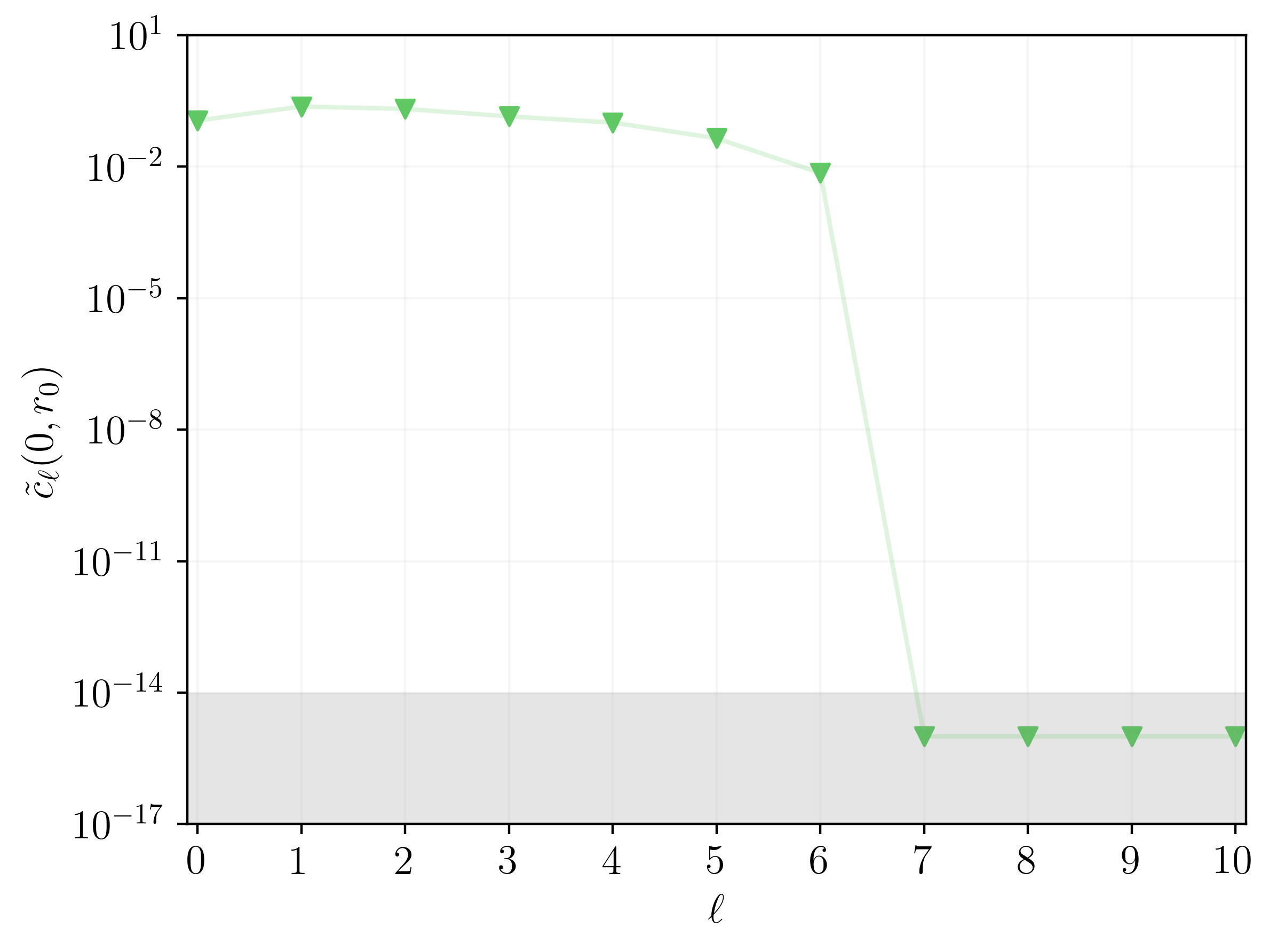}
    \caption{The $\ell$-spectrum of the cubic power $\phi^3$ of the solution at the initial time, as captured by the Fourier coefficients \eqref{Eq:SD_3}, with a logarithmic scale on the vertical axis. The grey shading indicates where the numerical floor of our simulations is. We note that the Fourier coefficients plotted are all positive, with no absolute value applied to the Fourier coefficients on the vertical axis.}
    \label{fig:SDIC_single}
\end{figure}

Given a (axisymmetric) solution to equation \eqref{WE_nonlin_th} arising from the initial data \eqref{proto_data} with the choices \eqref{choices_ID}-\eqref{Eq:piecewise_s} of Section \ref{sec_growth_derivatives}, we preliminarily observe that $$\partial_t c_{\ell}(t,r_0)|_{t=0}=0$$ and, using equation \eqref{WE_nonlin_th} and the form of the initial data considered, we compute
\begin{equation} \label{f_coeff_2nd_derivative}
    \partial_t^2 c_{\ell}(t,r_0)|_{t=0}= F_{r_0}\delta_{\ell}^1+H_{r_0}\delta_{\ell}^2 -f(r_0)\cdot\tilde{c}_{\ell}(0,r_0) \, ,
\end{equation}
with $\delta_i^j$ denoting the Kronecker delta, and the constants $F_{r_0}$ and $H_{r_0}$ are
\begin{align*}
    F_{r_0}&=2\pi r_0f(r_0)\cdot(\partial_r(rf(r)\partial_ru(r)))(r_0)+4\pi f(r_0) u(r_0) \, ,\\
    H_{r_0}&=2\pi r_0f(r_0)\cdot(\partial_r(rf(r)\partial_ru(r)))(r_0)+12\pi f(r_0) u(r_0) \, .
\end{align*}
Since the initial data are supported on $\ell \in [1,2]$, the non-trivial values of $\tilde{c}_{\ell}(0,r_0)$ in \eqref{f_coeff_2nd_derivative} correspond to $\ell \in [0,\dots,6]$ (see Figure \ref{fig:SDIC_single}), as one can compute using the Clebsch--Gordan coefficients. The quantity \eqref{f_coeff_2nd_derivative} is therefore non-trivial for $\ell \in [0,\dots,6]$.

\medskip

\begin{figure}
    \includegraphics[scale=.5]{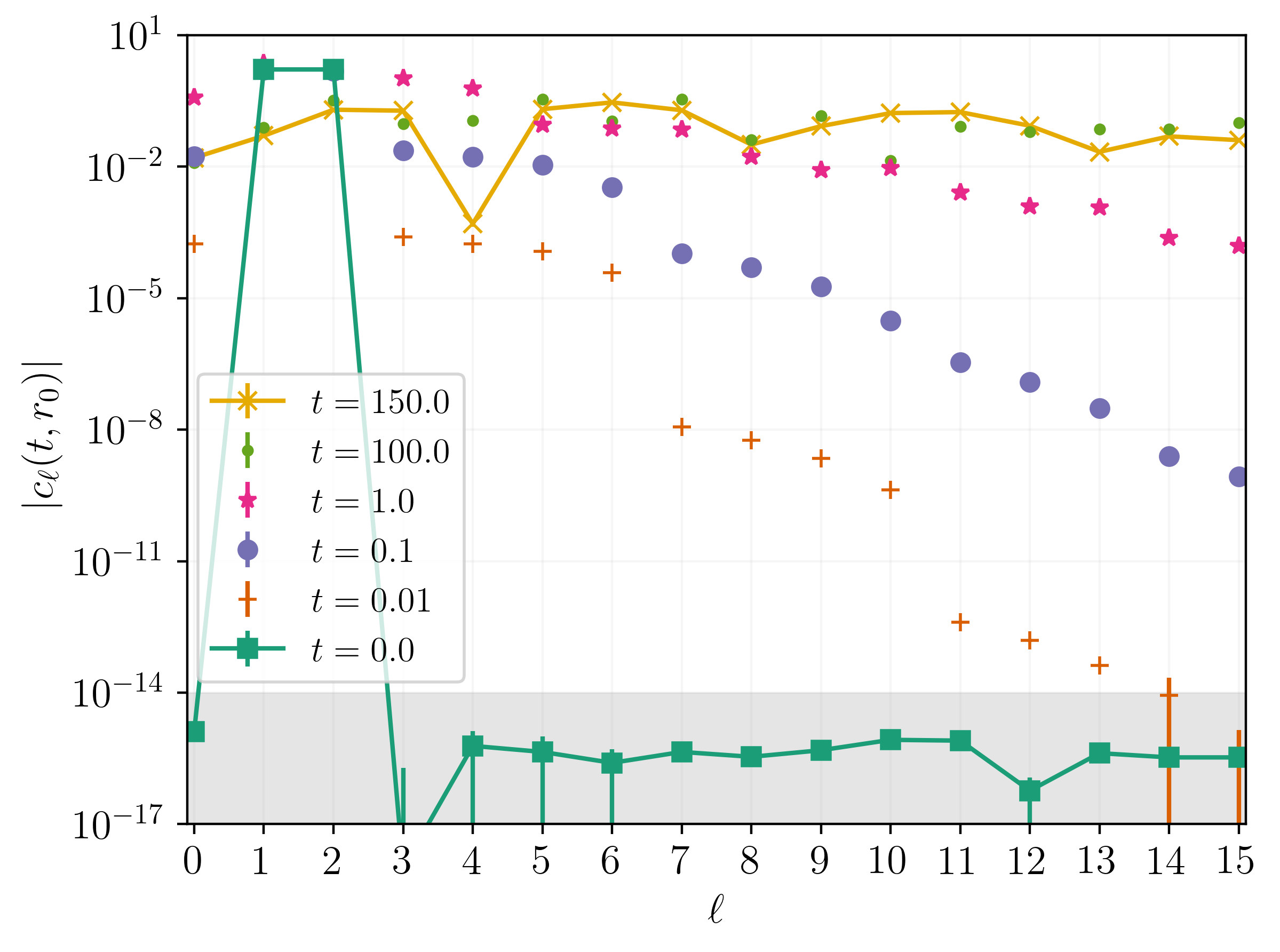}
    \caption{The $\ell$-spectrum of the solution at different times in the evolution, as captured by the Fourier coefficients \eqref{Eq:SD}, with a logarithmic scale on the vertical axis. Lines joining the points for the times $t=0$ and $t=150$ have been added to guide the reader. The grey shading indicates where the numerical floor of our simulations is. The numerical errors in the late time ($t=150$) quantities are on the order of $10^{-3}$ or less, as can be inferred from the resolution study shown in Figure \ref{fig:SDConvergence}. Note that the seemingly anomalous low value for the $l=4$-mode at $t=150$ is an artifact of us {\em not} averaging the amplitude coefficients in time over the local oscillation period of these modes, as illustrated in Figure \ref{fig:CascadeOfModes} below (i.e., the evolution is near a local minimum of the $l=4$-mode at that time).}
    \label{fig:SDAsTime}
\end{figure}

The time evolution of the $\ell$-spectrum of the numerical solution is depicted in Figure \ref{fig:SDAsTime}. In agreement with the prescribed initial data, the solution is only ``supported" on the $\ell\in [1,2]$-angular modes\footnote{Meaning that the numerical amplitude of the $\ell \in [1,2]$-angular modes is above our numerical floor (from double-precision floating point round-off), whereas the numerical amplitude of all other angular modes, though non-zero, is below our numerical floor. For the present discussion, the solution being ``supported" on a certain range of angular modes always takes this meaning.} at $t=0$, with $c_{1}(0,r_0)=c_{2}(0,r_0)\sim 3.2$. At time $t=0.01$, the numerical solution is supported on the wider range of angular modes with $\ell \in [0,\dots,14]$. At this time, the newly excited modes above the numerical floor appear to be grouped into sequences whose Fourier coefficients take approximately the same value. We shall refer to such sequences as \emph{generations}. The first generation of modes consists of modes with $\ell=0$ and $\ell \in [3,\dots,6]$, which are excited to larger amplitude than the modes with higher $\ell \in [7,\ldots,14]$. In fact, as encoded by equation \eqref{f_coeff_2nd_derivative}, the first generation of excited modes of Figure \ref{fig:SDAsTime} is driven, for small times (i.e., $t=0.01$ in Figure \ref{fig:SDAsTime}), by the $\ell$-spectrum of the cubic power $\phi^3(0,r_0,\vartheta)$ of the initial datum, which is indeed supported on angular modes with $\ell \in [0,\ldots, 6]$ (see Figure \ref{fig:SDIC_single}). By time $t=0.1$, all the angular modes with $\ell \in [0,\dots,15]$ are excited above the numerical floor. Approaching $t=100$, the curve connecting the points in Figure \ref{fig:SDAsTime} progressively flattens, with the excited $\ell\in [5,\dots,15]$-angular modes reaching their maximum amplitude (within our evolution) and attaining approximately the same amplitude as the $\ell\in [1,2]$-angular modes. After $t=100$, all the excited angular modes depicted appear to settle down to an almost stationary regime (see the small difference in amplitude of the modes between $t=100$ and $t=150$). Overall, we observe slow dispersion of the $\ell\in [1,2]$-angular modes over the entire evolution and slow dispersion of the excited $\ell\in [5,\dots,15]$-angular modes once they attain their maximum amplitude.

\medskip

\begin{figure}
    \includegraphics[width=\textwidth]{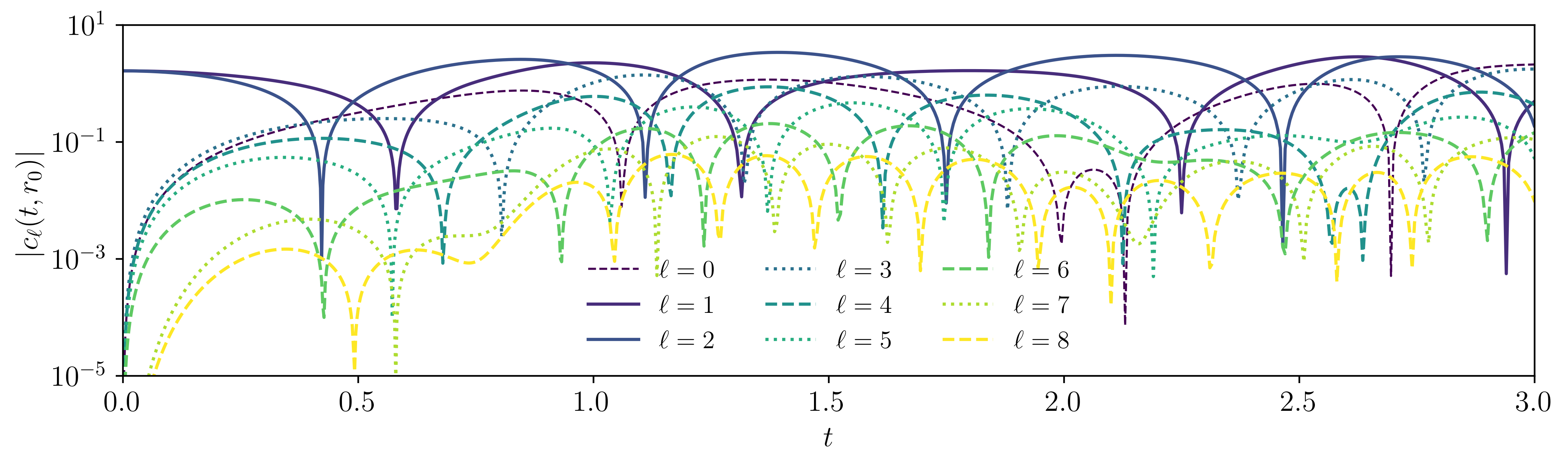}
    \includegraphics[width=\textwidth]{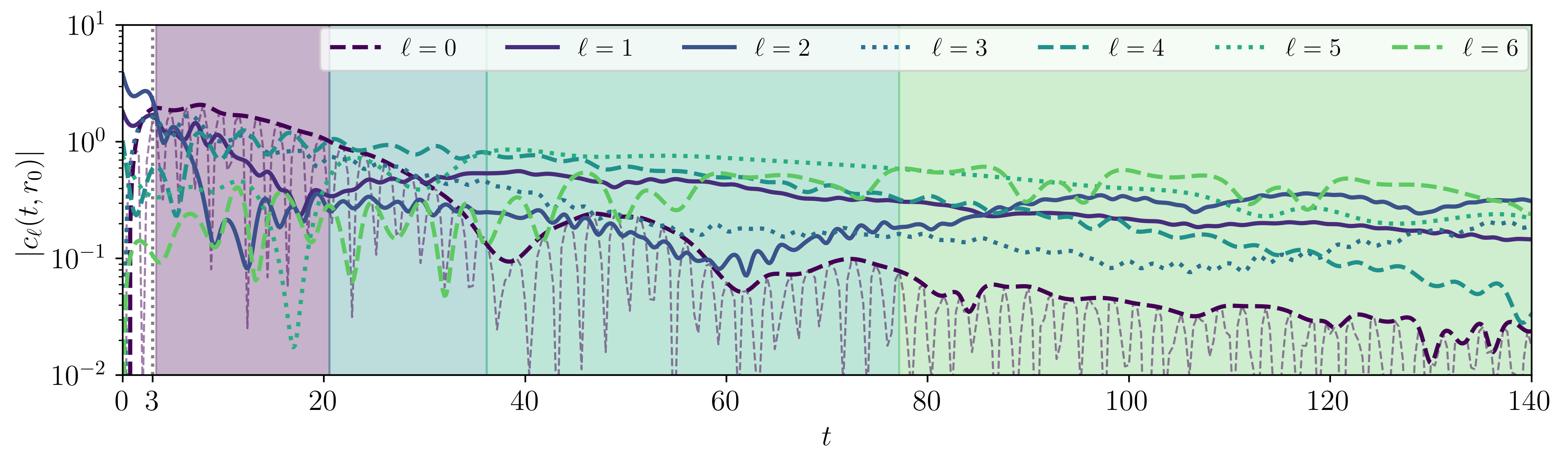}
    \caption{The time evolution of the Fourier coefficients \eqref{Eq:SD} for $\ell\in [0,\dots,8]$ ($[0,\dots,6]$) in the top (bottom) panel, with logarithmic scales on the vertical axes. The top panel shows the evolution over the initial time interval $t\in[0,3]$, whereas the bottom panel displays the envelope of the evolution over the entire time interval $t\in[0,140]$ (the vertical dotted line in the bottom panel marks the time $t=3$, i.e., the end of the initial interval of time blown-up in the top panel). For illustrative purposes, in the bottom panel we have also plotted the instantaneously measure amplitude of the $\ell=0$-mode (thinner line) in addition to its envelope (thicker line); similar behaviour occurs for all modes. Note the different ranges of the vertical axes in the two panels, and that both the lower ends of the two ranges are several orders of magnitude above the numerical floor. The curves in the panels correspond to single values of $\ell$, which are identified by different colors. In the top panel, the $\ell\in [1,2]$-angular modes (shown with solid lines) are contained in the prescribed initial data and are displayed with their initial amplitude at $t=0$. The growth of the excited angular modes with $\ell=0$ and $\ell \in [3,\dots,6]$ can be seen to precede (in time) the growth of the excited angular modes with $\ell \in [7,8]$ (cf. the first and second generations of excited modes in Figure \ref{fig:SDAsTime}). In the bottom panel, the multi-colored background identifies the time intervals over which the correspondingly colored angular mode dominates. For $t\in[0,3.3]$, the $\ell \in [1,2]$-angular modes dominate the evolution. For the later times in the plot, the dominant angular modes are (in order) $\ell \in [0,4,5,6]$.  The convergence of these coefficients at two different late times is shown in Figure \ref{fig:SDConvergence}.
    }
    \label{fig:CascadeOfModes}
\end{figure}

\begin{figure}
    \centering
    \includegraphics[scale=.5]{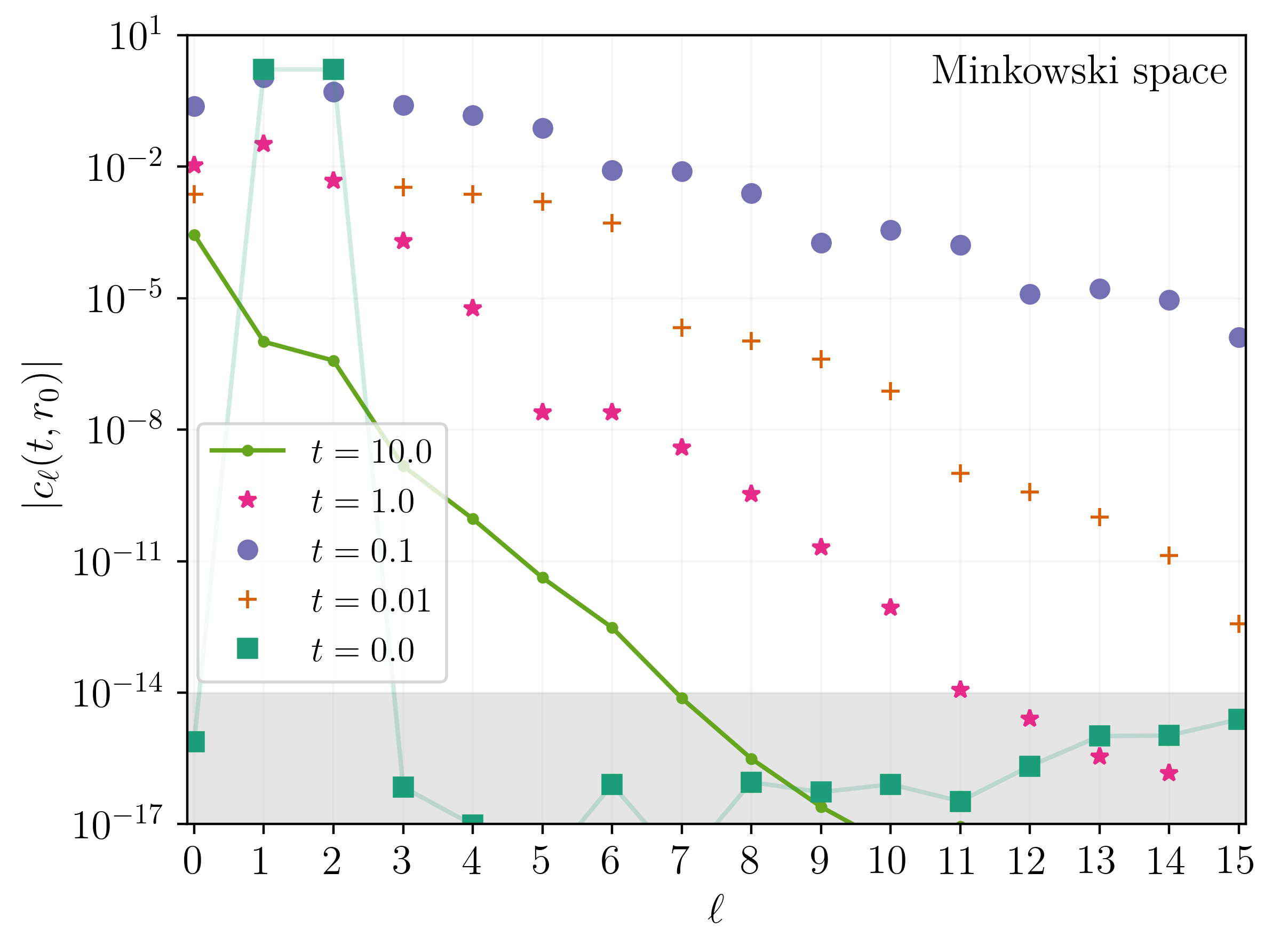}
    \caption{The $\ell$-spectrum of the solution to equation \eqref{eqn_intro_NLW} on Minkowski space, arising from the same initial data considered for Figure \ref{fig:SDAsTime}. The $\ell$-spectrum is depicted at different times in the evolution, as captured by the Fourier coefficients \eqref{Eq:SD}, with a logarithmic scale on the vertical axis. Lines joining the points for the times $t=0$ and $t=10$ have been added to guide the reader. The grey shading indicates where the numerical floor of our simulations is. The reader may compare this figure with Figure \ref{fig:SDAsTime}, noting however the shorter times plotted here.}
    \label{fig:l_modes_Minkowski}
\end{figure}

In Figure \ref{fig:CascadeOfModes}, we show a more detailed analysis of the time evolution of individual angular modes with $\ell\in [0,\dots,8]$ for the same numerical solution depicted in Figure \ref{fig:SDAsTime}. We observe that, not only do the excited angular modes attain approximately the same amplitude as the $\ell \in [1,2]$-modes over time (as already apparent from Figure \ref{fig:SDAsTime}), but progressively higher-$\ell$ excited angular modes come to \emph{dominate} in the evolution. Moreover, the time interval over which an individual higher-$\ell$ excited mode dominates appears to \emph{dilate} over the evolution (compare the different colored backgrounds in Figure \ref{fig:CascadeOfModes}).

\medskip

From Figures \ref{fig:SDAsTime} and \ref{fig:CascadeOfModes}, one can conclude that the numerical solution undergoes a \emph{low-to-high angular-mode cascade}, which we shall refer to as \emph{direct} or \emph{forward} cascade. We remark that the direct angular-mode cascade implies a \emph{transfer of energy} from low to high angular modes. Indeed, for $t=0$ we have 
\begin{equation*}
    \sum_{\ell}\mathbb{E}_{\ell}[\phi](0)=\mathbb{E}_{\ell=1}[\phi](0)+\mathbb{E}_{\ell=2}[\phi](0) = \mathbb{E}_{\text{nl}}[\phi](0)
\end{equation*}
(recall definition \eqref{nonlinear_energy_l_mode_b}), whereas for $t>0$, as the higher angular modes get excited, there are energies $\mathbb{E}_{\ell}[\phi](t)$ for $\ell\geq 3$ which become non-zero. By the conservation identity \eqref{conservation_fourier_energies}, the excitation of the higher-mode energies implies an energy transfer from the initial lower-mode energies $\mathbb{E}_{\ell=1}[\phi](0)$ and $\mathbb{E}_{\ell=2}[\phi](0)$ (i.e., the quantity  $\mathbb{E}_{\ell=1}[\phi](t)+\mathbb{E}_{\ell=2}[\phi](t)$ decreases over time).

\medskip

In Appendix \ref{app:ICs}, we show that a similar angular-mode cascade occurs for the (very) early time evolution of initial data of the same form as the ones considered in this section, but supported on different (finite) sums of low angular modes. There, in Figure \ref{fig:CoeffValsvsElls_ICs}, we observe that when the initial data contain more than two angular modes, the first generation consists of a correspondingly larger number of newly excited angular modes. Also, there are more excited modes in the first generation for initial data containing a single \emph{higher-frequency} mode than for initial data containing a single \emph{lower-frequency} mode.\footnote{Which can be understood by examining the spectrum of the corresponding cubic power $\phi^3$ of the initial data, as illustrated for the canonical case in Figure \ref{fig:SDIC_single}.} In addition to exciting a wider range of first-generation angular modes, we recall that higher-frequency angular modes are also characterised by a slower linear decay (see Section \ref{sec_lin_we}), and are thus expected to be longer-lived in the nonlinear evolution (this aspect is not captured by Figure \ref{fig:CoeffValsvsElls_ICs}, which displays the various $\ell$-spectra only for short times). The combination of these elements supports the expectation that more general, perhaps \emph{generic}, initial data, which are supported on all angular modes (in particular, on more angular modes and with higher frequency than the ones considered in this work), would still experience angular-mode dynamics with similar features to that described in this section. See also the related discussion in Section \ref{sec_discussion_beyond_scalar_model}.

\medskip

The accuracy of the values of the Fourier coefficients in Figures \ref{fig:SDIC_single}, \ref{fig:SDAsTime}, \ref{fig:CascadeOfModes} and \ref{fig:l_modes_Minkowski} (present section), Figure \ref{spectrum_epsilon_small_2_1} (following section), and Figures \ref{fig:CoeffValsvsElls_ICs} and \ref{fig:SDIC} (Appendix \ref{app:ICs}) depends on the choice of resolution for the numerical evolution of the solution and the precision of the interpolating procedure adopted to compute the integrals \eqref{Eq:SD} and \eqref{Eq:SD_3}. The details of the former are discussed in Appendix~\ref{app:convergence}. For the latter, we have used a fifth-degree spline interpolator and an adaptive quadrature with tolerance set to $10^{-12}$. The resulting errors are very small, i.e., they are typically at least three orders of magnitude smaller than the values of the Fourier coefficients plotted (Figure~\ref{fig:SDConvergence} in Appendix \ref{app:convergence} shows the numerical convergence of the Fourier coefficients for $\ell\in[1,\dots,15]$ for our canonical initial data case). We also note that the amplitude of the Fourier coefficients oscillate in time (as illustrated in Figure \ref{fig:CascadeOfModes}), and we have not performed any temporal averaging when presenting the data (nor, except for the bottom panel of Figure \ref{fig:CascadeOfModes}, do we show the envelope of the coefficients in time). 

\medskip

To conclude the section, we remark that the angular-mode cascade described is a genuinely \emph{nonlinear} phenomenon which is tied to the presence of \emph{stable} trapping. Indeed, the mixing of angular modes never occurs for \emph{linear} waves and, in particular, for the linear wave equation \eqref{WE_th} with initial data \eqref{proto_data}, each of the angular modes contained in the initial data decays rapidly in time (see Section \ref{sec_lin_solution_proto_data} and compare with the evolution of the $\ell\in[1,2]$-angular modes in Figures \ref{fig:SDAsTime} and \ref{fig:CascadeOfModes}). Moreover, although the mixing of angular modes typically occurs for nonlinear waves, the 
\begin{itemize}
    \item[(i)] slow dispersion of the angular modes contained in the initial data,
    \item[(ii)] slow dispersion of the excited angular modes, once these attain their maximum amplitude, and
    \item[(iii)] dominance of progressively higher angular modes over increasingly longer intervals of time in the evolution,
\end{itemize}
starting from initial data containing a small number of low angular modes, are more characteristic features of nonlinear waves \emph{on our model spacetime}. For example, solutions to equation \eqref{eqn_intro_NLW} on Minkowski space (no trapping) or Schwarzschild exterior spacetimes (\emph{un}stable trapping) would still exhibit the excitation of higher angular modes, but the excited modes, after attaining their maximum amplitude, would rapidly disperse; see Figure \ref{fig:l_modes_Minkowski} for an example on a Minkowski background. As our Fourier coefficients are numerically computed within the stable trapping region (i.e., at $r=r_0$), one can ascribe the characteristic features of our mode cascade to the presence of stable trapping.

\subsection{Small-data analysis} \label{sec_small_data}

In this section, we consider numerical solutions to equation \eqref{WE_nonlin_th} arising from the initial data prescribed in Sections \ref{sec_growth_derivatives} and \ref{sec_mode_cascade} (i.e., \eqref{proto_data} with the choices \eqref{Eq:piecewise_s}-\eqref{choices_ID}), but now with two smaller initial amplitudes $\epsilon=0.125$ and $\epsilon=0.25$. We recall that, for these two amplitudes, the evolution at early times appears to be dominated by the linear dynamics (at least for the quantity plotted in Figure \ref{fig:Amplitudes}). We view the dynamics of the numerical solutions analysed in this section as a preliminary hint at the dynamics of solutions to equation \eqref{WE_nonlin_th} arising from small (amplitude) initial data of the form considered. In light of the results that we shall describe, one may speculate that the characteristic features of the angular-mode cascade of Section \ref{sec_mode_cascade} would persist for \emph{arbitrarily small} initial data of the form considered.

\medskip

\begin{figure}
    \centering
    \includegraphics[scale=.45]{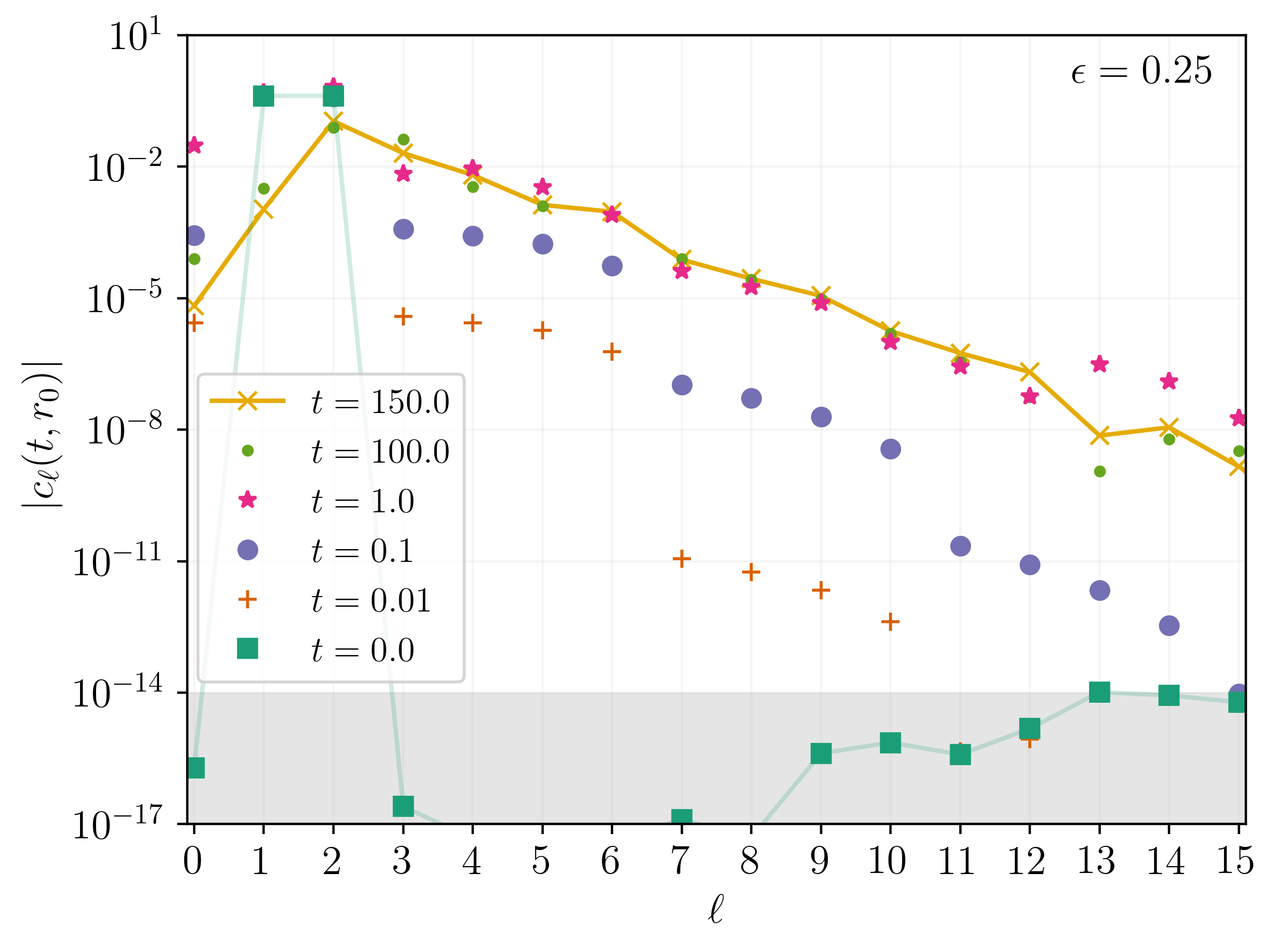}
    \includegraphics[scale=.45]{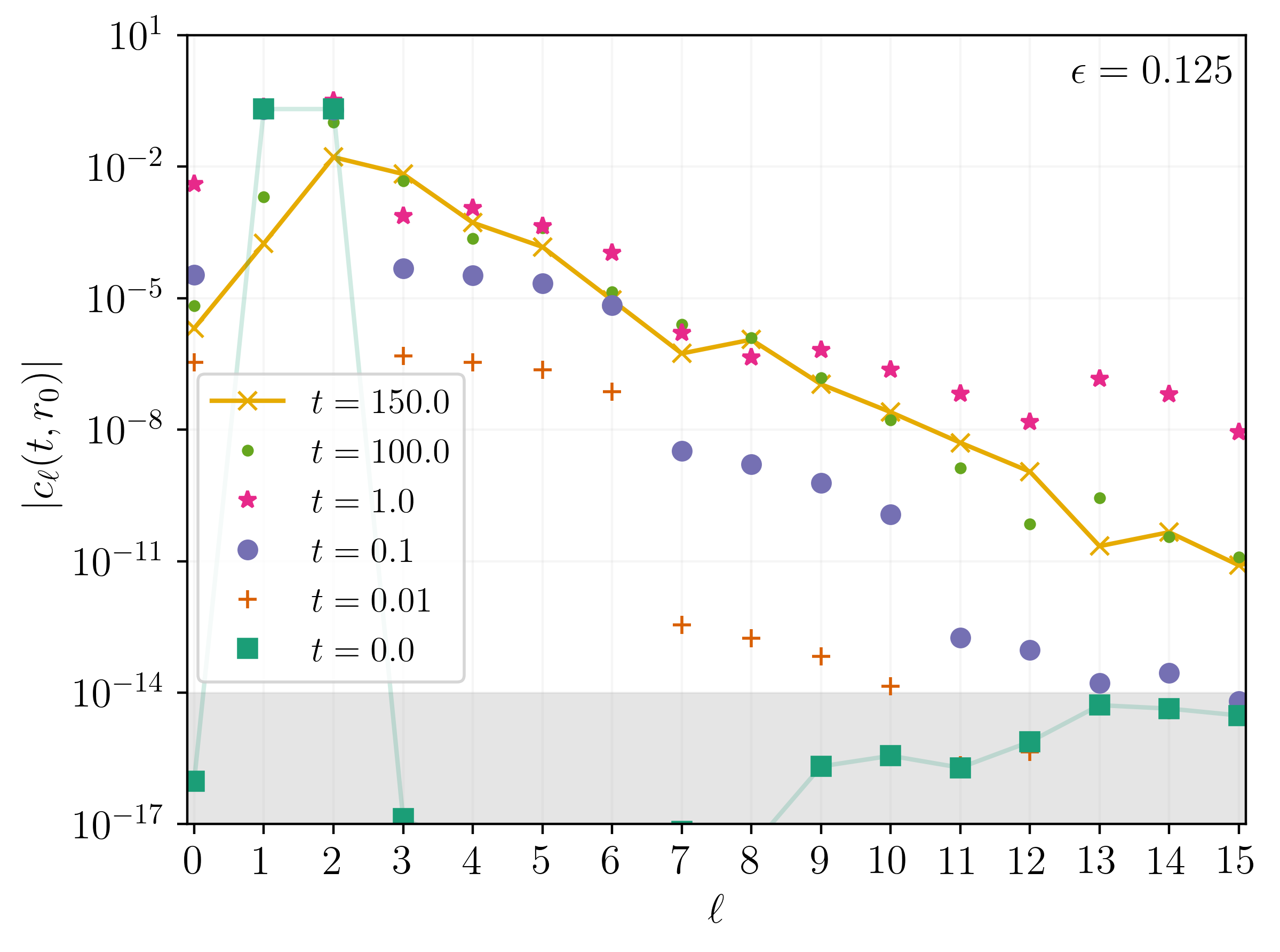}
    \caption{The $\ell$-spectrum of the solutions for $\epsilon=0.25$ (left panel) and $\epsilon=0.125$ (right panel) at different times in the evolution, as captured by the Fourier coefficients \eqref{Eq:SD}, with a logarithmic scale on the vertical axis. Lines joining the points for the times $t=0$ and $t=150$ have been added to guide the reader. The grey shading indicates where the numerical floor of our simulations is.}
    \label{spectrum_epsilon_small_2_1}
\end{figure}

Indeed, for both $\epsilon=0.125$ and $\epsilon=0.25$, we observe a direct angular-mode cascade characterised by (i) the slow dispersion of the angular modes contained in the initial data, (ii) the slow dispersion of the excited angular modes, once these attain their maximum amplitude, and (iii) the dominance of progressively higher angular modes over the evolution (see Figure \ref{spectrum_epsilon_small_2_1}). For what concerns the dominance of higher angular modes over the angular modes contained in the initial data, we point out that, strictly speaking, only the $\ell=1$-mode gets dominated within the time interval evolved (in particular, at $t=150$, the $\ell\in [3,4]$-modes dominate the $\ell=1$-mode in both cases shown in Figure \ref{spectrum_epsilon_small_2_1}). In fact, when comparing the angular-mode cascades for these two smaller amplitudes, we note that the excitation of the higher angular modes becomes \emph{less efficient} the smaller the amplitude of the initial data. For example, with the highest excited angular mode that we plot (i.e., $\ell=15$), the difference
\begin{equation*}
    |c_{\ell=2}(t,r_0)|-|c_{\ell=15}(t,r_0)| 
\end{equation*}
at $t=150$ is larger for smaller $\epsilon$ (see also Figure \ref{fig:SDAsTime} for a comparison with the $\epsilon=1$-case), resulting in steeper $t=150$-curves in Figure \ref{spectrum_epsilon_small_2_1} for smaller $\epsilon$. Despite the decreased efficiency of the mode cascade, we speculate that, over a longer interval of time than the one numerically evolved, the higher angular modes will eventually dominate \emph{both} the angular modes contained in the initial data, with the solution undergoing a qualitatively similar mode dynamics to the one illustrated in Figure \ref{fig:CascadeOfModes}\footnote{Recently, \cite{Redondo-Yuste:2025hlv} examined the same cubic equation as in this work, focusing on the regime of small initial data through a perturbative analysis of a dimensionally reduced model and found that  all higher-order energy norms grow, irrespective of the initial data's amplitude.}.

\medskip

In contrast with the $\epsilon=1$-case of Section \ref{sec_growth_derivatives}, we do not observe any growth of second derivatives of our numerical solutions for $\epsilon=0.25$ and $\epsilon=0.125$ within the time interval evolved. We argue that the absence of growth may be ascribed to the decreased efficiency of the angular-mode cascade: If the growth of second derivatives eventually occurs, it would become manifest over a longer interval of time than the one for which we can numerically evolve. Nonetheless, we expect sufficiently high-order derivatives (as well as sufficiently high-order energies) of the solutions to manifestly grow already over the time interval considered in this work. With the fourth-order finite difference methods we employ, computation of high-order derivatives becomes problematic in terms of the resolution that would be needed for accurate results, and so we leave to future investigations a thorough study of high-order quantities for small values of $\epsilon$, and in particular their potential growth.

\section{Discussion and outlook} \label{sec_discussion}

In this work, we have studied numerical solutions to the cubic (defocusing) wave equation \eqref{WE_nonlin_th} on a model geometry possessing stably trapped null geodesics. We have shown that initial data which are axisymmetric, radially localised within the stable trapping region, and supported on a small number of low-order multipole modes give rise to solutions whose high-order derivatives grow in time. The growth occurs within the stable trapping region and is induced by a direct angular-mode cascade, with higher-order modes being excited by the nonlinear interactions and eventually dominating the low-modes in the evolution. Moreover, we argued that the turbulent behavior of such solutions may persist for small (amplitude) initial data of the form considered.

\medskip

The growth of high-order derivatives hints at the possibility that higher-order quantities of our numerical solutions \emph{fail} to remain uniformly bounded in time. In particular, with reference to the growth of second-order derivatives of Figures \ref{Fig:GrowthOfPidot_Single} and \ref{Fig:GrowthOfTheta2_Single},\footnote{As noted at the end of Section \ref{sec_growth_derivatives}, the growth of high-order derivatives is in fact also observed for higher (than second) order. One may expect that the present considerations apply to higher-order energies $\mathbb{E}^k[\phi](t)$ for all $k\geq 2$.} there may \emph{not} exist a uniform bound for the second-order energy $\mathbb{E}^2[\phi](t)$ of the numerical solution in terms of its initial second-order energy $\mathbb{E}^2[\phi](0)$ (cf. inequality \eqref{unif_bddness_linear_waves} for $k=2$ in Theorem \ref{th_linear_waves} for \emph{linear} waves). The failure of uniform boundedness may originate from either an arbitrarily large (but finite) growth or an unbounded growth of the quantity $\mathbb{E}^2[\phi](t)$, and would be in stark contrast with known (mathematically rigorous) results on the decay of higher-order quantities for solutions to equation \eqref{WE_nonlin_th} on asymptotically flat spacetimes with \emph{un}stable trapping (e.g., on Schwarzschild exterior spacetimes). The uniform boundedness and, more generally, the asymptotic dynamics of our numerical solutions are left for future investigations.

\medskip

In the following, we will transition from our conclusions to discuss some open questions and contextualize our numerical results within the broader literature.

\subsection{Scalar turbulence and stable trapping beyond our model} \label{sec_discussion_beyond_scalar_model}

Extending our analysis to a broader class of wave models and more diverse initial data sets represents an exciting avenue for future research. However, due to the high computational cost and technical challenges in ensuring numerical convergence, a systematic study of these extensions lies beyond the scope of this manuscript. Nevertheless, we believe our current results provide a strong basis for future research aimed at refining these ideas within the broader context of nonlinear dispersive and gravitational wave phenomena.

In particular, based on the nature of the turbulent mechanism observed, we speculate on three aspects of our model problem which go beyond the numerical results presented in this work:
\begin{itemize}
    \item \emph{Generic initial data for equation \eqref{WE_nonlin_th} on our model spacetime.} For equation \eqref{WE_nonlin_th} on our model spacetime, the turbulent dynamics may arise for \emph{generic} smooth, compactly supported initial data (without any symmetry assumption). For initial data radially supported away from the stable trapping region, one expects that part of the solution would always reach and propagate into the effective potential well and trigger the turbulent evolution. See Sections \ref{sec_lin_solution_proto_data} and \ref{sec_mode_cascade} for some considerations on the persistence of the turbulent dynamics for initial data which are different from the ones considered in this work.
    \item \emph{Other nonlinear wave equations on our model spacetime.} The present work considers numerical solutions to a single nonlinear wave equation. Nonetheless, one would expect that the turbulent mechanism does \emph{not} rely on the precise structure of the nonlinearity, and thus the slow linear decay may lead to a turbulent dynamics of solutions to a large class of nonlinear wave equations on our model spacetime. The class may, in particular, include wave equations with derivative nonlinearities of the form \eqref{eqn_intro_NLW_null_condition}.
    \item \emph{Equation \eqref{WE_nonlin_th} on general spacetimes with stable trapping.} The turbulent dynamics of solutions to equation \eqref{WE_nonlin_th} may occur for more general spacetimes possessing stably trapped null geodesics,\footnote{We think of stationary spacetimes on which linear waves are uniformly bounded and uniformly decay in time.} meaning that the presence of stable trapping \emph{alone} may suffice to induce slow linear decay and the turbulent behaviour of nonlinear solutions. These more general spacetimes may be possibly characterised by different asymptotic structures, symmetries or number of dimensions from our model spacetime, or possess additional structures like an event horizon. Examples of such spacetimes may include the ones considered in \cite{SharpLogHolz, KeirLogLog, Benomio_Rings, Kunduri_slow_decay_solitons}.
\end{itemize}

\subsection{Turbulent gravitational waves} \label{sec_discussion_GW}

Whether turbulence (generically) persists for small \emph{gravitational} perturbations of our motivating spacetimes (i.e., ultracompact objects and black strings), and how it would affect their global dynamics, are questions which remain open. Nonetheless, there are two primary features of the scalar turbulent dynamics in our model problem which we argue already imply that the possible gravitational turbulent dynamics,\footnote{The considerations of these paragraphs focus on \emph{pure} gravitational wave perturbations, which are most relevant to the vacuum dynamics of black strings. With regards to ultracompact objects, if the {\em matter} comprising the object is governed by a nonlinear dynamics with similar properties to those discussed here, the evolution of these objects could be significantly impacted. The reason is that the turbulent cascade can then effectively ``drain'' the  core of the ultracompact object of matter, affecting its dynamics; we will discuss more of these arguments later in the section.} though likely leading to a similar turbulent cascade in the spectrum of partially trapped gravitational waves, would {\em not} have any dire consequences for the evolution of the corresponding spacetimes. First, the {\em physical} energy density of scalar waves, i.e., the (first-order) energy density of our numerical solutions, does {\em not} grow with time. Second, the effective potential barrier generating the stable trapping of scalar waves has finite height, and hence the (first-order) energy within the trapping region of initially trapped waves {\em decays} with time, albeit slowly. Moreover, we note that the direct cascade to arbitrarily small length-scales would {\em not} a-priori constitute an obstruction to considering the global gravitational turbulent dynamics of these spacetimes within the regime of classical general relativity. Indeed, where the classical regime fails as a theory predicting the structure of spacetime, and would need a theory of ``quantum gravity'' to take over, is when singularities (clothed or not) or Cauchy horizons form, neither of which is length-scale dependent in general relativity.\footnote{Even considering some hypothetical properties of ``quantum gravity,'' the cascade would not necessarily be problematic. For example, suppose that local Lorentz invariance fails below some minimum length-scale, say the Planck length $\ell_p$. That would imply that the classical description of the cascade fails once ``gravitons'' with wavelength $\lambda\sim\ell_p$ are produced. What happens then requires the dynamics of the quantum theory, but it does not seem implausible that the cascade is simply halted at this scale.}

\medskip

Beyond this, one can only speculate on the global dynamics of the possible gravitational turbulent dynamics, in that our model problem does not capture the back-reaction of partially trapped gravitational perturbations on the spacetime. In the following paragraphs, we discuss some lines of speculation in this direction.

\medskip

Suppose that a gravitational perturbation is initially localised within a stable trapping volume of characteristic size $R$ and its (first-order) energy is initially small. For such a perturbation, the energy within the stable trapping volume will remain small during evolution, meaning that the \emph{average} back-reaction over the trapping volume induced by the perturbation will remain small over time.\footnote{Leaving aside the issue of not being able to define a gauge-invariant quasi-local gravitational wave energy density.} While the average back-reaction remains small, the turbulent dynamics drives a direct energy cascade transferring part of the energy of the perturbation to progressively smaller wavelengths $\lambda$. In particular, one could envision a late-time state where part (or perhaps most) of the energy transfers to modes with very small characteristic wavelengths $\lambda\ll R$, raising the question of whether a black hole of radius of order $\lambda$ could form. Since the energy of the perturbation within the trapping volume remains small at all scales and no manifest focusing mechanism of the perturbation appears to be present, we argue that this would be very unlikely. Indeed, by hoop conjecture arguments, for a hyperspherical black hole of radius of order $\lambda$ to form, the energy density of the perturbation within the corresponding volume would need to be of order $\lambda^{-2}$ (and thus be very large for a black hole to form at very small scales). Reaching such a density would require a ``rogue wave'' : a constructive superposition of waves at a location resulting in an anomalously large amplitude, hence energy density there. The formation of rogue waves has been observed as a rare phenomenon in various wave and dispersive systems in physics, one notable example being the propagation of light in a reflective optical fiber \cite{rogue_waves_optics}. However, in our dynamics it appears unlikely that this type of rare phenomenon may {\em generically} lead to gravitational collapse. This is both because of the continuous transfer of energy to progressively smaller scales within the direct turbulent cascade, and that the effective confinement due to stable trapping is only partial, implying that the probability of black hole formation from such superpositions at any given scale decreases with time.
In the case of black strings, we remark that, even if the turbulent dynamics were to lead to a small black hole forming within the trapping region at some intermediate time, one would expect that the small black hole will eventually merge with the central horizon, thus not affecting the nature of the end state of the dynamics.

\medskip

We point out that, as opposed to complete dispersion, there may exist end states where a finite amount of energy remains within the trapping region, i.e., a kind of gravitational geon which condenses outside the central object. However, we would expect the eventual dispersion of the initially trapped, small gravitational perturbation as the \emph{generic} end state of the global dynamics.

\medskip

We observe that two factors, if they occur for our motivating spacetimes, could alter the picture described above:  
\begin{itemize}
    \item \emph{Rotating black strings.}~If the spacetime is a \emph{rotating} black string, the angular momentum of the spacetime is a potential source of energy to feed the turbulent dynamics of a trapped gravitational perturbation. Nonetheless, in view of the arguments already discussed, we argue that the additional source of energy would still unlikely lead to black hole formation through the turbulent cascade. On the other hand, rotating black strings are known to suffer from a superradiant instability \cite{Rosa_Black_String_Bomb}, which would likely dominate the global dynamics of these spacetimes.
    \item \emph{Ultracompact objects for certain matter models.}~Different considerations may apply to ultracompact objects when governed by nonlinear matter fields which undergo a similar turbulent dynamics (to the one described in vacuum) within the region of stable trapping (and for which other nonlinear dissipative mechanisms, such as viscosity in a fluid, are not strong enough to limit or halt the turbulent cascade). In particular, since the outer region of the central object in such spacetimes must necessarily be within the region of stable trapping, the central object may possibly provide a source of energy to feed the turbulent dynamics of trapped {\em matter} perturbations. Similar considerations to those above suggest it is unlikely that a large enough ``rogue'' matter wave could form in the trapped region to lead to small black hole formation there. Instead, for star-like compact objects, what could significantly affect the dynamics of the system comes from the transfer of energy from the core to the region of stable trapping, driven by the turbulent cascade in the latter. This would lead to an effective migration of the core along the star's mass-radius curve, with the two likely outcomes being either the star migrates to a larger, less dense configuration without a stable trapping region, or it evolves the opposite way and collapses to a black hole (see~\cite{Cardoso_et_al_light_ring_instability} for similar conjectures on the endstate, as well as how the presence of an ergoregion may influence the dynamics). These outcomes are consistent with what was shown to occur in numerical simulations of two classes of ultracompact bosonic stars in~\cite{Cunha_Herdeiro_Radu_fate_light_ring_instab_compact_objects}. For \emph{shell-like} ultracompact objects (such as AdS black bubbles or gravastars), the matter present in the spacetime is confined to the shell region, and thus the turbulent dynamics would \emph{not} be expected to significantly change the overall properties of the shell (especially if the shell material is viscous). The relevant questions on the global dynamics of these latter objects would then revert to those for partially trapped gravitational perturbations.
\end{itemize}

\subsection{Related works on turbulence} \label{sec_discussion_ads}

Wave turbulence is a classical theme in the analysis of model equations from physics \cite{Majda_McLaughlin_Tabak_wave_turbulence, Zakharov_wave_turbulence_1D_models}. Nonlinear models for which the existence of turbulent solutions has been widely studied include examples of nonlinear dispersive equations on compact domains, such as the nonlinear Schr\"{o}dinger equation on the torus \cite{kuksin_turbulence, Colliander_KSTT_growth_sobolev_nls} and the nonlinear Sz\"{e}go equation on the circle \cite{Gerard_Grellier_cubic_szego_equation}.

\medskip

In the realm of general relativity, turbulence due to perfect confinement arises in the nonlinear dynamics of Anti-de Sitter (AdS) space, when reflective boundary conditions are imposed at the conformal boundary.\footnote{For the present discussion, we always intend that \emph{reflective} boundary conditions are imposed. In this setting, the scalar dynamics of AdS space is already remarkably rich (see, for instance, \cite{Maliborski_thesis} and the references therein).} The numerical work \cite{Bizon_Rot_Instab_AdS} shows the nonlinear instability of AdS space as a solution to the Einstein--scalar field system (with negative cosmological constant) in spherical symmetry (see also the related works \cite{Maliborski_instability_minkowski_cavity, Okawa_Cardoso_Pani_cavity_problem}): There exist arbitrarily small perturbations which exhibit a turbulent dynamics leading to the formation of a black hole in finite time.\footnote{The dynamics of \emph{generic} small-data spherically symmetric solutions to the system is however unclear (see \cite{Dimitrakopoulos_et_al_physical_space_analysis_ads, Balasubramanian_et_al_stability_ads, Craps_Evnin_Vanhoof_1, Craps_Evnin_Vanhoof_2} for some examples).} More recently, works \cite{Moschidis_AdS, Moschidis_AdS_Vlasov} have given a first rigorous proof of instability of AdS space as a solution to the Einstein--null dust and Einstein--massless Vlasov systems (with negative cosmological constant) in spherical symmetry, in which turbulence is again the driving mechanism leading to black hole formation. Without any symmetry assumption, works \cite{Daf_Holz_Conj_AdS_instab, Daf_talk_AdS} and \cite{Anderson_AdS} have independently conjectured that AdS space remains nonlinearly unstable (to black hole formation) as a solution to the \emph{vacuum} Einstein equations (with negative cosmological constant). The mathematically rigorous resolution of the conjecture remains a major open problem in the subject (see \cite{Horowitz_Santos_geons_instability_ads, Dias_Santos_ads_instability_beyond_spherical_symmetry, Rostworowski_time_periodic_perturbations_ads} for some numerical investigations).

\medskip

The dynamics of asymptotically AdS \emph{black holes} with reflective boundary conditions may also exhibit some turbulent behaviour. Indeed, linear waves on Schwarzschild--AdS and Kerr--AdS black hole exteriors (the latter satisfying the so-called Hawking--Reall bound) only decay sharp-logarithmically in time \cite{Decay_KG_Kerr-AdS, SharpLogHolz}, due to the presence of stable trapping. For linearised gravitational perturbations, one thus expects these spacetimes to be only weakly stable (see \cite{Graf_Holzegel_mode_stability} and  \cite{Graf:2024nni, Graf:2024mui, Graf:2024yug} for rigorous results in this direction). The nonlinear (both scalar and gravitational) dynamics of these spacetimes remains, for the most part, to be understood.\footnote{For some numerical works on the nonlinear dynamics of asymptotically AdS black holes, see \cite{Dias_et_al_stability_ads_bhs, Ficek_Maliborski_nonlinear_scalar_field_SAdS, Figueras_Rossi_instability_kerr_AdS}.} Based on the numerical result presented in this work and in line with what discussed for the expected dynamics of our motivating spacetimes, one may speculate that these spacetimes similarly suffer from a turbulent (scalar and gravitational vacuum) dynamics.\footnote{It has been recently announced a first rigorous result on the existence of turbulent solutions to a model nonlinear wave equation on Schwarzschild--AdS black hole exteriors \cite{Kehle_Moschidis_talk_turbulence_SAdS}.} Examples of turbulent black holes already appear, and have been numerically investigated, in different scenarios \cite{Adams_chesler_liu_holographic_turbulence, Yang_Zimmerman_Lehner_turbulent_BHs}.

Recently, \cite{Redondo-Yuste:2025hlv} examined the same cubic equation as in this work, focusing on the regime of small initial data through a perturbative analysis of a dimensionally reduced model. Their study showed that the nonlinear wave equation on the sphere with added dissipation captures several key features of the full nonlinear problem presented here, accurately reproducing the dynamics of low $\ell$-modes observed in our numerical simulations. Since their perturbative approach grants access to high $\ell$-modes, dynamics that remain inaccessible in the four-dimensional problem due to numerical limitations, they were able to show that an inertial range emerges, characterized by a Kolmogorov--Zakharov-like spectrum, and that all higher-order energy norms grow, irrespective of the initial data's amplitude. These complementary findings not only reinforce our conjectures but also extend our understanding of the system beyond the reach of full numerical simulations.

\begin{acknowledgments}
We thank G. Holzegel, L. Lehner, J. Redondo-Yuste, and N. Siemonsen for their useful comments. G.B. was a Gravity Initiative Fellow at Princeton University when most of the research leading to this work was carried out. A.C.-A. acknowledges support from the Simons Foundation and the DOE through Los Alamos National Laboratory (LANL) Directed Research and Development, grant 20240748PRD1, as well as by the Center for Nonlinear Studies. This work is authorized for unlimited release under LA-UR-24-30368. FP acknowledges support from the NSF through the grant PHY-220728. Some of the simulations presented in this work were performed on computational resources managed and supported by Princeton Research Computing, a consortium of groups including the Princeton Institute for Computational Science and Engineering (PICSciE) and the Office of Information Technology's High Performance Computing Center and Visualization Laboratory at Princeton University, as well as resources provided by the LANL Institutional Computing Program. Map colors were based on www.ColorBrewer.org, by Cynthia A. Brewer, Penn State.
\end{acknowledgments}

\appendix

\section{Initial data and angular modes} 
\label{app:ICs}

In this appendix, we consider initial data for equation \eqref{WE_nonlin_th} of the form
\begin{align}
\phi|_{t=0}&=\frac{1}{\mathbb{E}_{\text{nl}}[\hat{\phi}](0)}\underbrace{\left(\epsilon \, u(r)\cdot \sum_{\left[\ell\right]}\ell^{-1}Y_{0}^{\ell}(\vartheta)\right)}_{:=\hat{\phi}} \, ,  & \partial_{t}\phi|_{t=0}&=0 \, ,  \label{proto_data_appendix}
\end{align}
with the choices 
\begin{align*}
\epsilon &= 1 \, , &  r_{0,1}&=0.02 \, ,  &  r_{0,2}&=0.6 
\end{align*}
and
\begin{equation*}
    u(r)=4000 \cdot \frac{\left(r-r_{0,1}\right)^{4}\left(r-r_{0,2}\right)^{4}}{\left(r_{0,1}-r_{0,2}\right)^{8}}
\end{equation*}
for $r\in [r_{0,1}, r_{0,2}]$ and identically vanishing otherwise. Up to the rescaling by the initial energy $\mathbb{E}_{\text{nl}}[\hat{\phi}](0)$ and by the factors $\ell^{-1}$, the initial data \eqref{proto_data_appendix} are of the same form as the initial data \eqref{proto_data} considered in Sections \ref{sec_growth_derivatives} and \ref{sec_mode_cascade}. 

\medskip

The aim of this appendix is to compare the numerical evolutions of the initial data \eqref{proto_data_appendix} for different choices of $[\ell]$. The additional rescalings in \eqref{proto_data_appendix} are introduced so that, \emph{for any choice of $[\ell]$}, the initial data possess the same energy (i.e., $\mathbb{E}_{\text{nl}}[\phi](0)=1$), which is moreover (approximately) equidistributed among the $\ell$-angular modes contained in the initial data.\footnote{Without the $\ell^{-1}$-rescaling in \eqref{proto_data_appendix}, the energy \eqref{nonlinear_energy_l_mode} of the individual $\ell$-angular modes contained in the initial data would scale like $\ell(\ell+1)$. A choice $[\ell]=[\ell_{\text{low}},\ell_{\text{high}}]$ pairing a low angular mode with a very high angular mode would result in initial data with $\mathbb{E}_{\text{nl}}[\phi](0)=1$ and $E_{\ell_{\text{low}}}[\phi](0)\ll E_{\ell_{\text{high}}}[\phi](0)$.} In our comparative analysis, this will be convenient for isolating properties of the solution which only depend on the choice of $[\ell]$.

\medskip

\begin{figure}
\includegraphics[width=\textwidth]{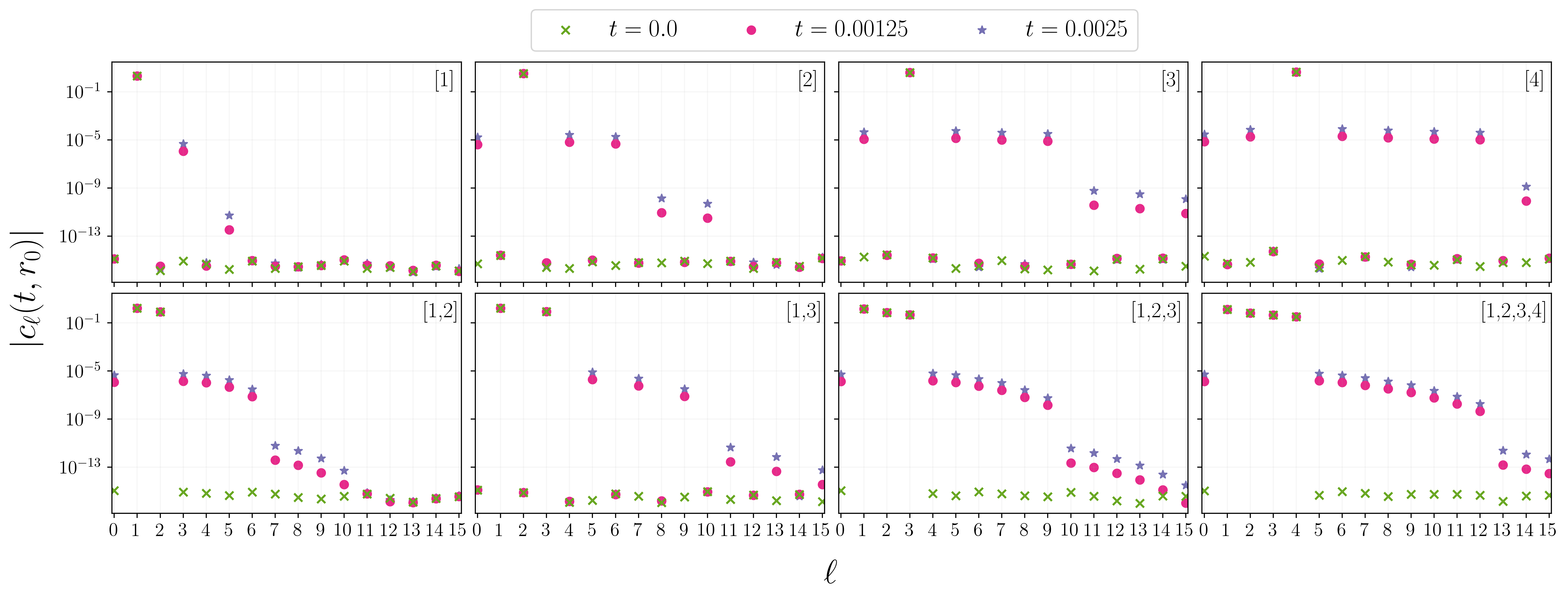}
    \caption{The $\ell$-spectrum of the solution at different times in the evolution, as captured by the Fourier coefficients \eqref{Eq:SD}, with a logarithmic scale on the vertical axis. Each panel corresponds to the choice of $[\ell]$ indicated in the top-right corner of the panel. The lowest value on the vertical axis corresponds to the numerical floor of our simulation. The top panels show initial data supported on a single angular mode, whereas the bottom panels show initial data supported on different combinations of angular modes. We note that the evolution times considered in the figure are much shorter than the ones in Figure \ref{fig:SDAsTime}.} 
    \label{fig:CoeffValsvsElls_ICs}
\end{figure}

In Figure \ref{fig:CoeffValsvsElls_ICs}, we examine the time evolution of the $\ell$-spectrum of the numerical solutions for different choices of $[\ell]$. As for the case $[\ell]=[1,2]$ analysed in Figure \ref{fig:SDAsTime}, one can see consecutive generations of angular modes getting excited in the evolution. Which excited modes are included in the first generation depends on the choice of $[\ell]$. Moreover, we note that single-mode initial data with higher $\ell$ yields a first generation of excited modes which is more populated (compare the first panel for $[\ell]=[1]$ with the fourth panel for $[\ell]=[4]$ in the figure). For two-mode initial data, the combination of an odd mode with an even mode yields a more populated first generation of excited modes than an odd-odd combination (compare the fifth panel for $[\ell]=[1,2]$ with the sixth panel for $[\ell]=[1,3]$ in the figure). Richer initial mode-configurations lead to more populated first generations of excited modes (see seventh and eight panel). We also note that single-odd-mode (single-even-mode) initial data and odd-odd (even-even) two-mode initial data only excite odd (even and $\ell=0$) modes.

\medskip

\begin{figure}
\includegraphics[scale=.5]{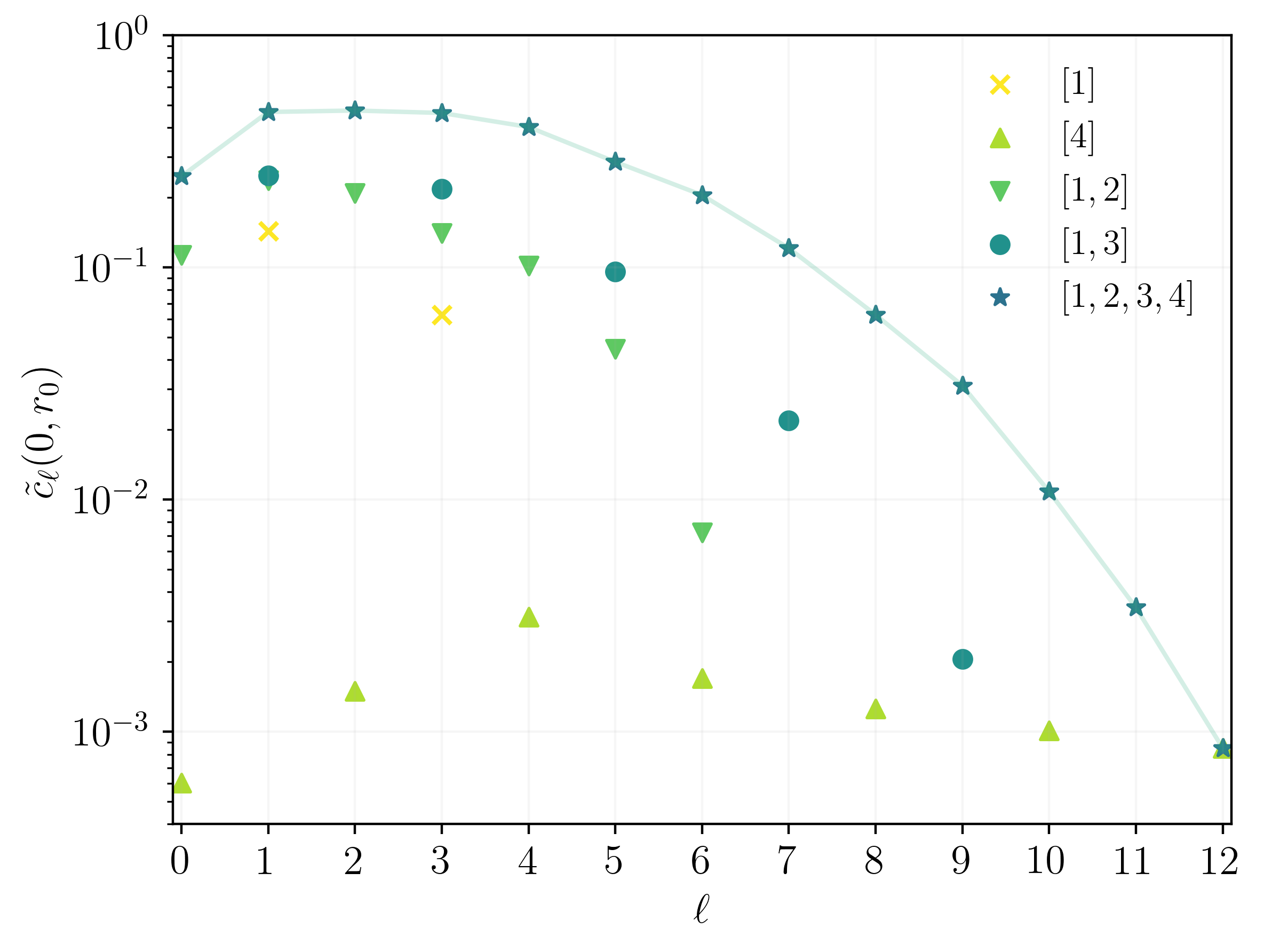}
    \caption{The $\ell$-spectrum of the cubic power $\phi^3$ of the solution at the initial time, as captured by the Fourier coefficients \eqref{Eq:SD_3}, with a logarithmic scale on the vertical axis. The range of the vertical axis is so as to include all the non-trivial values for the Fourier coefficients considered (i.e., no points appear between the lower end of the axis and the numerical floor of our simulation, cf. Figure \ref{fig:SDIC_single}). Each set of points corresponds to the initial data \eqref{proto_data_appendix} for the choice of $[\ell]$ indicated in the top-right corner of the panel. A line joining the points for the case $[\ell]=\left[1,2,3,4\right]$ has been added to guide the reader. We note that the Fourier coefficients plotted are all positive, with no absolute value applied to the Fourier coefficients on the vertical axis.}
    \label{fig:SDIC}
\end{figure}

The $\ell$-spectrum of the cubic power of the solution $\phi^3$ at the initial time is supported on a range of angular modes which depends on the choice of $[\ell]$ (see Figure \ref{fig:SDIC}, to be compared with Figure \ref{fig:SDIC_single}). The $\ell$-spectra of Figure \ref{fig:SDIC} drive the corresponding first generations of excited modes of Figure \ref{fig:CoeffValsvsElls_ICs}. In agreement with what is observed in Figure \ref{fig:CoeffValsvsElls_ICs}, the $\ell$-spectrum in Figure \ref{fig:SDIC} is richer for higher (rather than lower) single-mode initial data, for odd-even (rather than odd-odd) two-mode initial data and for multiple-mode (rather than single or two-mode) initial data. One can also read-off the parity of the excited modes described in Figure \ref{fig:CoeffValsvsElls_ICs}.

These results for different initial conditions indicate that, while we observe qualitative similarities in the energy-transfer process shown in the main text, a complete understanding of how to enhance or suppress the turbulent dynamics robustly remains an open question.

\section{Convergence tests} 
\label{app:convergence}

Let $h$ be a grid function from three levels of increasing resolutions $h^{\textrm{low}}$, $h^{\textrm{mid}}$ and $h^{\textrm{high}}$. We test the convergence of our numerical simulations by computing the convergence factor\footnote{The $\ell^2$-norm in \eqref{convergence_factor} is taken over the common grid points, and therefore we do not have to interpolate the solution to compute the convergence factor.}
\begin{equation} \label{convergence_factor}
    Q_{h}(t)=\frac{1}{\ln2}\ln\left(\frac{\left\lVert h^{\textrm{mid}}-h^{\textrm{low}}\right\rVert_{\ell^2}(t)}{\left\lVert h^{\textrm{high}}-h^{\textrm{mid}}\right\rVert_{\ell^2}(t)}\right) \, ,
\end{equation}
where the grid spacing of the coarsest resolution is $(\Delta_Y,\Delta_Z)=(\Delta^{\textrm{low}}_Y,\Delta^{\textrm{low}}_Z)$ and is decreased by a factor of $2$ for subsequent higher resolutions (e.g., $\Delta_Y^{\rm high}=\, \Delta^{\rm mid}_{Y}/2=\, \Delta^{\rm low}_{Y}/4$). For all our simulations, we have set the Courant factor to $C=0.5$ (i.e., $\Delta_t=0.5 \, \Delta_Y$) and applied a sixth-order Kreiss--Oliger dissipation, with $\epsilon_{\rm KO}=0.3$.

\medskip

\begin{figure}
    \includegraphics[scale=.5]{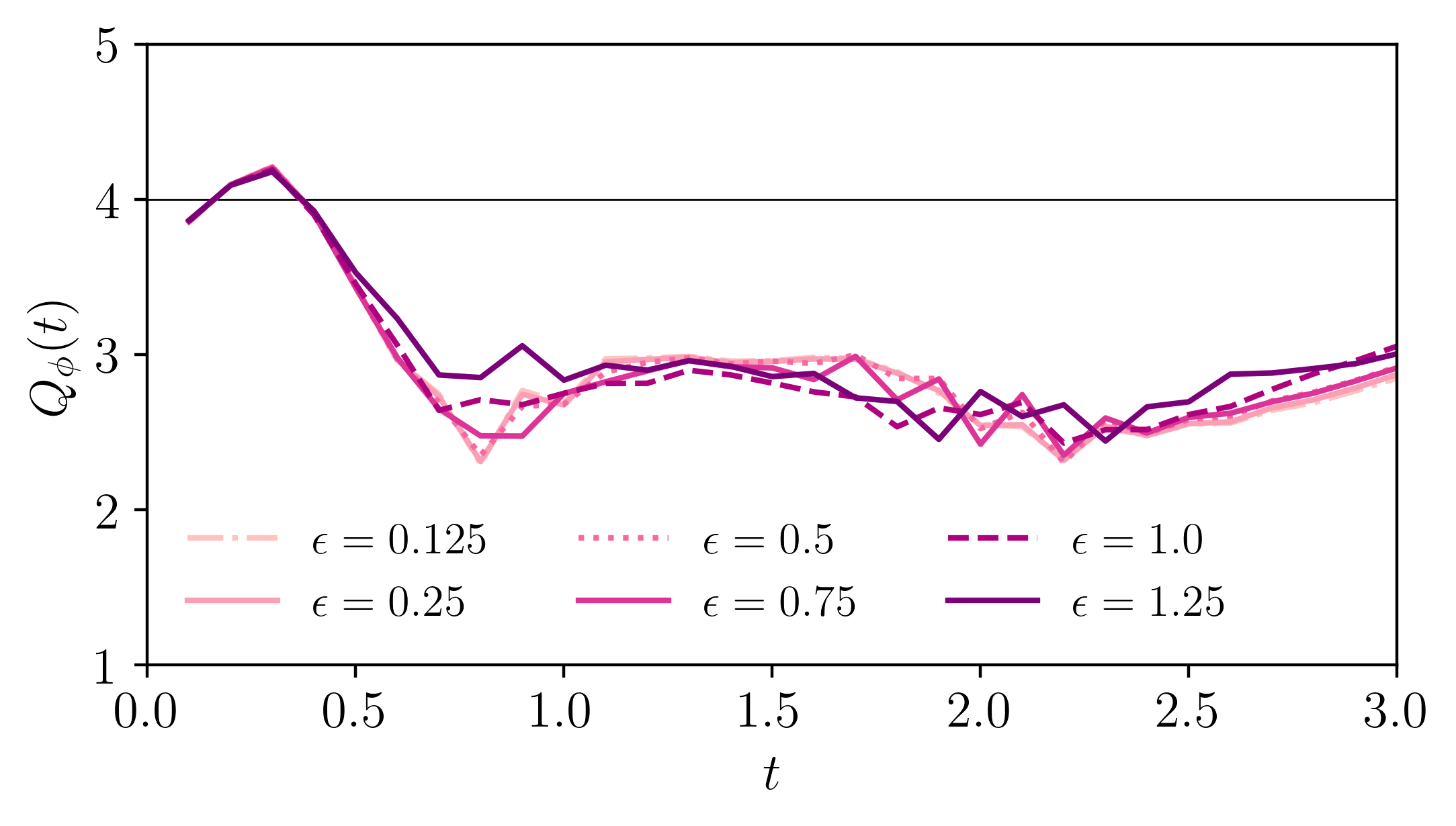}
    \caption{The evolution of the convergence factor \eqref{convergence_factor} over the time interval $t\in[0,3]$ for the numerical solutions analysed in Figure \ref{fig:Amplitudes}.}
    \label{fig:ConvergenceLinToNonLin}
\end{figure}

\begin{figure}
\includegraphics[scale=.5]{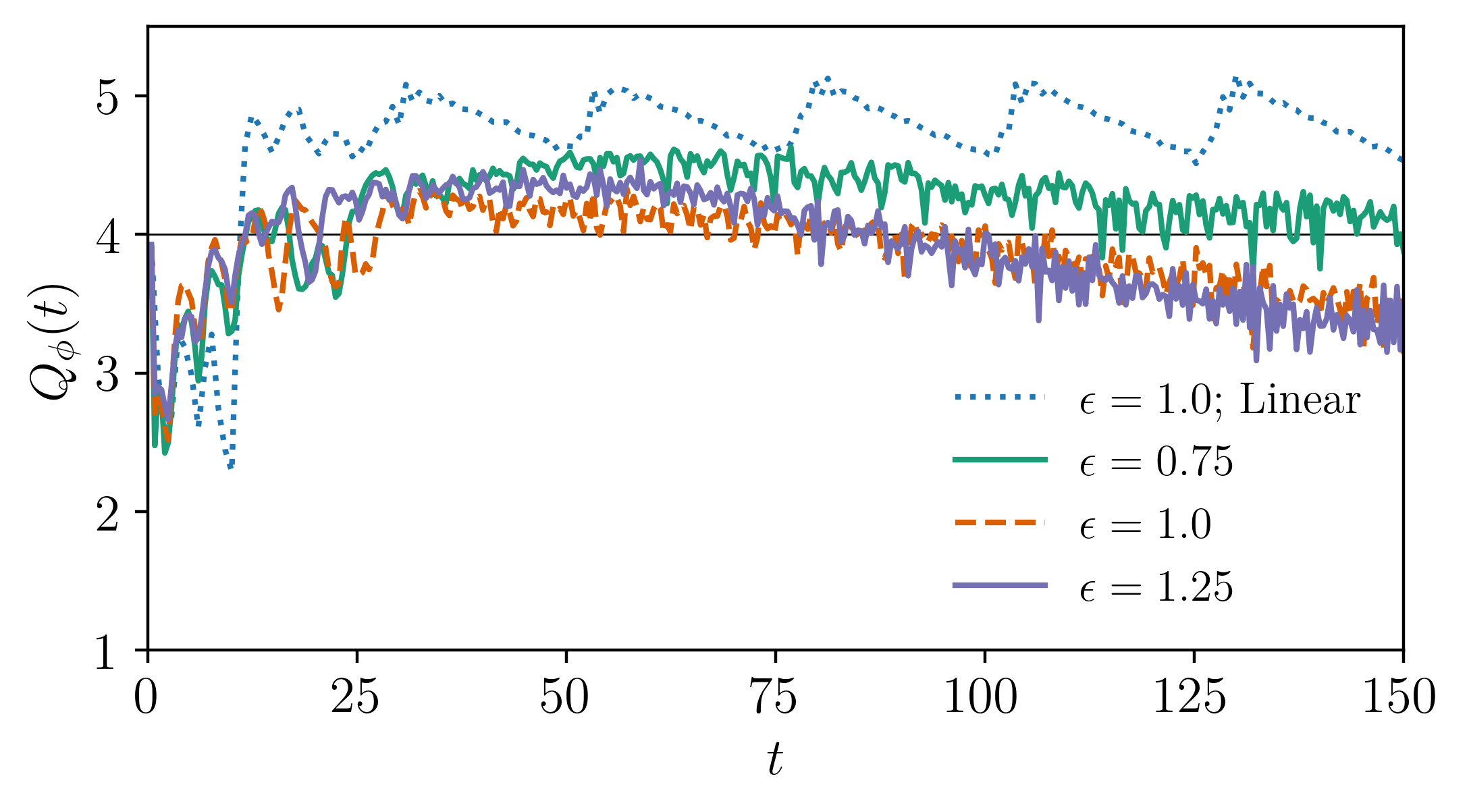}
    \caption{The evolution of the convergence factor \eqref{convergence_factor} over the time interval $t\in[0,150]$ for the numerical solutions analysed in Figures \ref{fig:Snapshots}, \ref{Fig:GrowthOfPidot_Single}, \ref{Fig:GrowthOfTheta2_Single}, \ref{fig:NlinVsLin}, \ref{Fig:GrowthOfPidot} and \ref{Fig:GrowthOfTheta2}.}
    \label{fig:Convergence}
\end{figure}

\begin{figure}
\includegraphics[scale=.5]{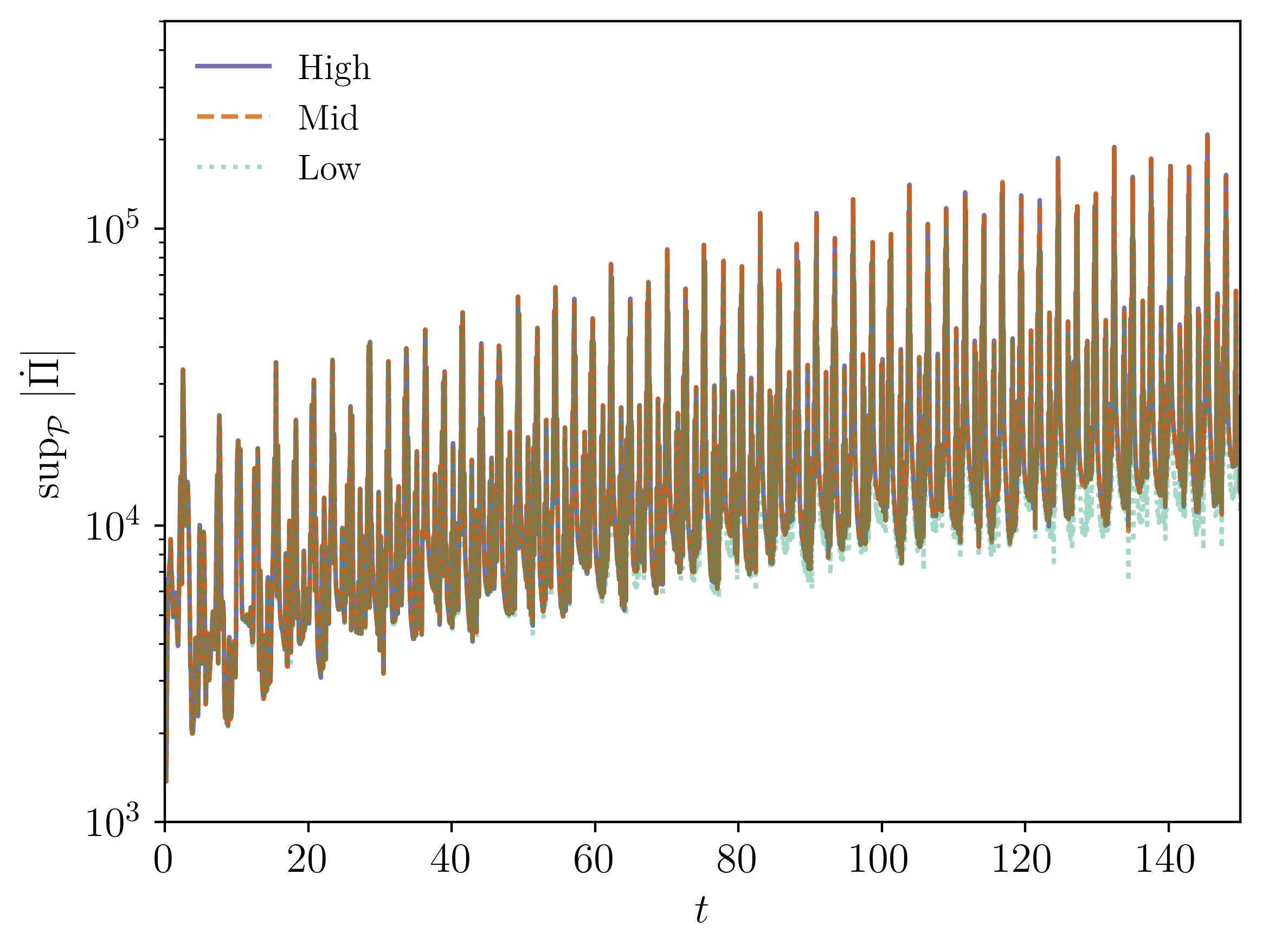}
    \caption{The evolution of the quantity \eqref{sup_pi_dot} over the time interval $t\in[0,150]$ for $\epsilon=1$ at three different (increasing by a factor of two) resolutions, starting with $(\Delta^{\rm low}_{Y},\Delta^{\rm low}_{Z})=(0.00125,0.00125)$. The numerical error in this quantity is very small. At $t=150$, for example, the relative error computed using the Mid and High resolutions is about $0.58\%$.}
    \label{fig:ConvergencePidot}
\end{figure}

\begin{figure}
\includegraphics[scale=.5]{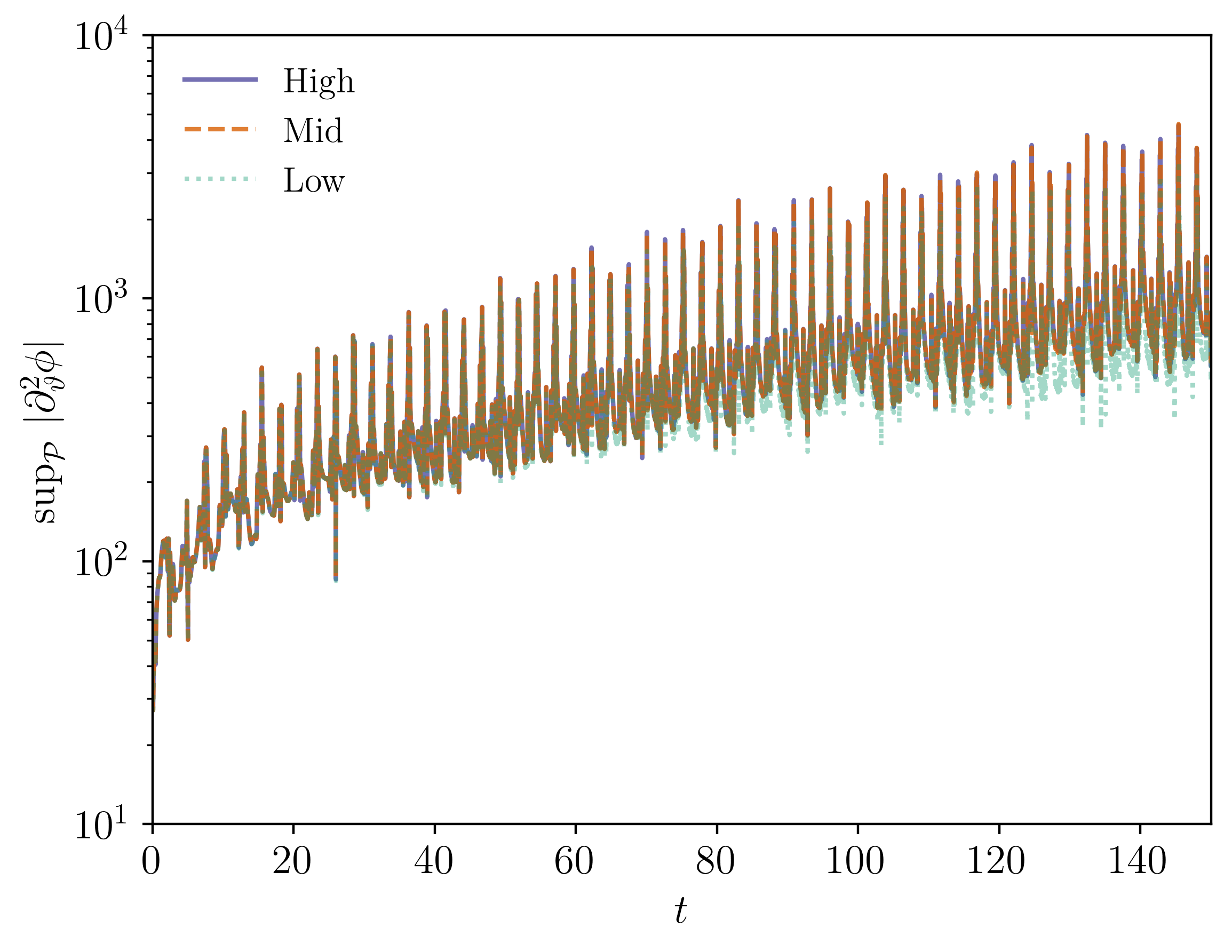}
    \caption{The evolution of the quantity \eqref{sup_2nd_deriv_theta} over the time interval $t\in[0,150]$ for $\epsilon=1$ at three different (increasing by a factor of two) resolutions, starting with $(\Delta^{\rm low}_{Y},\Delta^{\rm low}_{Z})=(0.00125,0.00125)$.The numerical error in this quantity is very small. At $t=150$, for example, the relative error computed from the Mid and High resolutions is about $0.55\%$.}
    \label{fig:Convergencetheta2}
\end{figure}

Figures \ref{fig:ConvergenceLinToNonLin} and \ref{fig:Convergence} show the time evolution of the convergence factor $Q_{\phi}(t)$ for the numerical solution to equation \eqref{WE_nonlin_th} arising from the initial data analysed in Sections \ref{sec_growth_derivatives} and \ref{sec_mode_cascade}. Analogously, Figures \ref{fig:ConvergencePidot} and \ref{fig:Convergencetheta2} show the values of the quantities \eqref{sup_pi_dot} and \eqref{sup_2nd_deriv_theta} respectively, for three increasing resolutions, illustrating how small the numerical error is for these quantities. Different choices of $\epsilon$ for the amplitude of the initial data are considered. For $\epsilon=1$, Figure \ref{fig:Convergence} also plots the time evolution of the convergence factor for the linear solution to equation \eqref{WE_th}. The resolution adopted for the numerical simulations is $(\Delta^{\textrm{low}}_Y,\Delta^{\textrm{low}}_Z)=(0.00125,0.00125)$, meaning that, for the highest resolution $(\Delta^{\textrm{high}}_Y,\Delta^{\textrm{high}}_Z)$, we evolve a grid size of $3201\times6401$. From Figure \ref{fig:ConvergenceLinToNonLin}, one sees that, at early times, the convergence factor is lower than fourth-order, with approximately third-order convergence at time $t=3$. Nonetheless, Figure \ref{fig:Convergence} shows that, by time $t=25$, we achieve approximately fourth-order convergence (recall that our numerical scheme is indeed forth-order, see Section \ref{sec_numerical_implementation}) for the cases $\epsilon=\lbrace 0.75, 1, 1.25\rbrace$ (which are the ones analysed in Sections \ref{sec_growth_derivatives} and \ref{sec_mode_cascade}). We remark that, in particular, Figure \ref{fig:Convergence} exhibits approximately fourth-order convergence over the time interval for which the growth of second-order derivatives of the numerical solutions is observed in Section \ref{sec_growth_derivatives}.

\medskip

\begin{figure}
    \includegraphics[scale=.5]{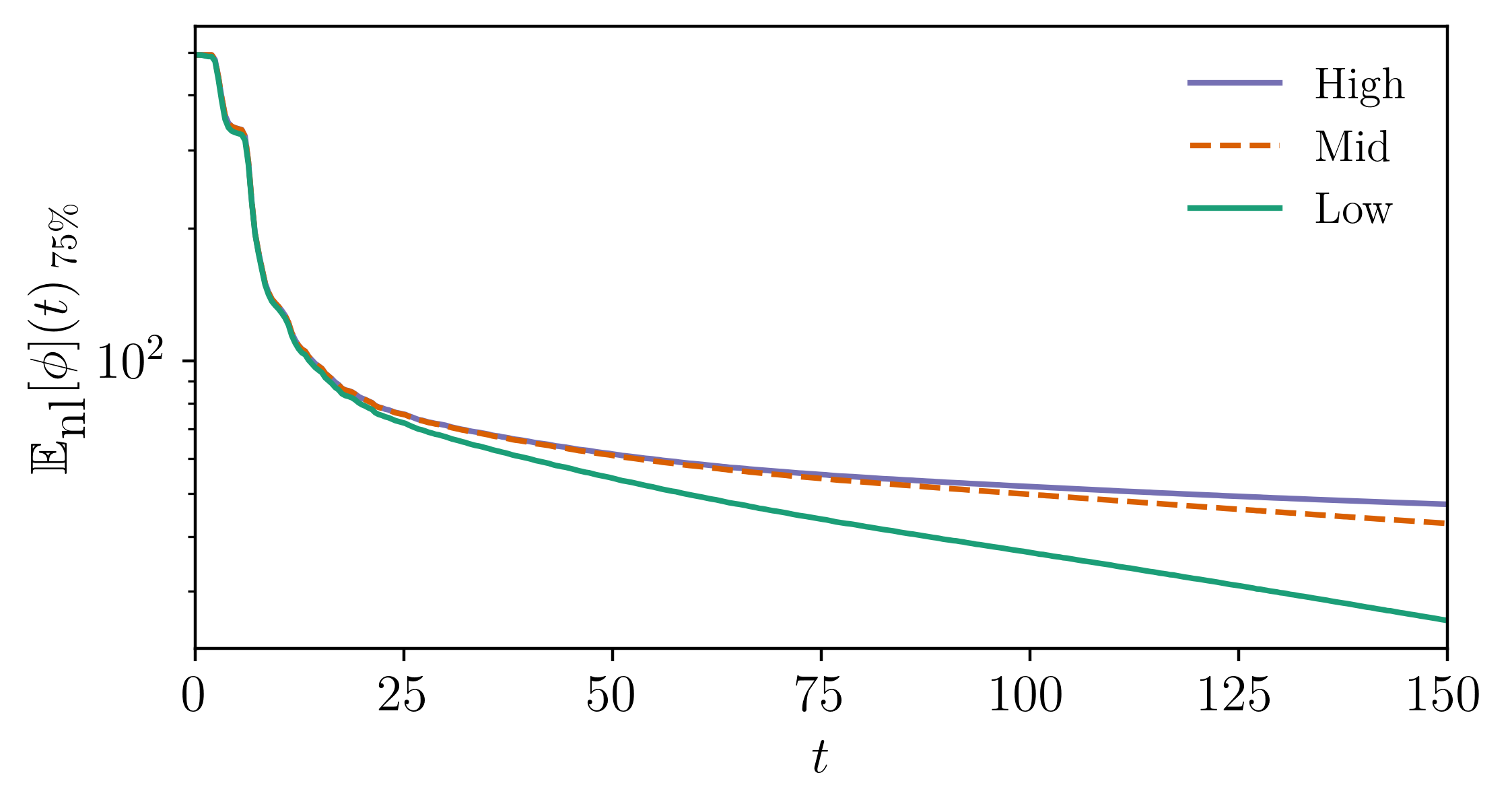}
    \caption{The evolution of the energy \eqref{energy_nonlin_waves} over the time interval $t\in [0,150]$ for the numerical solutions analysed in Figures \ref{fig:Snapshots}, \ref{Fig:GrowthOfPidot_Single} and \ref{Fig:GrowthOfTheta2_Single}. The energy is computed over $75\%$ of the computational domain centered about the origin, for three resolutions. The resolution is increasing by factors of $2$, starting with the lowest resolution given by $(\Delta^{\rm low}_{Y},\Delta^{\rm low}_{Z})=(0.00125,0.00125)$. The energy decreases over time as it leaks out of the central volume to which we limit its calculation (recall that the energy \eqref{energy_nonlin_waves} is conserved over time if computed over $100\%$ of the computational domain).}
    \label{fig:Energies}
\end{figure}

We test the convergence of the time evolution of the energy \eqref{energy_nonlin_waves} of the numerical solution arising from the initial data analysed in Sections \ref{sec_growth_derivatives} and \ref{sec_mode_cascade} (with $\epsilon=1$). For the sake of the convergence test, the energy \eqref{energy_nonlin_waves} is computed over $75\%$ of the computational domain (i.e., $X=0$, $Y\in \left[0,0.75\right]$ and $Z\in \left[-0.75,0.75\right]$), so as to exclude the relatively large numerical loss of energy coming from dissipation of outgoing waves as they become under-resolved propagating into the volume where the effects of spatial compactification dominate (i.e., we are restricting the domain of integration to a region which is well resolved in our compactified coordinates \eqref{Eq:Compactification}). See Figure \ref{fig:Energies}.

\medskip

\begin{figure}
    \includegraphics[scale=.5]{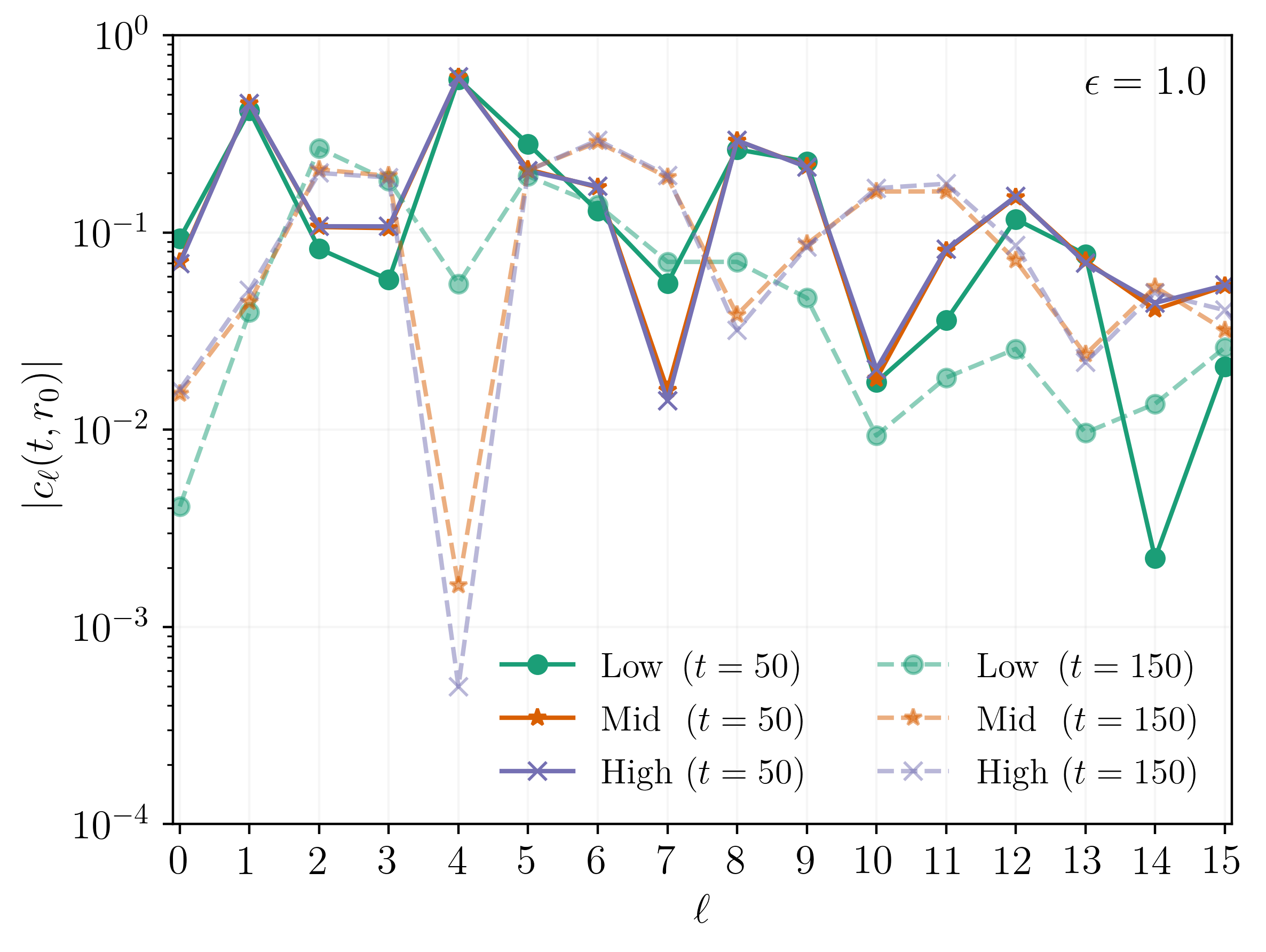}
    \caption{The $\ell$-spectrum of the numerical solution analysed in Figures \ref{fig:SDAsTime} and \ref{fig:CascadeOfModes} at two different times in the evolution. The times considered are $t=50$ (dark solid lines) and $t=150$ (shaded dashed lines). Three increasing resolutions are plotted, starting with $(\Delta^{\rm low}_{Y},\Delta^{\rm low}_{Z})=(0.00125,0.00125)$. Analogous convergence has been tested for the numerical solutions analysed in Figure \ref{spectrum_epsilon_small_2_1}, which are computationally less challenging than the higher-amplitude case shown in this figure.}
    \label{fig:SDConvergence}
\end{figure}

Lastly, Figure \ref{fig:SDConvergence} shows convergence for the Fourier coefficients of the numerical solution arising from the initial data analysed in Section \ref{sec_mode_cascade} (with $\epsilon=1$). For visualization purposes, we are only showing the convergence at two times ($t=50$ and $t=150$, the latter being the largest time we run) for all the coefficients. We have also tested the convergence for the lower amplitude cases, which are numerically less challenging.

\bibliography{black_string}

\begin{thebibliography}{82}%
\makeatletter
\providecommand \@ifxundefined [1]{%
 \@ifx{#1\undefined}
}%
\providecommand \@ifnum [1]{%
 \ifnum #1\expandafter \@firstoftwo
 \else \expandafter \@secondoftwo
 \fi
}%
\providecommand \@ifx [1]{%
 \ifx #1\expandafter \@firstoftwo
 \else \expandafter \@secondoftwo
 \fi
}%
\providecommand \natexlab [1]{#1}%
\providecommand \enquote  [1]{``#1''}%
\providecommand \bibnamefont  [1]{#1}%
\providecommand \bibfnamefont [1]{#1}%
\providecommand \citenamefont [1]{#1}%
\providecommand \href@noop [0]{\@secondoftwo}%
\providecommand \href [0]{\begingroup \@sanitize@url \@href}%
\providecommand \@href[1]{\@@startlink{#1}\@@href}%
\providecommand \@@href[1]{\endgroup#1\@@endlink}%
\providecommand \@sanitize@url [0]{\catcode `\\12\catcode `\$12\catcode
  `\&12\catcode `\#12\catcode `\^12\catcode `\_12\catcode `\%12\relax}%
\providecommand \@@startlink[1]{}%
\providecommand \@@endlink[0]{}%
\providecommand \url  [0]{\begingroup\@sanitize@url \@url }%
\providecommand \@url [1]{\endgroup\@href {#1}{\urlprefix }}%
\providecommand \urlprefix  [0]{URL }%
\providecommand \Eprint [0]{\href }%
\providecommand \doibase [0]{https://doi.org/}%
\providecommand \selectlanguage [0]{\@gobble}%
\providecommand \bibinfo  [0]{\@secondoftwo}%
\providecommand \bibfield  [0]{\@secondoftwo}%
\providecommand \translation [1]{[#1]}%
\providecommand \BibitemOpen [0]{}%
\providecommand \bibitemStop [0]{}%
\providecommand \bibitemNoStop [0]{.\EOS\space}%
\providecommand \EOS [0]{\spacefactor3000\relax}%
\providecommand \BibitemShut  [1]{\csname bibitem#1\endcsname}%
\let\auto@bib@innerbib\@empty
\bibitem [{\citenamefont {Courant}\ and\ \citenamefont
  {Hilbert}(1953)}]{Courant_Hilbert_pde_book}%
  \BibitemOpen
  \bibfield  {author} {\bibinfo {author} {\bibfnamefont {R.}~\bibnamefont
  {Courant}}\ and\ \bibinfo {author} {\bibfnamefont {D.}~\bibnamefont
  {Hilbert}},\ }\href {https://books.google.it/books?id=5qfvAAAAMAAJ} {\emph
  {\bibinfo {title} {{Methods of Mathematical Physics: Partial differential
  equations}}}},\ Methods of Mathematical Physics\ (\bibinfo  {publisher}
  {Interscience Publishers},\ \bibinfo {year} {1953})\BibitemShut {NoStop}%
\bibitem [{\citenamefont {Ruffini}\ and\ \citenamefont
  {Bonazzola}(1969)}]{Ruffini:1969qy}%
  \BibitemOpen
  \bibfield  {author} {\bibinfo {author} {\bibfnamefont {R.}~\bibnamefont
  {Ruffini}}\ and\ \bibinfo {author} {\bibfnamefont {S.}~\bibnamefont
  {Bonazzola}},\ }\bibfield  {title} {\bibinfo {title} {{Systems of
  selfgravitating particles in general relativity and the concept of an
  equation of state}},\ }\href {https://doi.org/10.1103/PhysRev.187.1767}
  {\bibfield  {journal} {\bibinfo  {journal} {Phys. Rev.}\ }\textbf {\bibinfo
  {volume} {187}},\ \bibinfo {pages} {1767} (\bibinfo {year}
  {1969})}\BibitemShut {NoStop}%
\bibitem [{\citenamefont {Colpi}\ \emph {et~al.}(1986)\citenamefont {Colpi},
  \citenamefont {Shapiro},\ and\ \citenamefont {Wasserman}}]{Colpi:1986ye}%
  \BibitemOpen
  \bibfield  {author} {\bibinfo {author} {\bibfnamefont {M.}~\bibnamefont
  {Colpi}}, \bibinfo {author} {\bibfnamefont {S.~L.}\ \bibnamefont {Shapiro}},\
  and\ \bibinfo {author} {\bibfnamefont {I.}~\bibnamefont {Wasserman}},\
  }\bibfield  {title} {\bibinfo {title} {{Boson Stars: Gravitational Equilibria
  of Selfinteracting Scalar Fields}},\ }\href
  {https://doi.org/10.1103/PhysRevLett.57.2485} {\bibfield  {journal} {\bibinfo
   {journal} {Phys. Rev. Lett.}\ }\textbf {\bibinfo {volume} {57}},\ \bibinfo
  {pages} {2485} (\bibinfo {year} {1986})}\BibitemShut {NoStop}%
\bibitem [{\citenamefont {Danielsson}\ \emph {et~al.}(2017)\citenamefont
  {Danielsson}, \citenamefont {Dibitetto},\ and\ \citenamefont
  {Giri}}]{Danielsson:2017riq}%
  \BibitemOpen
  \bibfield  {author} {\bibinfo {author} {\bibfnamefont {U.~H.}\ \bibnamefont
  {Danielsson}}, \bibinfo {author} {\bibfnamefont {G.}~\bibnamefont
  {Dibitetto}},\ and\ \bibinfo {author} {\bibfnamefont {S.}~\bibnamefont
  {Giri}},\ }\bibfield  {title} {\bibinfo {title} {{Black holes as bubbles of
  AdS}},\ }\href {https://doi.org/10.1007/JHEP10(2017)171} {\bibfield
  {journal} {\bibinfo  {journal} {JHEP}\ }\textbf {\bibinfo {volume} {10}},\
  \bibinfo {pages} {171}},\ \Eprint {https://arxiv.org/abs/1705.10172}
  {arXiv:1705.10172 [hep-th]} \BibitemShut {NoStop}%
\bibitem [{\citenamefont {Mathur}(2005)}]{Mathur:2005zp}%
  \BibitemOpen
  \bibfield  {author} {\bibinfo {author} {\bibfnamefont {S.~D.}\ \bibnamefont
  {Mathur}},\ }\bibfield  {title} {\bibinfo {title} {{The Fuzzball proposal for
  black holes: An Elementary review}},\ }\href
  {https://doi.org/10.1002/prop.200410203} {\bibfield  {journal} {\bibinfo
  {journal} {Fortsch. Phys.}\ }\textbf {\bibinfo {volume} {53}},\ \bibinfo
  {pages} {793} (\bibinfo {year} {2005})},\ \Eprint
  {https://arxiv.org/abs/hep-th/0502050} {arXiv:hep-th/0502050} \BibitemShut
  {NoStop}%
\bibitem [{\citenamefont {Mazur}\ and\ \citenamefont
  {Mottola}(2004)}]{Mazur:2004fk}%
  \BibitemOpen
  \bibfield  {author} {\bibinfo {author} {\bibfnamefont {P.~O.}\ \bibnamefont
  {Mazur}}\ and\ \bibinfo {author} {\bibfnamefont {E.}~\bibnamefont
  {Mottola}},\ }\bibfield  {title} {\bibinfo {title} {{Gravitational vacuum
  condensate stars}},\ }\href {https://doi.org/10.1073/pnas.0402717101}
  {\bibfield  {journal} {\bibinfo  {journal} {Proc. Nat. Acad. Sci.}\ }\textbf
  {\bibinfo {volume} {101}},\ \bibinfo {pages} {9545} (\bibinfo {year}
  {2004})},\ \Eprint {https://arxiv.org/abs/gr-qc/0407075}
  {arXiv:gr-qc/0407075} \BibitemShut {NoStop}%
\bibitem [{\citenamefont {Cardoso}\ and\ \citenamefont
  {Pani}(2019)}]{Cardoso:2019rvt}%
  \BibitemOpen
  \bibfield  {author} {\bibinfo {author} {\bibfnamefont {V.}~\bibnamefont
  {Cardoso}}\ and\ \bibinfo {author} {\bibfnamefont {P.}~\bibnamefont {Pani}},\
  }\bibfield  {title} {\bibinfo {title} {{Testing the nature of dark compact
  objects: a status report}},\ }\href
  {https://doi.org/10.1007/s41114-019-0020-4} {\bibfield  {journal} {\bibinfo
  {journal} {Living Rev. Rel.}\ }\textbf {\bibinfo {volume} {22}},\ \bibinfo
  {pages} {4} (\bibinfo {year} {2019})},\ \Eprint
  {https://arxiv.org/abs/1904.05363} {arXiv:1904.05363 [gr-qc]} \BibitemShut
  {NoStop}%
\bibitem [{\citenamefont {Benomio}(2020)}]{Benomio_thesis_doi_2}%
  \BibitemOpen
  \bibfield  {author} {\bibinfo {author} {\bibfnamefont {G.}~\bibnamefont
  {Benomio}},\ }\emph {\bibinfo {title} {{The wave equation on black rings and
  the linear stability of slowly rotating Kerr spacetimes}}},\ \href
  {https://drive.google.com/file/d/1rd7W8XR4U7oiUYkqqwD4Qn_ET-EfnxKY/view}
  {\bibinfo {type} {{PhD thesis}}},\ \bibinfo  {school} {Imperial College
  London} (\bibinfo {year} {2020})\BibitemShut {NoStop}%
\bibitem [{\citenamefont {Benomio}(2021)}]{Benomio_Rings}%
  \BibitemOpen
  \bibfield  {author} {\bibinfo {author} {\bibfnamefont {G.}~\bibnamefont
  {Benomio}},\ }\bibfield  {title} {\bibinfo {title} {{The stable trapping
  phenomenon for black strings and black rings and its obstructions on the
  decay of linear waves}},\ }\href {https://doi.org/10.2140/apde.2021.14.2427}
  {\bibfield  {journal} {\bibinfo  {journal} {Anal. Part. Diff. Eq.}\ }\textbf
  {\bibinfo {volume} {14}},\ \bibinfo {pages} {2427} (\bibinfo {year}
  {2021})},\ \Eprint {https://arxiv.org/abs/1809.07795} {arXiv:1809.07795
  [gr-qc]} \BibitemShut {NoStop}%
\bibitem [{\citenamefont {Gregory}\ and\ \citenamefont
  {Laflamme}(1993)}]{GLinstability}%
  \BibitemOpen
  \bibfield  {author} {\bibinfo {author} {\bibfnamefont {R.}~\bibnamefont
  {Gregory}}\ and\ \bibinfo {author} {\bibfnamefont {R.}~\bibnamefont
  {Laflamme}},\ }\bibfield  {title} {\bibinfo {title} {{Black strings and
  p-branes are unstable}},\ }\href
  {https://doi.org/10.1103/PhysRevLett.70.2837} {\bibfield  {journal} {\bibinfo
   {journal} {Phys. Rev. Lett.}\ }\textbf {\bibinfo {volume} {70}},\ \bibinfo
  {pages} {2837} (\bibinfo {year} {1993})},\ \Eprint
  {https://arxiv.org/abs/hep-th/9301052} {arXiv:hep-th/9301052} \BibitemShut
  {NoStop}%
\bibitem [{\citenamefont {Dafermos}\ \emph {et~al.}(2016)\citenamefont
  {Dafermos}, \citenamefont {Rodnianski},\ and\ \citenamefont
  {Shlapentokh-Rothman}}]{FullKerr}%
  \BibitemOpen
  \bibfield  {author} {\bibinfo {author} {\bibfnamefont {M.}~\bibnamefont
  {Dafermos}}, \bibinfo {author} {\bibfnamefont {I.}~\bibnamefont
  {Rodnianski}},\ and\ \bibinfo {author} {\bibfnamefont {Y.}~\bibnamefont
  {Shlapentokh-Rothman}},\ }\bibfield  {title} {\bibinfo {title} {Decay for
  solutions of the wave equation on {K}err exterior spacetimes {III}: {T}he
  full subextremal case {$|a|<M$}},\ }\href
  {https://doi.org/10.4007/annals.2016.183.3.2} {\bibfield  {journal} {\bibinfo
   {journal} {Ann. of Math. (2)}\ }\textbf {\bibinfo {volume} {183}},\ \bibinfo
  {pages} {787} (\bibinfo {year} {2016})}\BibitemShut {NoStop}%
\bibitem [{\citenamefont
  {Dyatlov}(2011)}]{Dyatlov_decay_linear_waves_kerr_deSitter}%
  \BibitemOpen
  \bibfield  {author} {\bibinfo {author} {\bibfnamefont {S.}~\bibnamefont
  {Dyatlov}},\ }\bibfield  {title} {\bibinfo {title} {{Quasi-normal modes and
  exponential energy decay for the Kerr-de Sitter black hole}},\ }\href
  {https://doi.org/10.1007/s00220-011-1286-x} {\bibfield  {journal} {\bibinfo
  {journal} {Commun. Math. Phys.}\ }\textbf {\bibinfo {volume} {306}},\
  \bibinfo {pages} {119} (\bibinfo {year} {2011})},\ \Eprint
  {https://arxiv.org/abs/1003.6128} {arXiv:1003.6128 [math.AP]} \BibitemShut
  {NoStop}%
\bibitem [{\citenamefont {Holzegel}\ and\ \citenamefont
  {Smulevici}(2014)}]{SharpLogHolz}%
  \BibitemOpen
  \bibfield  {author} {\bibinfo {author} {\bibfnamefont {G.}~\bibnamefont
  {Holzegel}}\ and\ \bibinfo {author} {\bibfnamefont {J.}~\bibnamefont
  {Smulevici}},\ }\bibfield  {title} {\bibinfo {title} {Quasimodes and a lower
  bound on the uniform energy decay rate for {K}err-{A}d{S} spacetimes},\
  }\href {https://doi.org/10.2140/apde.2014.7.1057} {\bibfield  {journal}
  {\bibinfo  {journal} {Anal. PDE}\ }\textbf {\bibinfo {volume} {7}},\ \bibinfo
  {pages} {1057} (\bibinfo {year} {2014})}\BibitemShut {NoStop}%
\bibitem [{\citenamefont {Keir}(2016)}]{KeirLogLog}%
  \BibitemOpen
  \bibfield  {author} {\bibinfo {author} {\bibfnamefont {J.}~\bibnamefont
  {Keir}},\ }\bibfield  {title} {\bibinfo {title} {Slowly decaying waves on
  spherically symmetric spacetimes and ultracompact neutron stars},\ }\href
  {https://doi.org/10.1088/0264-9381/33/13/135009} {\bibfield  {journal}
  {\bibinfo  {journal} {Classical Quantum Gravity}\ }\textbf {\bibinfo {volume}
  {33}},\ \bibinfo {pages} {135009, 42} (\bibinfo {year} {2016})}\BibitemShut
  {NoStop}%
\bibitem [{\citenamefont {Gunasekaran}\ and\ \citenamefont
  {Kunduri}(2021)}]{Kunduri_slow_decay_solitons}%
  \BibitemOpen
  \bibfield  {author} {\bibinfo {author} {\bibfnamefont {S.}~\bibnamefont
  {Gunasekaran}}\ and\ \bibinfo {author} {\bibfnamefont {H.~K.}\ \bibnamefont
  {Kunduri}},\ }\bibfield  {title} {\bibinfo {title} {{Slow decay of waves in
  gravitational solitons}},\ }\href
  {https://doi.org/10.1007/s00023-020-01010-3} {\bibfield  {journal} {\bibinfo
  {journal} {Annales Henri Poincare}\ }\textbf {\bibinfo {volume} {22}},\
  \bibinfo {pages} {821} (\bibinfo {year} {2021})},\ \Eprint
  {https://arxiv.org/abs/2007.04283} {arXiv:2007.04283 [gr-qc]} \BibitemShut
  {NoStop}%
\bibitem [{\citenamefont {Keir}(2019)}]{MicroKeir}%
  \BibitemOpen
  \bibfield  {author} {\bibinfo {author} {\bibfnamefont {J.}~\bibnamefont
  {Keir}},\ }\bibfield  {title} {\bibinfo {title} {{Wave propagation on
  microstate geometries}},\ }\href {https://doi.org/10.1007/s00023-019-00874-4}
  {\bibfield  {journal} {\bibinfo  {journal} {Annales Henri Poincare}\ }\textbf
  {\bibinfo {volume} {21}},\ \bibinfo {pages} {705} (\bibinfo {year} {2019})},\
  \Eprint {https://arxiv.org/abs/1609.01733} {arXiv:1609.01733 [gr-qc]}
  \BibitemShut {NoStop}%
\bibitem [{\citenamefont
  {{Idelon-Riton}}(2016)}]{Idelon-Riton_Lower_Bound_Dirac}%
  \BibitemOpen
  \bibfield  {author} {\bibinfo {author} {\bibfnamefont {G.}~\bibnamefont
  {{Idelon-Riton}}},\ }\bibfield  {title} {\bibinfo {title} {{Quasimodes and a
  lower bound for the local energy decay of the Dirac equation in
  Schwarzschild-Anti-de Sitter spacetime}},\ }\href@noop {} {\bibfield
  {journal} {\bibinfo  {journal} {ArXiv e-prints}\ } (\bibinfo {year}
  {2016})},\ \Eprint {https://arxiv.org/abs/1610.05542} {arXiv:1610.05542
  [math-ph]} \BibitemShut {NoStop}%
\bibitem [{\citenamefont {Luk}(2013)}]{Luk_Nonlinear_Wave_Kerr}%
  \BibitemOpen
  \bibfield  {author} {\bibinfo {author} {\bibfnamefont {J.}~\bibnamefont
  {Luk}},\ }\bibfield  {title} {\bibinfo {title} {The null condition and global
  existence for nonlinear wave equations on slowly rotating {K}err
  spacetimes},\ }\href {https://doi.org/10.4171/JEMS/400} {\bibfield  {journal}
  {\bibinfo  {journal} {J. Eur. Math. Soc. (JEMS)}\ }\textbf {\bibinfo {volume}
  {15}},\ \bibinfo {pages} {1629} (\bibinfo {year} {2013})}\BibitemShut
  {NoStop}%
\bibitem [{\citenamefont {Dafermos}\ \emph {et~al.}(2024)\citenamefont
  {Dafermos}, \citenamefont {Holzegel}, \citenamefont {Rodnianski},\ and\
  \citenamefont {Taylor}}]{DHRT_quasilinear_waves_full_kerr}%
  \BibitemOpen
  \bibfield  {author} {\bibinfo {author} {\bibfnamefont {M.}~\bibnamefont
  {Dafermos}}, \bibinfo {author} {\bibfnamefont {G.}~\bibnamefont {Holzegel}},
  \bibinfo {author} {\bibfnamefont {I.}~\bibnamefont {Rodnianski}},\ and\
  \bibinfo {author} {\bibfnamefont {M.}~\bibnamefont {Taylor}},\ }\bibfield
  {title} {\bibinfo {title} {{Quasilinear wave equations on Kerr black holes in
  the full subextremal range $|a|<M$}},\ }\href@noop {} {\  (\bibinfo {year}
  {2024})},\ \Eprint {https://arxiv.org/abs/2410.03639} {arXiv:2410.03639
  [gr-qc]} \BibitemShut {NoStop}%
\bibitem [{\citenamefont {Hintz}\ and\ \citenamefont
  {Vasy}(2016)}]{Hintz_Vasy_quasilinear_waves_Kerr_deSitter}%
  \BibitemOpen
  \bibfield  {author} {\bibinfo {author} {\bibfnamefont {P.}~\bibnamefont
  {Hintz}}\ and\ \bibinfo {author} {\bibfnamefont {A.}~\bibnamefont {Vasy}},\
  }\bibfield  {title} {\bibinfo {title} {{Global analysis of quasilinear wave
  equations on asymptotically Kerr-de Sitter spaces}},\ }\href
  {https://doi.org/10.1093/imrn/rnv311} {\bibfield  {journal} {\bibinfo
  {journal} {Int. Math. Res. Not.}\ }\textbf {\bibinfo {volume} {2016}},\
  \bibinfo {pages} {5355} (\bibinfo {year} {2016})},\ \Eprint
  {https://arxiv.org/abs/1404.1348} {arXiv:1404.1348 [math.AP]} \BibitemShut
  {NoStop}%
\bibitem [{\citenamefont {Klainerman}(1986)}]{Klainerman_null_cond}%
  \BibitemOpen
  \bibfield  {author} {\bibinfo {author} {\bibfnamefont {S.}~\bibnamefont
  {Klainerman}},\ }\bibfield  {title} {\bibinfo {title} {The null condition and
  global existence to nonlinear wave equations},\ }in\ \href@noop {} {\emph
  {\bibinfo {booktitle} {Nonlinear systems of partial differential equations in
  applied mathematics, {P}art 1 ({S}anta {F}e, {N}.{M}., 1984)}}},\ \bibinfo
  {series} {Lectures in Appl. Math.}, Vol.~\bibinfo {volume} {23}\ (\bibinfo
  {publisher} {Amer. Math. Soc., Providence, RI},\ \bibinfo {year} {1986})\
  pp.\ \bibinfo {pages} {293--326}\BibitemShut {NoStop}%
\bibitem [{\citenamefont {Christodoulou}(1986)}]{Christ_null_cond}%
  \BibitemOpen
  \bibfield  {author} {\bibinfo {author} {\bibfnamefont {D.}~\bibnamefont
  {Christodoulou}},\ }\bibfield  {title} {\bibinfo {title} {Global solutions of
  nonlinear hyperbolic equations for small initial data},\ }\href
  {https://doi.org/10.1002/cpa.3160390205} {\bibfield  {journal} {\bibinfo
  {journal} {Comm. Pure Appl. Math.}\ }\textbf {\bibinfo {volume} {39}},\
  \bibinfo {pages} {267} (\bibinfo {year} {1986})}\BibitemShut {NoStop}%
\bibitem [{\citenamefont {John}(1981)}]{John_blow_up_quasilinear_wave_eqns}%
  \BibitemOpen
  \bibfield  {author} {\bibinfo {author} {\bibfnamefont {F.}~\bibnamefont
  {John}},\ }\bibfield  {title} {\bibinfo {title} {{Blow-up for quasi-linear
  wave equations in three space dimensions}},\ }\href
  {https://doi.org/https://doi.org/10.1002/cpa.3160340103} {\bibfield
  {journal} {\bibinfo  {journal} {Communications on Pure and Applied
  Mathematics}\ }\textbf {\bibinfo {volume} {34}},\ \bibinfo {pages} {29}
  (\bibinfo {year} {1981})},\ \Eprint
  {https://arxiv.org/abs/https://onlinelibrary.wiley.com/doi/pdf/10.1002/cpa.3160340103}
  {https://onlinelibrary.wiley.com/doi/pdf/10.1002/cpa.3160340103} \BibitemShut
  {NoStop}%
\bibitem [{\citenamefont {Dafermos}\ \emph {et~al.}(2021)\citenamefont
  {Dafermos}, \citenamefont {Holzegel}, \citenamefont {Rodnianski},\ and\
  \citenamefont {Taylor}}]{DHRT}%
  \BibitemOpen
  \bibfield  {author} {\bibinfo {author} {\bibfnamefont {M.}~\bibnamefont
  {Dafermos}}, \bibinfo {author} {\bibfnamefont {G.}~\bibnamefont {Holzegel}},
  \bibinfo {author} {\bibfnamefont {I.}~\bibnamefont {Rodnianski}},\ and\
  \bibinfo {author} {\bibfnamefont {M.}~\bibnamefont {Taylor}},\ }\bibfield
  {title} {\bibinfo {title} {{The nonlinear stability of the Schwarzschild
  family of black holes}},\ }\href@noop {} {\bibfield  {journal} {\bibinfo
  {journal} {arXiv e-print 2104.08222}\ } (\bibinfo {year} {2021})}\BibitemShut
  {NoStop}%
\bibitem [{\citenamefont {Klainerman}\ and\ \citenamefont
  {Szeftel}(2023)}]{Klainerman_Szeftel_Kerr_small_a_1}%
  \BibitemOpen
  \bibfield  {author} {\bibinfo {author} {\bibfnamefont {S.}~\bibnamefont
  {Klainerman}}\ and\ \bibinfo {author} {\bibfnamefont {J.}~\bibnamefont
  {Szeftel}},\ }\bibfield  {title} {\bibinfo {title} {Kerr stability for small
  angular momentum},\ }\href@noop {} {\bibfield  {journal} {\bibinfo  {journal}
  {Pure and Applied Mathematics Quarterly}\ }\textbf {\bibinfo {volume} {19}}
  (\bibinfo {year} {2023})}\BibitemShut {NoStop}%
\bibitem [{\citenamefont {Hintz}\ and\ \citenamefont
  {Vasy}(2018)}]{StabKerrdS}%
  \BibitemOpen
  \bibfield  {author} {\bibinfo {author} {\bibfnamefont {P.}~\bibnamefont
  {Hintz}}\ and\ \bibinfo {author} {\bibfnamefont {A.}~\bibnamefont {Vasy}},\
  }\bibfield  {title} {\bibinfo {title} {The global non-linear stability of the
  {K}err–de {S}itter family of black holes},\ }\href
  {https://doi.org/10.4310/ACTA.2018.v220.n1.a1} {\bibfield  {journal}
  {\bibinfo  {journal} {Acta Math.}\ }\textbf {\bibinfo {volume} {220}},\
  \bibinfo {pages} {1} (\bibinfo {year} {2018})}\BibitemShut {NoStop}%
\bibitem [{\citenamefont {Bachelot}\ and\ \citenamefont
  {Nicolas}(1993)}]{Bachelot_Nicolas_nonlinear_KG_schwarzschild}%
  \BibitemOpen
  \bibfield  {author} {\bibinfo {author} {\bibfnamefont {A.}~\bibnamefont
  {Bachelot}}\ and\ \bibinfo {author} {\bibfnamefont {J.-P.}\ \bibnamefont
  {Nicolas}},\ }\bibfield  {title} {\bibinfo {title} {\'equation non lin\'eaire
  de {K}lein-{G}ordon dans des m\'etriques de type {S}chwarzschild},\
  }\href@noop {} {\bibfield  {journal} {\bibinfo  {journal} {C. R. Acad. Sci.
  Paris S\'er. I Math.}\ }\textbf {\bibinfo {volume} {316}},\ \bibinfo {pages}
  {1047} (\bibinfo {year} {1993})}\BibitemShut {NoStop}%
\bibitem [{\citenamefont {Nicolas}(1995)}]{Nicolas_nonlinear_KG_schwarzschild}%
  \BibitemOpen
  \bibfield  {author} {\bibinfo {author} {\bibfnamefont {J.-P.}\ \bibnamefont
  {Nicolas}},\ }\bibfield  {title} {\bibinfo {title} {Nonlinear
  {K}lein-{G}ordon equation on {S}chwarzschild-like metrics},\ }\href@noop {}
  {\bibfield  {journal} {\bibinfo  {journal} {J. Math. Pures Appl. (9)}\
  }\textbf {\bibinfo {volume} {74}},\ \bibinfo {pages} {35} (\bibinfo {year}
  {1995})}\BibitemShut {NoStop}%
\bibitem [{\citenamefont {Blue}\ and\ \citenamefont
  {Soffer}(2007)}]{Blue_Soffer_semilinear_waves_Schwarzschild_large_data}%
  \BibitemOpen
  \bibfield  {author} {\bibinfo {author} {\bibfnamefont {P.}~\bibnamefont
  {Blue}}\ and\ \bibinfo {author} {\bibfnamefont {A.}~\bibnamefont {Soffer}},\
  }\bibfield  {title} {\bibinfo {title} {A space-time integral estimate for a
  large data semi-linear wave equation on the {S}chwarzschild manifold},\
  }\href {https://doi.org/10.1007/s11005-007-0177-8} {\bibfield  {journal}
  {\bibinfo  {journal} {Lett. Math. Phys.}\ }\textbf {\bibinfo {volume} {81}},\
  \bibinfo {pages} {227} (\bibinfo {year} {2007})}\BibitemShut {NoStop}%
\bibitem [{\citenamefont {Blue}\ and\ \citenamefont
  {Soffer}(2003)}]{Blue_Soffer_semilinear_waves_Schwarzschild}%
  \BibitemOpen
  \bibfield  {author} {\bibinfo {author} {\bibfnamefont {P.}~\bibnamefont
  {Blue}}\ and\ \bibinfo {author} {\bibfnamefont {A.}~\bibnamefont {Soffer}},\
  }\bibfield  {title} {\bibinfo {title} {{Semilinear wave equations on the
  Schwarzschild manifold. 1. Local decay estimates}},\ }\href@noop {}
  {\bibfield  {journal} {\bibinfo  {journal} {Adv. Diff. Eq.}\ }\textbf
  {\bibinfo {volume} {8}},\ \bibinfo {pages} {595} (\bibinfo {year} {2003})},\
  \Eprint {https://arxiv.org/abs/gr-qc/0310091} {arXiv:gr-qc/0310091}
  \BibitemShut {NoStop}%
\bibitem [{\citenamefont {Horowitz}\ and\ \citenamefont
  {Strominger}(1991)}]{Horowitz_Strominger_strings_branes}%
  \BibitemOpen
  \bibfield  {author} {\bibinfo {author} {\bibfnamefont {G.~T.}\ \bibnamefont
  {Horowitz}}\ and\ \bibinfo {author} {\bibfnamefont {A.}~\bibnamefont
  {Strominger}},\ }\bibfield  {title} {\bibinfo {title} {{Black strings and
  P-branes}},\ }\href {https://doi.org/10.1016/0550-3213(91)90440-9} {\bibfield
   {journal} {\bibinfo  {journal} {Nucl. Phys. B}\ }\textbf {\bibinfo {volume}
  {360}},\ \bibinfo {pages} {197} (\bibinfo {year} {1991})}\BibitemShut
  {NoStop}%
\bibitem [{\citenamefont {Collingbourne}(2021)}]{Collingbourne_GL}%
  \BibitemOpen
  \bibfield  {author} {\bibinfo {author} {\bibfnamefont {S.~C.}\ \bibnamefont
  {Collingbourne}},\ }\bibfield  {title} {\bibinfo {title} {{The
  Gregory\textendash{}Laflamme instability of the Schwarzschild black string
  exterior}},\ }\href {https://doi.org/10.1063/5.0043059} {\bibfield  {journal}
  {\bibinfo  {journal} {J. Math. Phys.}\ }\textbf {\bibinfo {volume} {62}},\
  \bibinfo {pages} {032502} (\bibinfo {year} {2021})},\ \Eprint
  {https://arxiv.org/abs/2007.08441} {arXiv:2007.08441 [gr-qc]} \BibitemShut
  {NoStop}%
\bibitem [{\citenamefont {Gubser}\ and\ \citenamefont
  {Mitra}(2002)}]{Gubser_mitra_black_string_1}%
  \BibitemOpen
  \bibfield  {author} {\bibinfo {author} {\bibfnamefont {S.~S.}\ \bibnamefont
  {Gubser}}\ and\ \bibinfo {author} {\bibfnamefont {I.}~\bibnamefont {Mitra}},\
  }\bibfield  {title} {\bibinfo {title} {{Instability of charged black holes in
  Anti-de Sitter space}},\ }\href@noop {} {\bibfield  {journal} {\bibinfo
  {journal} {Clay Math. Proc.}\ }\textbf {\bibinfo {volume} {1}},\ \bibinfo
  {pages} {221} (\bibinfo {year} {2002})},\ \Eprint
  {https://arxiv.org/abs/hep-th/0009126} {arXiv:hep-th/0009126} \BibitemShut
  {NoStop}%
\bibitem [{\citenamefont {Gubser}\ and\ \citenamefont
  {Mitra}(2001)}]{Gubser_mitra_black_string_2}%
  \BibitemOpen
  \bibfield  {author} {\bibinfo {author} {\bibfnamefont {S.~S.}\ \bibnamefont
  {Gubser}}\ and\ \bibinfo {author} {\bibfnamefont {I.}~\bibnamefont {Mitra}},\
  }\bibfield  {title} {\bibinfo {title} {{The Evolution of unstable black holes
  in anti-de Sitter space}},\ }\href
  {https://doi.org/10.1088/1126-6708/2001/08/018} {\bibfield  {journal}
  {\bibinfo  {journal} {JHEP}\ }\textbf {\bibinfo {volume} {08}},\ \bibinfo
  {pages} {018}},\ \Eprint {https://arxiv.org/abs/hep-th/0011127}
  {arXiv:hep-th/0011127} \BibitemShut {NoStop}%
\bibitem [{\citenamefont {Hovdebo}\ and\ \citenamefont
  {Myers}(2006)}]{Hovd_Myers_String_Ring_GL}%
  \BibitemOpen
  \bibfield  {author} {\bibinfo {author} {\bibfnamefont {J.~L.}\ \bibnamefont
  {Hovdebo}}\ and\ \bibinfo {author} {\bibfnamefont {R.~C.}\ \bibnamefont
  {Myers}},\ }\bibfield  {title} {\bibinfo {title} {{Black rings, boosted
  strings and Gregory-Laflamme}},\ }\href
  {https://doi.org/10.1103/PhysRevD.73.084013} {\bibfield  {journal} {\bibinfo
  {journal} {Phys. Rev.}\ }\textbf {\bibinfo {volume} {D73}},\ \bibinfo {pages}
  {084013} (\bibinfo {year} {2006})},\ \Eprint
  {https://arxiv.org/abs/hep-th/0601079} {arXiv:hep-th/0601079 [hep-th]}
  \BibitemShut {NoStop}%
\bibitem [{\citenamefont {Reall}(2001)}]{Reall_stability_black_branes}%
  \BibitemOpen
  \bibfield  {author} {\bibinfo {author} {\bibfnamefont {H.~S.}\ \bibnamefont
  {Reall}},\ }\bibfield  {title} {\bibinfo {title} {{Classical and
  thermodynamic stability of black branes}},\ }\href
  {https://doi.org/10.1103/PhysRevD.64.044005} {\bibfield  {journal} {\bibinfo
  {journal} {Phys. Rev. D}\ }\textbf {\bibinfo {volume} {64}},\ \bibinfo
  {pages} {044005} (\bibinfo {year} {2001})},\ \Eprint
  {https://arxiv.org/abs/hep-th/0104071} {arXiv:hep-th/0104071} \BibitemShut
  {NoStop}%
\bibitem [{\citenamefont {Figueras}\ \emph {et~al.}(2011)\citenamefont
  {Figueras}, \citenamefont {Murata},\ and\ \citenamefont
  {Reall}}]{Figueras_murata_reall_penrose_inequalities}%
  \BibitemOpen
  \bibfield  {author} {\bibinfo {author} {\bibfnamefont {P.}~\bibnamefont
  {Figueras}}, \bibinfo {author} {\bibfnamefont {K.}~\bibnamefont {Murata}},\
  and\ \bibinfo {author} {\bibfnamefont {H.~S.}\ \bibnamefont {Reall}},\
  }\bibfield  {title} {\bibinfo {title} {{Black hole instabilities and local
  Penrose inequalities}},\ }\href
  {https://doi.org/10.1088/0264-9381/28/22/225030} {\bibfield  {journal}
  {\bibinfo  {journal} {Class. Quant. Grav.}\ }\textbf {\bibinfo {volume}
  {28}},\ \bibinfo {pages} {225030} (\bibinfo {year} {2011})},\ \Eprint
  {https://arxiv.org/abs/1107.5785} {arXiv:1107.5785 [gr-qc]} \BibitemShut
  {NoStop}%
\bibitem [{\citenamefont {Hollands}\ and\ \citenamefont
  {Wald}(2013)}]{Hollands_wald_black_branes}%
  \BibitemOpen
  \bibfield  {author} {\bibinfo {author} {\bibfnamefont {S.}~\bibnamefont
  {Hollands}}\ and\ \bibinfo {author} {\bibfnamefont {R.~M.}\ \bibnamefont
  {Wald}},\ }\bibfield  {title} {\bibinfo {title} {{Stability of Black Holes
  and Black Branes}},\ }\href {https://doi.org/10.1007/s00220-012-1638-1}
  {\bibfield  {journal} {\bibinfo  {journal} {Commun. Math. Phys.}\ }\textbf
  {\bibinfo {volume} {321}},\ \bibinfo {pages} {629} (\bibinfo {year}
  {2013})},\ \Eprint {https://arxiv.org/abs/1201.0463} {arXiv:1201.0463
  [gr-qc]} \BibitemShut {NoStop}%
\bibitem [{\citenamefont {Choptuik}\ \emph {et~al.}(2003)\citenamefont
  {Choptuik}, \citenamefont {Lehner}, \citenamefont {Olabarrieta},
  \citenamefont {Petryk}, \citenamefont {Pretorius},\ and\ \citenamefont
  {Villegas}}]{Choptuik_et_al_black_string}%
  \BibitemOpen
  \bibfield  {author} {\bibinfo {author} {\bibfnamefont {M.~W.}\ \bibnamefont
  {Choptuik}}, \bibinfo {author} {\bibfnamefont {L.}~\bibnamefont {Lehner}},
  \bibinfo {author} {\bibfnamefont {I.}~\bibnamefont {Olabarrieta}}, \bibinfo
  {author} {\bibfnamefont {R.}~\bibnamefont {Petryk}}, \bibinfo {author}
  {\bibfnamefont {F.}~\bibnamefont {Pretorius}},\ and\ \bibinfo {author}
  {\bibfnamefont {H.}~\bibnamefont {Villegas}},\ }\bibfield  {title} {\bibinfo
  {title} {{Towards the final fate of an unstable black string}},\ }\href
  {https://doi.org/10.1103/PhysRevD.68.044001} {\bibfield  {journal} {\bibinfo
  {journal} {Phys. Rev. D}\ }\textbf {\bibinfo {volume} {68}},\ \bibinfo
  {pages} {044001} (\bibinfo {year} {2003})},\ \Eprint
  {https://arxiv.org/abs/gr-qc/0304085} {arXiv:gr-qc/0304085} \BibitemShut
  {NoStop}%
\bibitem [{\citenamefont {Garfinkle}\ \emph {et~al.}(2005)\citenamefont
  {Garfinkle}, \citenamefont {Lehner},\ and\ \citenamefont
  {Pretorius}}]{Garfinkle_et_al_black_string}%
  \BibitemOpen
  \bibfield  {author} {\bibinfo {author} {\bibfnamefont {D.}~\bibnamefont
  {Garfinkle}}, \bibinfo {author} {\bibfnamefont {L.}~\bibnamefont {Lehner}},\
  and\ \bibinfo {author} {\bibfnamefont {F.}~\bibnamefont {Pretorius}},\
  }\bibfield  {title} {\bibinfo {title} {{A Numerical examination of an
  evolving black string horizon}},\ }\href
  {https://doi.org/10.1103/PhysRevD.71.064009} {\bibfield  {journal} {\bibinfo
  {journal} {Phys. Rev. D}\ }\textbf {\bibinfo {volume} {71}},\ \bibinfo
  {pages} {064009} (\bibinfo {year} {2005})},\ \Eprint
  {https://arxiv.org/abs/gr-qc/0412014} {arXiv:gr-qc/0412014} \BibitemShut
  {NoStop}%
\bibitem [{\citenamefont {Lehner}\ and\ \citenamefont
  {Pretorius}(2010)}]{Lehner_Pret_Instab_String}%
  \BibitemOpen
  \bibfield  {author} {\bibinfo {author} {\bibfnamefont {L.}~\bibnamefont
  {Lehner}}\ and\ \bibinfo {author} {\bibfnamefont {F.}~\bibnamefont
  {Pretorius}},\ }\bibfield  {title} {\bibinfo {title} {{Black Strings, Low
  Viscosity Fluids, and Violation of Cosmic Censorship}},\ }\href
  {https://doi.org/10.1103/PhysRevLett.105.101102} {\bibfield  {journal}
  {\bibinfo  {journal} {Phys. Rev. Lett.}\ }\textbf {\bibinfo {volume} {105}},\
  \bibinfo {pages} {101102} (\bibinfo {year} {2010})},\ \Eprint
  {https://arxiv.org/abs/1006.5960} {arXiv:1006.5960 [hep-th]} \BibitemShut
  {NoStop}%
\bibitem [{\citenamefont {Cagnac}\ and\ \citenamefont
  {Choquet-Bruhat}(1984)}]{Cagnac_choquet_bruhat_global_wp_nlw}%
  \BibitemOpen
  \bibfield  {author} {\bibinfo {author} {\bibfnamefont {F.}~\bibnamefont
  {Cagnac}}\ and\ \bibinfo {author} {\bibfnamefont {Y.}~\bibnamefont
  {Choquet-Bruhat}},\ }\bibfield  {title} {\bibinfo {title} {Solution globale
  d'une \'{e}quation non lin\'{e}aire sur une vari\'{e}t\'{e} hyperbolique},\
  }\href@noop {} {\bibfield  {journal} {\bibinfo  {journal} {J. Math. Pures
  Appl. (9)}\ }\textbf {\bibinfo {volume} {63}},\ \bibinfo {pages} {377}
  (\bibinfo {year} {1984})}\BibitemShut {NoStop}%
\bibitem [{\citenamefont {Kreiss}\ and\ \citenamefont
  {Oliger}(1973)}]{kreiss1973methods}%
  \BibitemOpen
  \bibfield  {author} {\bibinfo {author} {\bibfnamefont {H.}~\bibnamefont
  {Kreiss}}\ and\ \bibinfo {author} {\bibfnamefont {J.}~\bibnamefont
  {Oliger}},\ }\href@noop {} {\emph {\bibinfo {title} {Methods for the
  approximate solution of time dependent problems}}},\ \bibinfo {number} {10}\
  (\bibinfo  {publisher} {International Council of Scientific Unions, World
  Meteorological Organization},\ \bibinfo {year} {1973})\BibitemShut {NoStop}%
\bibitem [{\citenamefont {Pretorius}(2005)}]{Pretorius:2004jg}%
  \BibitemOpen
  \bibfield  {author} {\bibinfo {author} {\bibfnamefont {F.}~\bibnamefont
  {Pretorius}},\ }\bibfield  {title} {\bibinfo {title} {{Numerical relativity
  using a generalized harmonic decomposition}},\ }\href
  {https://doi.org/10.1088/0264-9381/22/2/014} {\bibfield  {journal} {\bibinfo
  {journal} {Class. Quant. Grav.}\ }\textbf {\bibinfo {volume} {22}},\ \bibinfo
  {pages} {425} (\bibinfo {year} {2005})},\ \Eprint
  {https://arxiv.org/abs/gr-qc/0407110} {arXiv:gr-qc/0407110} \BibitemShut
  {NoStop}%
\bibitem [{\citenamefont {Alcubierre}\ \emph {et~al.}(2001)\citenamefont
  {Alcubierre}, \citenamefont {Brandt}, \citenamefont {Bruegmann},
  \citenamefont {Holz}, \citenamefont {Seidel}, \citenamefont {Takahashi},\
  and\ \citenamefont {Thornburg}}]{Alcubierre:1999ab}%
  \BibitemOpen
  \bibfield  {author} {\bibinfo {author} {\bibfnamefont {M.}~\bibnamefont
  {Alcubierre}}, \bibinfo {author} {\bibfnamefont {S.}~\bibnamefont {Brandt}},
  \bibinfo {author} {\bibfnamefont {B.}~\bibnamefont {Bruegmann}}, \bibinfo
  {author} {\bibfnamefont {D.}~\bibnamefont {Holz}}, \bibinfo {author}
  {\bibfnamefont {E.}~\bibnamefont {Seidel}}, \bibinfo {author} {\bibfnamefont
  {R.}~\bibnamefont {Takahashi}},\ and\ \bibinfo {author} {\bibfnamefont
  {J.}~\bibnamefont {Thornburg}},\ }\bibfield  {title} {\bibinfo {title}
  {{Symmetry without symmetry: Numerical simulation of axisymmetric systems
  using Cartesian grids}},\ }\href {https://doi.org/10.1142/S0218271801000834}
  {\bibfield  {journal} {\bibinfo  {journal} {Int. J. Mod. Phys. D}\ }\textbf
  {\bibinfo {volume} {10}},\ \bibinfo {pages} {273} (\bibinfo {year} {2001})},\
  \Eprint {https://arxiv.org/abs/gr-qc/9908012} {arXiv:gr-qc/9908012}
  \BibitemShut {NoStop}%
\bibitem [{\citenamefont {Redondo-Yuste}\ and\ \citenamefont
  {C\'ardenas-Avenda\~no}(2025)}]{Redondo-Yuste:2025hlv}%
  \BibitemOpen
  \bibfield  {author} {\bibinfo {author} {\bibfnamefont {J.}~\bibnamefont
  {Redondo-Yuste}}\ and\ \bibinfo {author} {\bibfnamefont {A.}~\bibnamefont
  {C\'ardenas-Avenda\~no}},\ }\bibfield  {title} {\bibinfo {title}
  {{Perturbative and non-linear analyses of gravitational turbulence in
  spacetimes with stable light rings}},\ }\href@noop {} {\  (\bibinfo {year}
  {2025})},\ \Eprint {https://arxiv.org/abs/2502.18643} {arXiv:2502.18643
  [gr-qc]} \BibitemShut {NoStop}%
\bibitem [{\citenamefont {Solli}\ \emph {et~al.}(2007)\citenamefont {Solli},
  \citenamefont {Ropers}, \citenamefont {Koonath},\ and\ \citenamefont
  {Jalali}}]{rogue_waves_optics}%
  \BibitemOpen
  \bibfield  {author} {\bibinfo {author} {\bibfnamefont {D.~R.}\ \bibnamefont
  {Solli}}, \bibinfo {author} {\bibfnamefont {C.}~\bibnamefont {Ropers}},
  \bibinfo {author} {\bibfnamefont {P.}~\bibnamefont {Koonath}},\ and\ \bibinfo
  {author} {\bibfnamefont {B.}~\bibnamefont {Jalali}},\ }\bibfield  {title}
  {\bibinfo {title} {Optical rogue waves},\ }\href
  {https://doi.org/10.1038/nature06402} {\bibfield  {journal} {\bibinfo
  {journal} {Nature}\ }\textbf {\bibinfo {volume} {450(7172)}},\ \bibinfo
  {pages} {1054} (\bibinfo {year} {2007})}\BibitemShut {NoStop}%
\bibitem [{\citenamefont {Rosa}(2013)}]{Rosa_Black_String_Bomb}%
  \BibitemOpen
  \bibfield  {author} {\bibinfo {author} {\bibfnamefont {J.~G.}\ \bibnamefont
  {Rosa}},\ }\bibfield  {title} {\bibinfo {title} {{Boosted black string
  bombs}},\ }\href {https://doi.org/10.1007/JHEP02(2013)014} {\bibfield
  {journal} {\bibinfo  {journal} {JHEP}\ }\textbf {\bibinfo {volume} {02}},\
  \bibinfo {pages} {014}},\ \Eprint {https://arxiv.org/abs/1209.4211}
  {arXiv:1209.4211 [hep-th]} \BibitemShut {NoStop}%
\bibitem [{\citenamefont {Cardoso}\ \emph {et~al.}(2014)\citenamefont
  {Cardoso}, \citenamefont {Crispino}, \citenamefont {Macedo}, \citenamefont
  {Okawa},\ and\ \citenamefont {Pani}}]{Cardoso_et_al_light_ring_instability}%
  \BibitemOpen
  \bibfield  {author} {\bibinfo {author} {\bibfnamefont {V.}~\bibnamefont
  {Cardoso}}, \bibinfo {author} {\bibfnamefont {L.~C.~B.}\ \bibnamefont
  {Crispino}}, \bibinfo {author} {\bibfnamefont {C.~F.~B.}\ \bibnamefont
  {Macedo}}, \bibinfo {author} {\bibfnamefont {H.}~\bibnamefont {Okawa}},\ and\
  \bibinfo {author} {\bibfnamefont {P.}~\bibnamefont {Pani}},\ }\bibfield
  {title} {\bibinfo {title} {{Light rings as observational evidence for event
  horizons: long-lived modes, ergoregions and nonlinear instabilities of
  ultracompact objects}},\ }\href {https://doi.org/10.1103/PhysRevD.90.044069}
  {\bibfield  {journal} {\bibinfo  {journal} {Phys. Rev. D}\ }\textbf {\bibinfo
  {volume} {90}},\ \bibinfo {pages} {044069} (\bibinfo {year} {2014})},\
  \Eprint {https://arxiv.org/abs/1406.5510} {arXiv:1406.5510 [gr-qc]}
  \BibitemShut {NoStop}%
\bibitem [{\citenamefont {Cunha}\ \emph {et~al.}(2023)\citenamefont {Cunha},
  \citenamefont {Herdeiro}, \citenamefont {Radu},\ and\ \citenamefont
  {Sanchis-Gual}}]{Cunha_Herdeiro_Radu_fate_light_ring_instab_compact_objects}%
  \BibitemOpen
  \bibfield  {author} {\bibinfo {author} {\bibfnamefont {P.~V.~P.}\
  \bibnamefont {Cunha}}, \bibinfo {author} {\bibfnamefont {C.}~\bibnamefont
  {Herdeiro}}, \bibinfo {author} {\bibfnamefont {E.}~\bibnamefont {Radu}},\
  and\ \bibinfo {author} {\bibfnamefont {N.}~\bibnamefont {Sanchis-Gual}},\
  }\bibfield  {title} {\bibinfo {title} {{Exotic Compact Objects and the Fate
  of the Light-Ring Instability}},\ }\href
  {https://doi.org/10.1103/PhysRevLett.130.061401} {\bibfield  {journal}
  {\bibinfo  {journal} {Phys. Rev. Lett.}\ }\textbf {\bibinfo {volume} {130}},\
  \bibinfo {pages} {061401} (\bibinfo {year} {2023})},\ \Eprint
  {https://arxiv.org/abs/2207.13713} {arXiv:2207.13713 [gr-qc]} \BibitemShut
  {NoStop}%
\bibitem [{\citenamefont {Majda}\ \emph {et~al.}(1997)\citenamefont {Majda},
  \citenamefont {McLaughlin},\ and\ \citenamefont
  {Tabak}}]{Majda_McLaughlin_Tabak_wave_turbulence}%
  \BibitemOpen
  \bibfield  {author} {\bibinfo {author} {\bibfnamefont {A.~J.}\ \bibnamefont
  {Majda}}, \bibinfo {author} {\bibfnamefont {D.~W.}\ \bibnamefont
  {McLaughlin}},\ and\ \bibinfo {author} {\bibfnamefont {E.~G.}\ \bibnamefont
  {Tabak}},\ }\bibfield  {title} {\bibinfo {title} {A one-dimensional model for
  dispersive wave turbulence},\ }\href
  {https://api.semanticscholar.org/CorpusID:7208900} {\bibfield  {journal}
  {\bibinfo  {journal} {Journal of Nonlinear Science}\ }\textbf {\bibinfo
  {volume} {7}},\ \bibinfo {pages} {9} (\bibinfo {year} {1997})}\BibitemShut
  {NoStop}%
\bibitem [{\citenamefont {Zakharov}\ \emph {et~al.}(2001)\citenamefont
  {Zakharov}, \citenamefont {Guyenne}, \citenamefont {Pushkarev},\ and\
  \citenamefont {Dias}}]{Zakharov_wave_turbulence_1D_models}%
  \BibitemOpen
  \bibfield  {author} {\bibinfo {author} {\bibfnamefont {V.}~\bibnamefont
  {Zakharov}}, \bibinfo {author} {\bibfnamefont {P.}~\bibnamefont {Guyenne}},
  \bibinfo {author} {\bibfnamefont {A.}~\bibnamefont {Pushkarev}},\ and\
  \bibinfo {author} {\bibfnamefont {F.}~\bibnamefont {Dias}},\ }\bibfield
  {title} {\bibinfo {title} {Wave turbulence in one-dimensional models},\
  }\href {https://doi.org/https://doi.org/10.1016/S0167-2789(01)00194-4}
  {\bibfield  {journal} {\bibinfo  {journal} {Physica D: Nonlinear Phenomena}\
  }\textbf {\bibinfo {volume} {152-153}},\ \bibinfo {pages} {573} (\bibinfo
  {year} {2001})},\ \bibinfo {note} {advances in Nonlinear Mathematics and
  Science: A Special Issue to Honor Vladimir Zakharov}\BibitemShut {NoStop}%
\bibitem [{\citenamefont {Kuksin}(1997)}]{kuksin_turbulence}%
  \BibitemOpen
  \bibfield  {author} {\bibinfo {author} {\bibfnamefont {S.~B.}\ \bibnamefont
  {Kuksin}},\ }\bibfield  {title} {\bibinfo {title} {On turbulence in nonlinear
  {S}chr\"odinger equations},\ }\href {https://doi.org/10.1007/s000390050026}
  {\bibfield  {journal} {\bibinfo  {journal} {Geom. Funct. Anal.}\ }\textbf
  {\bibinfo {volume} {7}},\ \bibinfo {pages} {783} (\bibinfo {year}
  {1997})}\BibitemShut {NoStop}%
\bibitem [{\citenamefont {Colliander}\ \emph {et~al.}(2010)\citenamefont
  {Colliander}, \citenamefont {Keel}, \citenamefont {Staffilani}, \citenamefont
  {Takaoka},\ and\ \citenamefont {Tao}}]{Colliander_KSTT_growth_sobolev_nls}%
  \BibitemOpen
  \bibfield  {author} {\bibinfo {author} {\bibfnamefont {J.}~\bibnamefont
  {Colliander}}, \bibinfo {author} {\bibfnamefont {M.}~\bibnamefont {Keel}},
  \bibinfo {author} {\bibfnamefont {G.}~\bibnamefont {Staffilani}}, \bibinfo
  {author} {\bibfnamefont {H.}~\bibnamefont {Takaoka}},\ and\ \bibinfo {author}
  {\bibfnamefont {T.}~\bibnamefont {Tao}},\ }\bibfield  {title} {\bibinfo
  {title} {Transfer of energy to high frequencies in the cubic defocusing
  nonlinear {S}chr\"odinger equation},\ }\href
  {https://doi.org/10.1007/s00222-010-0242-2} {\bibfield  {journal} {\bibinfo
  {journal} {Invent. Math.}\ }\textbf {\bibinfo {volume} {181}},\ \bibinfo
  {pages} {39} (\bibinfo {year} {2010})}\BibitemShut {NoStop}%
\bibitem [{\citenamefont {G\'erard}\ and\ \citenamefont
  {Grellier}(2010)}]{Gerard_Grellier_cubic_szego_equation}%
  \BibitemOpen
  \bibfield  {author} {\bibinfo {author} {\bibfnamefont {P.}~\bibnamefont
  {G\'erard}}\ and\ \bibinfo {author} {\bibfnamefont {S.}~\bibnamefont
  {Grellier}},\ }\bibfield  {title} {\bibinfo {title} {The cubic {S}zeg{\"o}
  equation},\ }\href {https://doi.org/10.24033/asens.2133} {\bibfield
  {journal} {\bibinfo  {journal} {Ann. Sci. \'Ec. Norm. Sup\'er. (4)}\ }\textbf
  {\bibinfo {volume} {43}},\ \bibinfo {pages} {761} (\bibinfo {year}
  {2010})}\BibitemShut {NoStop}%
\bibitem [{\citenamefont {Maliborski}(2014)}]{Maliborski_thesis}%
  \BibitemOpen
  \bibfield  {author} {\bibinfo {author} {\bibfnamefont {M.}~\bibnamefont
  {Maliborski}},\ }\emph {\bibinfo {title} {{Dynamics of Nonlinear Waves on
  Bounded Domains}}},\ \href@noop {} {Ph.D. thesis},\ \bibinfo  {school}
  {Jagiellonian U.} (\bibinfo {year} {2014}),\ \Eprint
  {https://arxiv.org/abs/1603.00935} {arXiv:1603.00935 [gr-qc]} \BibitemShut
  {NoStop}%
\bibitem [{\citenamefont {Bizon}\ and\ \citenamefont
  {Rostworowski}(2011)}]{Bizon_Rot_Instab_AdS}%
  \BibitemOpen
  \bibfield  {author} {\bibinfo {author} {\bibfnamefont {P.}~\bibnamefont
  {Bizon}}\ and\ \bibinfo {author} {\bibfnamefont {A.}~\bibnamefont
  {Rostworowski}},\ }\bibfield  {title} {\bibinfo {title} {{On weakly turbulent
  instability of anti-de Sitter space}},\ }\href
  {https://doi.org/10.1103/PhysRevLett.107.031102} {\bibfield  {journal}
  {\bibinfo  {journal} {Phys. Rev. Lett.}\ }\textbf {\bibinfo {volume} {107}},\
  \bibinfo {pages} {031102} (\bibinfo {year} {2011})},\ \Eprint
  {https://arxiv.org/abs/1104.3702} {arXiv:1104.3702 [gr-qc]} \BibitemShut
  {NoStop}%
\bibitem [{\citenamefont
  {Maliborski}(2012)}]{Maliborski_instability_minkowski_cavity}%
  \BibitemOpen
  \bibfield  {author} {\bibinfo {author} {\bibfnamefont {M.}~\bibnamefont
  {Maliborski}},\ }\bibfield  {title} {\bibinfo {title} {{Instability of Flat
  Space Enclosed in a Cavity}},\ }\href
  {https://doi.org/10.1103/PhysRevLett.109.221101} {\bibfield  {journal}
  {\bibinfo  {journal} {Phys. Rev. Lett.}\ }\textbf {\bibinfo {volume} {109}},\
  \bibinfo {pages} {221101} (\bibinfo {year} {2012})},\ \Eprint
  {https://arxiv.org/abs/1208.2934} {arXiv:1208.2934 [gr-qc]} \BibitemShut
  {NoStop}%
\bibitem [{\citenamefont {Okawa}\ \emph {et~al.}(2014)\citenamefont {Okawa},
  \citenamefont {Cardoso},\ and\ \citenamefont
  {Pani}}]{Okawa_Cardoso_Pani_cavity_problem}%
  \BibitemOpen
  \bibfield  {author} {\bibinfo {author} {\bibfnamefont {H.}~\bibnamefont
  {Okawa}}, \bibinfo {author} {\bibfnamefont {V.}~\bibnamefont {Cardoso}},\
  and\ \bibinfo {author} {\bibfnamefont {P.}~\bibnamefont {Pani}},\ }\bibfield
  {title} {\bibinfo {title} {{Study of the nonlinear instability of confined
  geometries}},\ }\href {https://doi.org/10.1103/PhysRevD.90.104032} {\bibfield
   {journal} {\bibinfo  {journal} {Phys. Rev. D}\ }\textbf {\bibinfo {volume}
  {90}},\ \bibinfo {pages} {104032} (\bibinfo {year} {2014})},\ \Eprint
  {https://arxiv.org/abs/1409.0533} {arXiv:1409.0533 [gr-qc]} \BibitemShut
  {NoStop}%
\bibitem [{\citenamefont {Dimitrakopoulos}\ \emph {et~al.}(2015)\citenamefont
  {Dimitrakopoulos}, \citenamefont {Freivogel}, \citenamefont {Lippert},\ and\
  \citenamefont {Yang}}]{Dimitrakopoulos_et_al_physical_space_analysis_ads}%
  \BibitemOpen
  \bibfield  {author} {\bibinfo {author} {\bibfnamefont {F.~V.}\ \bibnamefont
  {Dimitrakopoulos}}, \bibinfo {author} {\bibfnamefont {B.}~\bibnamefont
  {Freivogel}}, \bibinfo {author} {\bibfnamefont {M.}~\bibnamefont {Lippert}},\
  and\ \bibinfo {author} {\bibfnamefont {I.-S.}\ \bibnamefont {Yang}},\
  }\bibfield  {title} {\bibinfo {title} {{Position space analysis of the AdS
  (in)stability problem}},\ }\href {https://doi.org/10.1007/JHEP08(2015)077}
  {\bibfield  {journal} {\bibinfo  {journal} {JHEP}\ }\textbf {\bibinfo
  {volume} {08}},\ \bibinfo {pages} {077}},\ \Eprint
  {https://arxiv.org/abs/1410.1880} {arXiv:1410.1880 [hep-th]} \BibitemShut
  {NoStop}%
\bibitem [{\citenamefont {Balasubramanian}\ \emph {et~al.}(2014)\citenamefont
  {Balasubramanian}, \citenamefont {Buchel}, \citenamefont {Green},
  \citenamefont {Lehner},\ and\ \citenamefont
  {Liebling}}]{Balasubramanian_et_al_stability_ads}%
  \BibitemOpen
  \bibfield  {author} {\bibinfo {author} {\bibfnamefont {V.}~\bibnamefont
  {Balasubramanian}}, \bibinfo {author} {\bibfnamefont {A.}~\bibnamefont
  {Buchel}}, \bibinfo {author} {\bibfnamefont {S.~R.}\ \bibnamefont {Green}},
  \bibinfo {author} {\bibfnamefont {L.}~\bibnamefont {Lehner}},\ and\ \bibinfo
  {author} {\bibfnamefont {S.~L.}\ \bibnamefont {Liebling}},\ }\bibfield
  {title} {\bibinfo {title} {{Holographic Thermalization, Stability of
  Anti\textendash{}de Sitter Space, and the Fermi-Pasta-Ulam Paradox}},\ }\href
  {https://doi.org/10.1103/PhysRevLett.113.071601} {\bibfield  {journal}
  {\bibinfo  {journal} {Phys. Rev. Lett.}\ }\textbf {\bibinfo {volume} {113}},\
  \bibinfo {pages} {071601} (\bibinfo {year} {2014})},\ \Eprint
  {https://arxiv.org/abs/1403.6471} {arXiv:1403.6471 [hep-th]} \BibitemShut
  {NoStop}%
\bibitem [{\citenamefont {Craps}\ \emph {et~al.}(2014)\citenamefont {Craps},
  \citenamefont {Evnin},\ and\ \citenamefont
  {Vanhoof}}]{Craps_Evnin_Vanhoof_1}%
  \BibitemOpen
  \bibfield  {author} {\bibinfo {author} {\bibfnamefont {B.}~\bibnamefont
  {Craps}}, \bibinfo {author} {\bibfnamefont {O.}~\bibnamefont {Evnin}},\ and\
  \bibinfo {author} {\bibfnamefont {J.}~\bibnamefont {Vanhoof}},\ }\bibfield
  {title} {\bibinfo {title} {{Renormalization group, secular term resummation
  and AdS (in)stability}},\ }\href {https://doi.org/10.1007/JHEP10(2014)048}
  {\bibfield  {journal} {\bibinfo  {journal} {JHEP}\ }\textbf {\bibinfo
  {volume} {10}},\ \bibinfo {pages} {048}},\ \Eprint
  {https://arxiv.org/abs/1407.6273} {arXiv:1407.6273 [gr-qc]} \BibitemShut
  {NoStop}%
\bibitem [{\citenamefont {Craps}\ \emph {et~al.}(2015)\citenamefont {Craps},
  \citenamefont {Evnin},\ and\ \citenamefont
  {Vanhoof}}]{Craps_Evnin_Vanhoof_2}%
  \BibitemOpen
  \bibfield  {author} {\bibinfo {author} {\bibfnamefont {B.}~\bibnamefont
  {Craps}}, \bibinfo {author} {\bibfnamefont {O.}~\bibnamefont {Evnin}},\ and\
  \bibinfo {author} {\bibfnamefont {J.}~\bibnamefont {Vanhoof}},\ }\bibfield
  {title} {\bibinfo {title} {{Renormalization, averaging, conservation laws and
  AdS (in)stability}},\ }\href {https://doi.org/10.1007/JHEP01(2015)108}
  {\bibfield  {journal} {\bibinfo  {journal} {JHEP}\ }\textbf {\bibinfo
  {volume} {01}},\ \bibinfo {pages} {108}},\ \Eprint
  {https://arxiv.org/abs/1412.3249} {arXiv:1412.3249 [gr-qc]} \BibitemShut
  {NoStop}%
\bibitem [{\citenamefont {Moschidis}(2020)}]{Moschidis_AdS}%
  \BibitemOpen
  \bibfield  {author} {\bibinfo {author} {\bibfnamefont {G.}~\bibnamefont
  {Moschidis}},\ }\bibfield  {title} {\bibinfo {title} {A proof of the
  instability of {A}d{S} for the {E}instein-null dust system with an inner
  mirror},\ }\href {https://doi.org/10.2140/apde.2020.13.1671} {\bibfield
  {journal} {\bibinfo  {journal} {Anal. PDE}\ }\textbf {\bibinfo {volume}
  {13}},\ \bibinfo {pages} {1671} (\bibinfo {year} {2020})}\BibitemShut
  {NoStop}%
\bibitem [{\citenamefont {Moschidis}(2023)}]{Moschidis_AdS_Vlasov}%
  \BibitemOpen
  \bibfield  {author} {\bibinfo {author} {\bibfnamefont {G.}~\bibnamefont
  {Moschidis}},\ }\bibfield  {title} {\bibinfo {title} {A proof of the
  instability of {A}d{S} for the {E}instein-massless {V}lasov system},\ }\href
  {https://doi.org/10.1007/s00222-022-01152-7} {\bibfield  {journal} {\bibinfo
  {journal} {Invent. Math.}\ }\textbf {\bibinfo {volume} {231}},\ \bibinfo
  {pages} {467} (\bibinfo {year} {2023})}\BibitemShut {NoStop}%
\bibitem [{\citenamefont {Dafermos}\ and\ \citenamefont
  {Holzegel}(2006)}]{Daf_Holz_Conj_AdS_instab}%
  \BibitemOpen
  \bibfield  {author} {\bibinfo {author} {\bibfnamefont {M.}~\bibnamefont
  {Dafermos}}\ and\ \bibinfo {author} {\bibfnamefont {G.}~\bibnamefont
  {Holzegel}},\ }\bibfield  {title} {\bibinfo {title} {{Dynamic instability of
  solitons in 4 + 1-dimensional gravity with negative cosmological constant}},\
  }\href@noop {} {\bibfield  {journal} {\bibinfo  {journal} {Unpublished}\ }
  (\bibinfo {year} {2006})}\BibitemShut {NoStop}%
\bibitem [{\citenamefont {Dafermos}(2006)}]{Daf_talk_AdS}%
  \BibitemOpen
  \bibfield  {author} {\bibinfo {author} {\bibfnamefont {M.}~\bibnamefont
  {Dafermos}},\ }\bibfield  {title} {\bibinfo {title} {The black hole stability
  problem},\ }\href@noop {} {\bibfield  {journal} {\bibinfo  {journal}
  {http://www.newton.ac.uk/webseminars/pg+ws /2006/gmx/1010/dafermos/}\ }
  (\bibinfo {year} {2006})}\BibitemShut {NoStop}%
\bibitem [{\citenamefont {Anderson}(2006)}]{Anderson_AdS}%
  \BibitemOpen
  \bibfield  {author} {\bibinfo {author} {\bibfnamefont {M.~T.}\ \bibnamefont
  {Anderson}},\ }\bibfield  {title} {\bibinfo {title} {{On the uniqueness and
  global dynamics of AdS spacetimes}},\ }\href
  {https://doi.org/10.1088/0264-9381/23/23/021} {\bibfield  {journal} {\bibinfo
   {journal} {Class. Quant. Grav.}\ }\textbf {\bibinfo {volume} {23}},\
  \bibinfo {pages} {6935} (\bibinfo {year} {2006})},\ \Eprint
  {https://arxiv.org/abs/hep-th/0605293} {arXiv:hep-th/0605293 [hep-th]}
  \BibitemShut {NoStop}%
\bibitem [{\citenamefont {Horowitz}\ and\ \citenamefont
  {Santos}(2015)}]{Horowitz_Santos_geons_instability_ads}%
  \BibitemOpen
  \bibfield  {author} {\bibinfo {author} {\bibfnamefont {G.~T.}\ \bibnamefont
  {Horowitz}}\ and\ \bibinfo {author} {\bibfnamefont {J.~E.}\ \bibnamefont
  {Santos}},\ }\bibfield  {title} {\bibinfo {title} {{Geons and the Instability
  of Anti-de Sitter Spacetime}},\ }\href
  {https://doi.org/10.4310/SDG.2015.v20.n1.a13} {\bibfield  {journal} {\bibinfo
   {journal} {Surveys Diff. Geom.}\ }\textbf {\bibinfo {volume} {20}},\
  \bibinfo {pages} {321} (\bibinfo {year} {2015})},\ \Eprint
  {https://arxiv.org/abs/1408.5906} {arXiv:1408.5906 [gr-qc]} \BibitemShut
  {NoStop}%
\bibitem [{\citenamefont {Dias}\ and\ \citenamefont
  {Santos}(2016)}]{Dias_Santos_ads_instability_beyond_spherical_symmetry}%
  \BibitemOpen
  \bibfield  {author} {\bibinfo {author} {\bibfnamefont {O.~J.~C.}\
  \bibnamefont {Dias}}\ and\ \bibinfo {author} {\bibfnamefont {J.~E.}\
  \bibnamefont {Santos}},\ }\bibfield  {title} {\bibinfo {title} {{AdS
  nonlinear instability: moving beyond spherical symmetry}},\ }\href
  {https://doi.org/10.1088/0264-9381/33/23/23LT01} {\bibfield  {journal}
  {\bibinfo  {journal} {Class. Quant. Grav.}\ }\textbf {\bibinfo {volume}
  {33}},\ \bibinfo {pages} {23LT01} (\bibinfo {year} {2016})},\ \Eprint
  {https://arxiv.org/abs/1602.03890} {arXiv:1602.03890 [hep-th]} \BibitemShut
  {NoStop}%
\bibitem [{\citenamefont
  {Rostworowski}(2017)}]{Rostworowski_time_periodic_perturbations_ads}%
  \BibitemOpen
  \bibfield  {author} {\bibinfo {author} {\bibfnamefont {A.}~\bibnamefont
  {Rostworowski}},\ }\bibfield  {title} {\bibinfo {title} {{Higher order
  perturbations of anti\textendash{}de Sitter space and time-periodic solutions
  of vacuum Einstein equations}},\ }\href
  {https://doi.org/10.1103/PhysRevD.95.124043} {\bibfield  {journal} {\bibinfo
  {journal} {Phys. Rev. D}\ }\textbf {\bibinfo {volume} {95}},\ \bibinfo
  {pages} {124043} (\bibinfo {year} {2017})},\ \Eprint
  {https://arxiv.org/abs/1701.07804} {arXiv:1701.07804 [gr-qc]} \BibitemShut
  {NoStop}%
\bibitem [{\citenamefont {Holzegel}\ and\ \citenamefont
  {Smulevici}(2013)}]{Decay_KG_Kerr-AdS}%
  \BibitemOpen
  \bibfield  {author} {\bibinfo {author} {\bibfnamefont {G.}~\bibnamefont
  {Holzegel}}\ and\ \bibinfo {author} {\bibfnamefont {J.}~\bibnamefont
  {Smulevici}},\ }\bibfield  {title} {\bibinfo {title} {Decay properties of
  {K}lein-{G}ordon fields on {K}err-{A}d{S} spacetimes},\ }\href
  {https://doi.org/10.1002/cpa.21470} {\bibfield  {journal} {\bibinfo
  {journal} {Comm. Pure Appl. Math.}\ }\textbf {\bibinfo {volume} {66}},\
  \bibinfo {pages} {1751} (\bibinfo {year} {2013})}\BibitemShut {NoStop}%
\bibitem [{\citenamefont {Graf}\ and\ \citenamefont
  {Holzegel}(2022)}]{Graf_Holzegel_mode_stability}%
  \BibitemOpen
  \bibfield  {author} {\bibinfo {author} {\bibfnamefont {O.}~\bibnamefont
  {Graf}}\ and\ \bibinfo {author} {\bibfnamefont {G.}~\bibnamefont
  {Holzegel}},\ }\bibfield  {title} {\bibinfo {title} {Mode stability for the
  {T}eukolsky equations on {K}err-anti-de {S}itter spacetimes},\ }\bibfield
  {journal} {\bibinfo  {journal} {arXiv e-print 2205.02801}\ }\href
  {https://doi.org/10.48550/ARXIV.2205.02801} {10.48550/ARXIV.2205.02801}
  (\bibinfo {year} {2022})\BibitemShut {NoStop}%
\bibitem [{\citenamefont {Graf}\ and\ \citenamefont
  {Holzegel}(2024{\natexlab{a}})}]{Graf:2024nni}%
  \BibitemOpen
  \bibfield  {author} {\bibinfo {author} {\bibfnamefont {O.}~\bibnamefont
  {Graf}}\ and\ \bibinfo {author} {\bibfnamefont {G.}~\bibnamefont
  {Holzegel}},\ }\bibfield  {title} {\bibinfo {title} {{Linear Stability of
  Schwarzschild-Anti-de Sitter spacetimes I: The system of gravitational
  perturbations}},\ }\href@noop {} {\  (\bibinfo {year}
  {2024}{\natexlab{a}})},\ \Eprint {https://arxiv.org/abs/2408.02251}
  {arXiv:2408.02251 [gr-qc]} \BibitemShut {NoStop}%
\bibitem [{\citenamefont {Graf}\ and\ \citenamefont
  {Holzegel}(2024{\natexlab{b}})}]{Graf:2024mui}%
  \BibitemOpen
  \bibfield  {author} {\bibinfo {author} {\bibfnamefont {O.}~\bibnamefont
  {Graf}}\ and\ \bibinfo {author} {\bibfnamefont {G.}~\bibnamefont
  {Holzegel}},\ }\bibfield  {title} {\bibinfo {title} {{Linear Stability of
  Schwarzschild-Anti-de Sitter spacetimes II: Logarithmic decay of solutions to
  the Teukolsky system}},\ }\href@noop {} {\  (\bibinfo {year}
  {2024}{\natexlab{b}})},\ \Eprint {https://arxiv.org/abs/2408.02252}
  {arXiv:2408.02252 [gr-qc]} \BibitemShut {NoStop}%
\bibitem [{\citenamefont {Graf}\ and\ \citenamefont
  {Holzegel}(2024{\natexlab{c}})}]{Graf:2024yug}%
  \BibitemOpen
  \bibfield  {author} {\bibinfo {author} {\bibfnamefont {O.}~\bibnamefont
  {Graf}}\ and\ \bibinfo {author} {\bibfnamefont {G.}~\bibnamefont
  {Holzegel}},\ }\bibfield  {title} {\bibinfo {title} {{Linear Stability of
  Schwarzschild-Anti-de Sitter spacetimes III: Quasimodes and sharp decay of
  gravitational perturbations}},\ }\href@noop {} {\  (\bibinfo {year}
  {2024}{\natexlab{c}})},\ \Eprint {https://arxiv.org/abs/2410.21994}
  {arXiv:2410.21994 [gr-qc]} \BibitemShut {NoStop}%
\bibitem [{\citenamefont {Dias}\ \emph {et~al.}(2012)\citenamefont {Dias},
  \citenamefont {Horowitz}, \citenamefont {Marolf},\ and\ \citenamefont
  {Santos}}]{Dias_et_al_stability_ads_bhs}%
  \BibitemOpen
  \bibfield  {author} {\bibinfo {author} {\bibfnamefont {O.~J.~C.}\
  \bibnamefont {Dias}}, \bibinfo {author} {\bibfnamefont {G.~T.}\ \bibnamefont
  {Horowitz}}, \bibinfo {author} {\bibfnamefont {D.}~\bibnamefont {Marolf}},\
  and\ \bibinfo {author} {\bibfnamefont {J.~E.}\ \bibnamefont {Santos}},\
  }\bibfield  {title} {\bibinfo {title} {{On the Nonlinear Stability of
  Asymptotically Anti-de Sitter Solutions}},\ }\href
  {https://doi.org/10.1088/0264-9381/29/23/235019} {\bibfield  {journal}
  {\bibinfo  {journal} {Class. Quant. Grav.}\ }\textbf {\bibinfo {volume}
  {29}},\ \bibinfo {pages} {235019} (\bibinfo {year} {2012})},\ \Eprint
  {https://arxiv.org/abs/1208.5772} {arXiv:1208.5772 [gr-qc]} \BibitemShut
  {NoStop}%
\bibitem [{\citenamefont {Ficek}\ and\ \citenamefont
  {Maliborski}(2024)}]{Ficek_Maliborski_nonlinear_scalar_field_SAdS}%
  \BibitemOpen
  \bibfield  {author} {\bibinfo {author} {\bibfnamefont {F.}~\bibnamefont
  {Ficek}}\ and\ \bibinfo {author} {\bibfnamefont {M.}~\bibnamefont
  {Maliborski}},\ }\bibfield  {title} {\bibinfo {title} {{Dynamics of nonlinear
  scalar field with Robin boundary condition on the
  Schwarzschild\textendash{}anti\textendash{}de~Sitter background}},\ }\href
  {https://doi.org/10.1103/PhysRevD.109.044015} {\bibfield  {journal} {\bibinfo
   {journal} {Phys. Rev. D}\ }\textbf {\bibinfo {volume} {109}},\ \bibinfo
  {pages} {044015} (\bibinfo {year} {2024})},\ \Eprint
  {https://arxiv.org/abs/2312.02760} {arXiv:2312.02760 [gr-qc]} \BibitemShut
  {NoStop}%
\bibitem [{\citenamefont {Figueras}\ and\ \citenamefont
  {Rossi}(2023)}]{Figueras_Rossi_instability_kerr_AdS}%
  \BibitemOpen
  \bibfield  {author} {\bibinfo {author} {\bibfnamefont {P.}~\bibnamefont
  {Figueras}}\ and\ \bibinfo {author} {\bibfnamefont {L.}~\bibnamefont
  {Rossi}},\ }\bibfield  {title} {\bibinfo {title} {{Non-linear instability of
  slowly rotating Kerr-AdS black holes}},\ }\href@noop {} {\  (\bibinfo {year}
  {2023})},\ \Eprint {https://arxiv.org/abs/2311.14167} {arXiv:2311.14167
  [hep-th]} \BibitemShut {NoStop}%
\bibitem [{\citenamefont {Kehle}\ and\ \citenamefont
  {Moschidis}(2023)}]{Kehle_Moschidis_talk_turbulence_SAdS}%
  \BibitemOpen
  \bibfield  {author} {\bibinfo {author} {\bibfnamefont {C.}~\bibnamefont
  {Kehle}}\ and\ \bibinfo {author} {\bibfnamefont {G.}~\bibnamefont
  {Moschidis}},\ }\bibfield  {title} {\bibinfo {title} {{Weak turbulence on
  Schwarzschild-AdS spacetime}},\ }\href@noop {} {\bibfield  {journal}
  {\bibinfo  {journal}
  {\url{https://www.dpmms.cam.ac.uk/~rbdt2/NAGR/NAGR_17_Moschidis.pdf}}\ }
  (\bibinfo {year} {2023})}\BibitemShut {NoStop}%
\bibitem [{\citenamefont {Adams}\ \emph {et~al.}(2014)\citenamefont {Adams},
  \citenamefont {Chesler},\ and\ \citenamefont
  {Liu}}]{Adams_chesler_liu_holographic_turbulence}%
  \BibitemOpen
  \bibfield  {author} {\bibinfo {author} {\bibfnamefont {A.}~\bibnamefont
  {Adams}}, \bibinfo {author} {\bibfnamefont {P.~M.}\ \bibnamefont {Chesler}},\
  and\ \bibinfo {author} {\bibfnamefont {H.}~\bibnamefont {Liu}},\ }\bibfield
  {title} {\bibinfo {title} {{Holographic turbulence}},\ }\href
  {https://doi.org/10.1103/PhysRevLett.112.151602} {\bibfield  {journal}
  {\bibinfo  {journal} {Phys. Rev. Lett.}\ }\textbf {\bibinfo {volume} {112}},\
  \bibinfo {pages} {151602} (\bibinfo {year} {2014})},\ \Eprint
  {https://arxiv.org/abs/1307.7267} {arXiv:1307.7267 [hep-th]} \BibitemShut
  {NoStop}%
\bibitem [{\citenamefont {Yang}\ \emph {et~al.}(2015)\citenamefont {Yang},
  \citenamefont {Zimmerman},\ and\ \citenamefont
  {Lehner}}]{Yang_Zimmerman_Lehner_turbulent_BHs}%
  \BibitemOpen
  \bibfield  {author} {\bibinfo {author} {\bibfnamefont {H.}~\bibnamefont
  {Yang}}, \bibinfo {author} {\bibfnamefont {A.}~\bibnamefont {Zimmerman}},\
  and\ \bibinfo {author} {\bibfnamefont {L.}~\bibnamefont {Lehner}},\
  }\bibfield  {title} {\bibinfo {title} {{Turbulent Black Holes}},\ }\href
  {https://doi.org/10.1103/PhysRevLett.114.081101} {\bibfield  {journal}
  {\bibinfo  {journal} {Phys. Rev. Lett.}\ }\textbf {\bibinfo {volume} {114}},\
  \bibinfo {pages} {081101} (\bibinfo {year} {2015})},\ \Eprint
  {https://arxiv.org/abs/1402.4859} {arXiv:1402.4859 [gr-qc]} \BibitemShut
  {NoStop}%
\end{thebibliography}%

\end{document}